\documentclass[%
 reprint,
 amsmath,amssymb,
 aps,
]{revtex4-2}

\usepackage{graphicx}
\usepackage{dcolumn}
\usepackage{bm}

\usepackage{amssymb}
\usepackage{amsmath}
\usepackage{hyperref}
\usepackage{physics}
\usepackage{booktabs}
\usepackage{esint}
\usepackage{soul}
\usepackage{cancel}
\usepackage{kotex}
\usepackage[margin=1in]{geometry}
\usepackage{array}
\usepackage{siunitx}
\usepackage{subcaption}  
\usepackage{tikz}
\usetikzlibrary{positioning, shapes.geometric}
\begin{document}

\preprint{APS/123-QED}

\title{Computational framework for non-Markovian multi-emitter dynamics beyond the single-excitation limit}

\author{%
  Hyunwoo Choi$^{1}$, 
  Weng Cho Chew$^{2}$, 
  and Dong-Yeop Na$^{1,2,\dagger}$
}

\affiliation{%
  $^{1}$Department of Electrical Engineering, Pohang University of Science and Technology (POSTECH),\\
  77 Cheongam-ro, Nam-gu, Pohang 37673, Gyeongsangbuk-do, Republic of Korea\\
  $^{2}$Elmore Family School of Electrical and Computer Engineering, Purdue University,\\
  610 Purdue Mall, West Lafayette, Indiana 47907, United States of America
}

\email{dyna22@postech.ac.kr}

\date{\today}

\begin{abstract}
While non-Markovian dynamics have been extensively studied in the single-excitation limit to predict non-trivial phenomena, this limit remains a specific idealization. Moving beyond this constraint is essential because optical nonlinearities and phase error accumulation in multi-photon processes make the Markovian approximation fragile. In this article, we present a Green's function-based framework for modeling non-Markovian multi-emitter quantum electrodynamics within the two-excitation manifold. We employ the modified Langevin noise (M-LN) formalism for first-principles modeling of dissipative environments and utilize the emitter-centered mode (ECM) framework to render the extensive degrees of freedom computationally feasible. Unlike conventional approaches that formally integrate out the reservoir, we establish a non-Markovian hierarchy of coupled differential equations by explicitly retaining the photonic amplitudes. Within the present two-excitation hierarchy, the formulation preserves total probability and retains the phase information needed to describe multi-photon interference. As numerical demonstrations, we investigate non-Markovian atom-field interactions in structured semi-infinite waveguide environments. We first consider a homogeneous waveguide as a representative baseline with enhanced Bell-state fidelity in selected configurations. Second, we investigate the collective decay of symmetric Dicke states in a waveguide with an embedded lossy dielectric slab, revealing the selective stabilization and delayed excitation transfer induced by the structured reservoir. Third, we analyze the entanglement dynamics within the same environment, highlighting the emergence of entanglement sudden birth and oscillatory revivals. In principle, the formulation is applicable to electromagnetic environments for which the required dyadic Green’s function can be obtained numerically. This makes it a versatile tool for investigating complex non-Markovian multi-photon phenomena across diverse dissipative photonic platforms, well beyond the single-excitation limit.
\end{abstract}
\maketitle

\section{Introduction}
Structured photonic environments such as plasmonic nanostructures, optical cavities, and photonic waveguides~\cite{Lodahl_RevModPhy_nanostructure_2015,Tame2013,Dowran_Laserphotonics_nanocavity_2025,Ye_PRL_Coldatom_2023,Sheremet_RMP_waveguideQED_2023,Mahmoodian_PRL_waveguide_2018}, have established photons as promising quantum sources, enabling light–matter interactions far beyond those achievable in free space~\cite{Slussarenko_APL_Photonic_2019,Thomas2022,psiquantum_2025}.
Such platforms enable effective optical nonlinearities through the saturation of quantum emitters, converting otherwise non-interacting bosons into strongly correlated quantum excitations. However, this compels the environment to act as more than an innocuous background, introducing complex effects such as dispersion, dissipation, and re-interaction~\cite{Gonz_PRA_nonMarkovian_2017,de_RMP_nonMarkovian_2018}.
In this situation, a significant gap hinders the translation between idealized theory and the implementation of many-body quantum systems. While theoretical frameworks continue to push the boundaries of large-scale entanglement, realistic systems often fall short of these predictions. Such processes are inextricably coupled to environmental effects, which not only degrade coherence but also fundamentally alter the underlying dynamics through non-Markovian memory effects~\cite{Scully1997quantum,Mandel1995optical}.
Conventional studies of open quantum systems have focused on the universal dynamical laws using canonical interaction models, typically through master equation~\cite{Vidal_PRL_fewmode_2021,Franke_PRL_QNMQuantization_2019} or stochastic Schrödinger equation~\cite{Bouten2004} formalisms. While these methods provide profound insights into dissipative processes, they often rely on generic bath models rather than account for the fine-grained features of the electromagnetic environment. Consequently, the specific details of realistic systems such as material dispersion, non-local losses, and the spatial inhomogeneity of structured photonic environments are often simplified into phenomenological kernels. 

In parallel, advances in computational electromagnetics (CEM)~\cite{Jin2014FEM,taflove_fdtd} have enabled the high-fidelity modeling of electromagnetic systems involving complex geometries. By numerically solving Maxwell’s equations either in the time or frequency domain, these methods provide a rigorous and detailed description of scattering, absorption, and boundary-induced interference within inhomogeneous, dispersive and lossy media. By leveraging these capabilities, the application of CEM to quantum electrodynamic (QED) frameworks has been steadily pursued to bridge the gap between idealized models and realistic EM environments through the dyadic Green’s function or advanced modal expansion techniques. Recently, these efforts have consolidated into the specialized field of computational quantum electromagnetics (CQEM)~\cite{Chew2024, Moon2025CQEM,elkin_CQEMreview_2025}, providing a rigorous platform for analyzing quantum dynamics within nanophotonic structures of arbitrary complexity~\cite{Roth2021JMMCT,2022RothJMMCT,Na2021cqNMD,Na2023quantumEMLossy,Moon2024TAP,Forestiere2022PRA,Forestiere2023PRA,Ryu2023Dicke,Ryu2023MPS,huang2025VIE,seo_nano_2025}.

In line with these efforts, we have recently developed a numerical framework for non-Markovian atom-field interactions in structured environments within the single-excitation manifold~\cite{Choi_PRApplied, Choi_FDTD_QE}. These studies leveraged the finite element method (FEM) and the finite-difference time-domain (FDTD) method to bridge the gap between rigorous EM modeling and quantum dynamical evolution. However, this regime remains  limited to linear response, where at most single quanta of excitation is present in the combined atom–field system. 
Extending the description to the two-excitation manifold constitutes the minimal step beyond the linear regime, where genuine photon–photon correlations and saturation-induced nonlinearities first emerge. In this regime, the system dynamics involve a nontrivial coupling between doubly excited atomic states, intermediate one-photon sectors, and the continuum of two-photon states, leading to rich phenomena such as bound-state formation, correlated emission, and entanglement generation~\cite{PRA_twophoton_2014,PRA_multiphoton_2008,Reiserer_gate_nature_2014}.

Unfortunately, exact treatments of two-excitation dynamics remain highly challenging. While a number of studies have explored dynamics beyond the single-excitation regime, most existing approaches rely on formulations that are closely tied to specific model systems. In particular, non-perturbative treatments of two-photon processes have been developed primarily in the context of waveguide QED, where early works demonstrated strongly correlated two-photon transport and were later extended to few-emitter configurations~\cite{Shen_PRL_twophoton_2007,Roy_PRL_twophoton_2011}. More recent efforts have incorporated non-Markovian effects in such systems, including giant-atom configurations~\cite{Gu_PRA_twophoton_2024}, further elucidating the role of retardation and photon-mediated correlations~\cite{PRA_twophoton_2025}.
However, these approaches are typically not formulated to retain a closed dyadic-Green’s function description, which limits their ability to incorporate material dispersion and dissipation in a first-principles and self-consistent manner when moving beyond idealized settings.

In this work, we present a Green's function-based computational framework for modeling non-Markovian multi-emitter QED within the two-excitation manifold. We employ the modified Langevin noise (M-LN) formalism~\cite{DiStefano2001ModeExpansion,Drezet2017QuantizingPolaritons,Na2023quantumEMLossy,ciatonni_MLN_2024,ciattoni_MLN_2026}, a generalized theory of macroscopic QED~\cite{Gruner1996QEDEvanescent,Dung2000QEDLocalized,Scheel2008MacroscopicQED}, to provide a first-principles description of open and dissipative EM environments. By incorporating both boundary-assisted (BA) and medium-assisted (MA) continuum contributions, our approach ensures theoretical self-consistency and preserves the fluctuation-dissipation theorem while accounting for not only medium dissipation but also radiation leakage. This is integrated with the emitter-centered mode (ECM) framework~\cite{ECM_nanophotonics_2021,Monica2022ECM,Forestire_nanophotonics_2025,Forestire_PRA_2025,miano_ECM_PRA_2026}, which renders the extensive degrees of freedom (DoFs) of the polaritonic reservoir computationally feasible. We develop an ECM-based two-quantum hierarchy that resolves the doubly excited atomic sector, the intermediate one-photon sector, and the pure two-photon sector. The core of this formulation lies in the determination of the environment’s Green’s function spectrum at the emitter positions. Once the Green’s-function spectrum at the emitter positions is obtained, the subsequent dynamics can be solved within the same emitter-centered hierarchy, irrespective of how that spectrum was generated. This modularity allows the framework to incorporate Green's function data from complex EM environments, thereby enabling an accurate description of many-body interactions in structured photonic media.

To demonstrate the scope of the present framework, we investigate structured one-dimensional waveguide-QED environments through three distinct cases.
First, a homogeneous semi-infinite waveguide with a perfect mirror~\cite{lambda_advanced_2025} is employed as a representative baseline to ensure the numerical consistency of our formulation. Second, we introduce an embedded lossy dielectric slab to investigate the selective stabilization of symmetric Dicke states. Third, within the same slab environment, we analyze entanglement dynamics, highlighting non-Markovian effects such as sudden birth and oscillatory revivals~\cite{PRA_ESBESD_2009}. These examples illustrate how reservoir memory, dissipation, and multi-photon interference jointly influence collective quantum dynamics beyond the single-excitation approximation.

The rest of this paper is organized as follows. Section II establishes the theoretical foundation, detailing the first-principles quantization of dissipative environments via M-LN and ECM formalisms. Section III formulates the non-Markovian hierarchy for multi-emitter dynamics within the two-excitation manifold. Numerical demonstrations in structured 1D waveguide-QED environments are provided in Section IV, followed by a summary of key findings and future research directions in Section V.
\section{Theory}
\subsection{Hamiltonian for multiple emitters}
Consider $N$ emitters at positions $\{\mathbf{R}_a\}$ interacting with the quantized EM field. Within the electric-dipole approximation and a two-level description of each emitter, we use the Hamiltonian as~\cite{Scully1997quantum} 
\begin{align}
\hat{H}_{\mathrm{tot}} &= \hat{H}_{\mathrm{atoms}} + \hat{H}_{\mathrm{field}} + \hat{H}_{\mathrm{int}},
\label{eq:H_eff_main}
\end{align}
where
\begin{align}
\hat{H}_{\mathrm{atoms}} &= \sum_{p=1}^{N} \hbar \omega_a \hat{\sigma}_{+}^{(p)} \hat{\sigma}_{-}^{(p)}, \\
\hat{H}_{\mathrm{field}} &= \int d^3 \mathbf{r} \left[ \frac{\epsilon_0}{2} \hat{\mathbf{E}}_{}^2(\mathbf{r}) + \frac{1}{2\mu_0} \hat{\mathbf{B}}^2(\mathbf{r}) \right], \label{eq:Hamiltonian_em} \\
\hat{H}_{\mathrm{int}} &= -\sum_{a=1}^{N} \hat{\mathbf{d}}_a \cdot \hat{\mathbf{E}}(\mathbf{R}_a), \label{eq:H_int_sum_dE}
\end{align}
Here, the dipole operator projected onto the two-level subspace is expressed as
\begin{align}
\hat{\mathbf{d}}_a = \mathbf{d}_a \hat{\sigma}_{-}^{(a)} + \mathbf{d}_a^* \hat{\sigma}_{+}^{(a)}.
\end{align}
The EM field energy in Eq.~\eqref{eq:Hamiltonian_em} is quantized by introducing bosonic creation and annihilation operators. In the conventional multimode Tavis-Cummings model, which describes the closed and lossless limit, the field is expanded in terms of discrete cavity modes as
\begin{align}
    \hat{H}_{\mathrm{field}}^{\mathrm{closed}} = \sum_{k} \hbar \omega_k \hat{a}_k^\dagger \hat{a}_k,
    \label{eq:H_field_closed}
\end{align}
where $\hat{a}_k$ ($\hat{a}_k^\dagger$) denotes the annihilation (creation) operator for the $k$-th resonant normal mode. In general open and dissipative systems, however, the electromagnetic field must be considered coupled to a continuum reservoir. In this context, the fundamental excitations are polaritonic in nature, representing dressed states of the radiation and the lossy medium. While certain phenomenological approaches allow for quantization without an explicit reservoir, a more rigorous treatment requires quantizing the field Hamiltonian in a way that fully incorporates the reservoir's DoFs. This is typically expressed as a continuum of harmonic oscillators as
\begin{align}
    \hat{H}_{\mathrm{field}}^{\mathrm{open}} =  \int_{0}^{\infty}d\omega \int_\mathcal{D_\lambda} d\lambda \, \hbar\omega \, \hat{\mathbf{f}}_{\omega,\lambda}^{\dagger}\, \hat{\mathbf{f}}_{\omega,\lambda}
    \label{eq:H_field_open}
\end{align}
where $\hat{\mathbf{f}}_{\omega,\lambda}$ is the bosonic field operator representing the polariton at frequency $\omega$. Here, $\lambda$ characterizes the continuous set of DoFs of the reservoir over the domain $\mathcal{D}_\lambda$.

\subsection{Modified Langevin noise formalism}
While several methods exist for diagonalizing the coupled EM-reservoir polaritons~\cite{Na2021cqNMD, Huttner1992Quantization}, we adopt the M-LN formalism. This framework has been recently developed by incorporating previously missing terms into the standard macroscopic QED approach, thereby providing a more rigorous description of structured dissipative environments.
In this formalism, two distinct polaritons are introduced, each originating from a separate reservoir that accounts for (i) radiative loss and (ii) medium loss, respectively. While including a radiative reservoir in the definition of a polariton may appear unconventional, we use the term consistently with recent M-LN formalisms to represent the total energy quanta of the field with reservoir. The following positive-frequency part of the electric field operator can be written as 
\begin{align}
\hat{\mathbf{E}}^{(+)}(\mathbf{r})
=
\hat{\mathbf{E}}^{(+)}_{\mathrm{BA}}(\mathbf{r})
+
\hat{\mathbf{E}}^{(+)}_{\mathrm{MA}}(\mathbf{r}),
\label{eq:E_BA--MA_decomp}
\end{align}
where
\begin{align}
\hat{\mathbf{E}}^{(+)}_{\mathrm{BA}}(\mathbf{r})
&= \int_{0}^{\infty} d\omega \sum_{s=1,2} \int_{\mathbb{S}^2} k^2 d\Omega \nonumber \\
&\quad \times \mathbf{E}^{(\mathrm{BA})}(\mathbf{r};\Omega,s,\omega) \hat{a}^{(\mathrm{BA})}(\Omega,s,\omega), \\[2mm]
\hat{\mathbf{E}}^{(+)}_{\mathrm{MA}}(\mathbf{r})
&= \int_{0}^{\infty} d\omega \sum_{\xi=x,y,z} \int_{V_{m}} d\mathbf{r}' \nonumber \\
&\quad \times \mathbf{E}^{(\mathrm{MA})}(\mathbf{r};\mathbf{r}',\xi,\omega) \hat{a}^{(\mathrm{MA})}(\mathbf{r}',\xi,\omega).
\end{align}
We define the BA and MA modes through the spatial profiles $\mathbf{E}^{(\mathrm{BA})}$ and $\mathbf{E}^{(\mathrm{MA})}$, with their corresponding annihilation operators $\hat{a}^{(\mathrm{BA})}$ and $\hat{a}^{(\mathrm{MA})}$ representing the boundary-assisted and medium-assisted contributions to the field operator.
The BA and MA modes are given by~\cite{Na2023quantumEMLossy,ciatonni_MLN_2024}
\begin{align}
    \mathbf{E}_{\text{B}}(\mathbf{r};\Omega,s,\omega) &= \mathcal{A}(\omega) \lim_{R \to \infty} \Big[ 4\pi R e^{-ikR} \nonumber \\ 
    &\quad \times \overline{\mathbf{G}}_{E}(\mathbf{r}, R\hat{\mathbf{n}}; \omega) \cdot \hat{\mathbf{e}}_{\Omega, s} \Big] \\
    \mathbf{E}_{\text{M}}(\mathbf{r};\mathbf{r}',\xi,\omega) &= \mathcal{A}(\omega)\sqrt{\mathrm{Im}[\epsilon(\mathbf{r}',\omega)]}\, \overline{\mathbf{G}}_{E}(\mathbf{r}, \mathbf{r}'; \omega) \cdot \hat{\mathbf{e}}_{\xi} \label{eqn:E_m_GF}
\end{align}
Here, the normalization constant is defined as $\mathcal{A}(\omega) = \sqrt{\hbar \mu_0 \omega^2/\pi}$, while $\hat{\mathbf{n}}$ denotes the direction vector corresponding to $\Omega$. The term $4\pi R e^{-ikR}$ is included to renormalize the amplitude decay and phase accumulation characteristic of spherical waves in the far field.
To simplify the notation, we combine both the BA and MA contributions into a unified continuum, labeled by a generalized index $\lambda$ that spans both branches at each frequency $\omega$.
Accordingly, the positive-frequency electric-field operator is expanded in terms of the unified BA and MA modes as
\begin{align}
\hat{\mathbf{E}}^{(+)}(\mathbf{r})
=\int_0^\infty\,d\omega\,  \hat{\mathbf{E}}^{(+)}(\mathbf{r,\omega})
\end{align}
where,
\begin{align}
\hat{\mathbf{E}}^{(+)}(\mathbf{r,\omega})
=
\int_{\mathcal{D}_{\lambda}} d\lambda\;
\mathbf{E}_{\omega,\lambda}(\mathbf{r})\,
\hat{a}_{\omega,\lambda},
\label{eq:Epos_multi_combined}
\end{align}
The M-LN framework exhaustively maps every dissipation channel within the system, spanning all spatial points within the lossy material ($V_m$) and all radiative directions at the infinite boundary ($S_\infty$). By invoking the fluctuation-dissipation theorem (FDT), this approach identifies the fluctuation reactions that are exactly reciprocal to these dissipative channels, thereby ensuring the Hermiticity of the total system Hamiltonian. The following BA-MA transverse modal completeness relation, which we have recently validated numerically~\cite{Choi_FDTD_QE}, ensures that the BA-MA modes account for all radiative and dissipative channels
\begin{align}
&\mathcal{A}(\omega)^2\,
\Im\!\big[
\overline{\mathbf{G}}_E(\mathbf{r};\mathbf{r}',\omega)
\big]
=
\int_{\mathcal{D}_{\lambda}} d\lambda\,
\mathbf{E}_{\omega,\lambda}(\mathbf{r})
\otimes
\mathbf{E}^{*}_{\omega,\lambda}(\mathbf{r}').
\label{eq:BA--MA_completeness_TMC}
\end{align}
Finally, the field Hamiltonian in Eq.~\eqref{eq:Hamiltonian_em} can be quantized by decomposing the contributions into BA and MA polaritonic branches, as follows
\begin{align}
\hat{H}_{\mathrm{field}} 
= \int_{0}^{\infty} d\omega\int_{\mathcal{D}_\lambda} d\lambda\, \hbar \omega \hat{a}_{\omega, \lambda}^\dagger \hat{a}_{\omega, \lambda}
\label{eq:H_field_final_sum}
\end{align}

\subsection{M-LN formalism with emitter-centered mode framework}
Substituting the BA--MA field operator~\eqref{eq:E_BA--MA_decomp} into the interaction Hamiltonian~\eqref{eq:H_int_sum_dE} yields
\begin{align}
\hat{H}_{\mathrm{int}}
&=
- \sum_{a=1}^{N}
\hat{\mathbf{d}}_a \cdot
\left[\hat{\mathbf{E}}_{(\mathrm{BA})}(\mathbf{R}_a)+\hat{\mathbf{E}}_{(\mathrm{MA})}(\mathbf{R}_a)\right].
\end{align}
Crucially, the interaction Hamiltonian governing the system dynamics involves the electric field operator solely at the emitter position, $\hat{\mathbf{E}}(\mathbf{R}_a)$. This implies that only the components of the BA and MA modes parallel to the electric field operator at the atomic positions $\mathbf{R}_a$ contribute to the interaction Hamiltonian. The field components projected along the atomic dipoles at the emitter positions are given by
\begin{align}
    \hat{E}^{(+)}_{a}(\omega) 
    &\equiv \mathbf{n}_a \cdot \hat{\mathbf{E}}^{(+)}(\mathbf{R}_a, \omega) \nonumber \\
    &= \int_{\mathcal{D}_\lambda} d\lambda \, \left( \mathbf{n}_a \cdot \mathbf{E}_{\omega, \lambda}(\mathbf{R}_a) \right) \hat{a}_{\omega, \lambda},
\end{align}
The commutation relation between the physical field operators projected at emitter positions involves the cross-correlation of the structured environment as (see Appendix~\ref{app:Gamma_matrix})
\begin{align}
    [\hat{E}^{(+)}_i(\omega), \hat{E}^{(-)}_j(\omega')] 
    &= \hat{I}\frac{\hbar \omega^2}{\pi \epsilon_0 c^2} \delta(\omega - \omega') \nonumber \\
    & \times \left( \mathbf{n}_i \cdot \Im [\overline{\mathbf{G}}(\mathbf{R}_i, \mathbf{R}_j, \omega)] \cdot \mathbf{n}_j \right).
    \label{eq:commutation_relation}
\end{align}
In the case of multiple emitters, the set of $N$ field operators $\{ \hat{E}^{(+)}_{a}(\omega) \}$ are neither orthogonal nor normalized with respect to each other. To construct a new set of operators that satisfy the standard bosonic commutation relations, the overlap matrix $\mathbf{\Gamma}(\omega)$ is introduced, whose elements $\Gamma_{ij}(\omega)$ are given by $\mathbf{n}_i \cdot \Im [\overline{\mathbf{G}}(\mathbf{R}_i, \mathbf{R}_j, \omega)] \cdot \mathbf{n}_j$.
Finally, by diagonalizing the overlap matrix $\mathbf{\Gamma}$, we can construct a set of collective bright modes ($\hat{c}_a$) 
by performing spectral decomposition, $\mathbf{\Gamma} = \mathbf{V} \mathbf{\Lambda} \mathbf{V}^\dagger$, where $\mathbf{\Lambda} = \mathrm{diag}(\gamma_1, \dots, \gamma_N)$ contains the eigenvalues. 
Then each $\hat{c}_a$ satisfy the standard bosonic commutation relations,
\begin{align}
    [\hat{c}_i(\omega), \hat{c}_{j}^\dagger(\omega')] = \delta_{ij} 
\delta(\omega - \omega')
\end{align}
The corresponding orthonormal bosonic operators are then defined as
\begin{align}
\hat{c}_k(\omega) = \frac{1}{\mathcal{A} (\omega)(\sqrt{\gamma_k(\omega)}} \sum_{a=1}^N V_{ak}^*(\omega) \hat{E}^{(+)}_a(\omega),
\label{eq:bright_mode_operators}
\end{align}
The coupling coefficient between the $i$-th emitter and the $k$-th bright mode is defined as
\begin{align}
g_{ik}(\omega) = d_i \mathcal{A}(\omega) \sqrt{\gamma_k(\omega)} V_{ik}(\omega).
\label{eq:ECM_coupling_coefficient}
\end{align}
Consequently, the field and interaction terms of the Hamiltonian are rewritten as
\begin{align}
\hat{H}_{\mathrm{field}} = \sum_{k=1}^N \int_{0}^{\infty} d\omega \, \hbar \omega \hat{c}_{k}^\dagger(\omega) \hat{c}_{k}(\omega)
\end{align}
under the rotating wave approximation (RWA), the Equation\eqref{eq:H_int_sum_dE} can be simplified as
\begin{align}
\hat{H}_{\mathrm{int}} = -\sum_{k=1}^N \sum_{i=1}^N \int_{0}^{\infty} d\omega \left( g_{ik}(\omega) \hat{\sigma}_i^+ \hat{c}_k(\omega) + \text{H.c.} \right).
\end{align}
The complete positive-frequency electric field operator $\hat{\mathbf{E}}^{(+)}(\mathbf{r})$ at an arbitrary observation point $\mathbf{r}$ is reconstructed using the minimal basis set $\{\hat{c}_k(\omega)\}$. This formulation represents the total field as a superposition of bright mode profiles $\mathbf{\Phi}_k$, where the transformation accounts for the normalization of the bosonic operators. The essential coupling information and environmental response are encapsulated in the imaginary part of the Green's function,
\begin{align}
    \hat{\mathbf{E}}^{(+)}(\mathbf{r}) &= \sum_{k=1}^{N} \int_{0}^{\infty} d\omega \, \mathbf{\Phi}_{k}(\mathbf{r}, \omega) \, \hat{c}_k(\omega), \label{eq:new_field_reconstruction} \\
    \mathbf{\Phi}_{k}(\mathbf{r}, \omega) &= \mathcal{A}(\omega)\sqrt{ \gamma_k(\omega)} \sum_{j=1}^{N} V_{jk}(\omega) \mathbf{\Psi}_{j}(\mathbf{r}, \omega), \label{eq:new_mode_profile} \\
    \mathbf{\Psi}_{j}(\mathbf{r}, \omega) &= \frac{\Im[\overline{\mathbf{G}}(\mathbf{r}, \mathbf{R}_j, \omega)] \cdot \mathbf{n}_j}{\Gamma_{jj}(\omega)}. \label{eq:new_raw_profile}
\end{align}
In Appendix~\ref{app:ecm_field_profile}, the derivation of Eq.~\eqref{eq:new_raw_profile} within the M-LN framework is provided. It should be noted that while earlier studies reached formally identical expressions using the conventional macroscopic QED framework~\cite{ECM_nanophotonics_2021,Vidal_PRL_fewmode_2021,Vidal_PRB_fewmode_2025}, those approaches remain conceptually incomplete for open systems as they neglect the essential BA contributions required for self-consistent field reconstruction.

\section{Non-Markovian Dynamics in the Two-Excitation Manifold}
\subsection{Motivation}
While the ECM description renders the temporal evolution of the quantum state more accessible, each emitter remains coupled to a continuum of bright modes over a broad frequency range. A common strategy is to eliminate these environmental DoFs, leading to integro-differential or master-equation descriptions. However, such reduced formulations generally do not retain the explicit amplitude- and phase-resolved structure of the emitted field, making it more difficult to directly capture multi-photon interference and spatiotemporal field dynamics in strongly non-Markovian regimes.
In contrast, the emitter-centered M-LN framework retains the explicit amplitude structure of the intermediate and pure two-photon sectors. This leads to a closed hierarchical system within the two-excitation manifold, where retardation and memory effects are preserved directly at the level of the dynamical variables. As a result, the formulation provides a transparent description of multi-emitter correlations and collective non-Markovian dynamics while maintaining direct access to the emitted-field amplitudes.

\subsection{Model}
The dynamics of the system are governed by the Schrödinger equation $i \partial_t |\Psi(t)\rangle = \hat{H} |\Psi(t)\rangle$, where $\hat{H}$ is the effective Hamiltonian considering only bright modes. First, we expand the system wavefunction within the truncated two-excitation manifold. The symmetrized ansatz is given by
\begin{align}
    |\Psi(t)\rangle
    &= \sum_{a<b} C_{ab}(t) |e_a, e_b; \{0\}\rangle \nonumber\\
    & + \sum_{a=1}^N \sum_{k=1}^N \int_0^\infty d\omega\, B_{a, k\omega}(t) |e_a; 1_{k\omega}\rangle \nonumber\\
    & + \frac{1}{2} \sum_{k,l=1}^N \iint_0^\infty d\omega d\omega'\, D_{kl, \omega\omega'}(t) |\{g\}; 1_{k\omega}, 1_{l\omega'}\rangle,
    \label{eq:Main_Ansatz_Revised}
\end{align}
where $C_{ab}(t)$ is the amplitude of the doubly excited atomic sector, 
$B_{a,k\omega}(t)$ is the amplitude of the intermediate one-photon, one-atom-excited sector, and 
$D_{k\omega,k'\omega'}(t)$ is the amplitude of the pure two-photon sector. 
Under the rotating wave approximation (RWA), the interaction Hamiltonian contains only excitation conserving terms. Consequently, the total excitation operator 
$\hat{N} = \sum_a \hat{\sigma}^\dagger_a \hat{\sigma}_a + \sum_k \int d\omega \hat{c}^\dagger_k \hat{c}_k$
satisfies $[\hat{H}, \hat{N}] = 0$. The Hilbert space decomposes into invariant sectors labeled by the eigenvalues of $\hat{N}$. Any state initialized within a given $n$-excitation manifold evolves entirely within that manifold for all time. In this work, we focus on the two-excitation sector, where the truncation in Eq.~\eqref{eq:Main_Ansatz_Revised} represents an exact decomposition rather than an approximation.
The bosonic symmetry of the photonic continuum implies
\begin{equation}
    D_{k\omega,k'\omega'}(t)=D_{k'\omega',k\omega}(t).
\end{equation}
The interaction with the structured electromagnetic environment is fully encoded in the ECM couplings $g_{ak}(\omega)$ in Eq.~\eqref{eq:ECM_coupling_coefficient} as
\begin{equation}
    \sum_k g_{ak}(\omega) g_{bk}^*(\omega)
    =
    \frac{\mu_0}{\pi\hbar}\,\omega^2\,
    \mathbf d_a \cdot
    \mathrm{Im}\!\left[\overline{\mathbf G}(\mathbf R_a,\mathbf R_b,\omega)\right]
    \cdot \mathbf d_b^* .
    \label{eq:ECM_spectral_decomp}
\end{equation}
Therefore, the entire dynamics is determined by the dyadic Green's function through the bright-mode couplings. Substituting Eq.~\eqref{eq:Main_Ansatz_Revised} into the Schrödinger equation yields the following coupled hierarchy of equations
\begin{widetext}
\begin{align}
    i\dot C_{ab}(t) &= (\omega_a+\omega_b)\,C_{ab}(t) + \sum_{k} \int_0^\infty d\omega\, \Big[ g_{bk}(\omega)\,B_{a,k\omega}(t) + g_{ak}(\omega)\,B_{b,k\omega}(t) \Big] \label{eq:C_sector_EOM} \\
    i\dot B_{a,k\omega}(t) &= (\omega_a+\omega)\,B_{a,k\omega}(t) + \sum_{b\neq a} g_{bk}^*(\omega)\,C_{ab}(t) + \sum_{k'} \int_0^\infty d\omega'\, g_{ak'}(\omega')\,D_{k\omega,k'\omega'}(t) \label{eq:B_sector_EOM} \\
    i\dot D_{k\omega,k'\omega'}(t) &= (\omega+\omega')\,D_{k\omega,k'\omega'}(t) + \sum_a \Big[ g_{ak'}^*(\omega')\,B_{a,k\omega}(t) + g_{ak}^*(\omega)\,B_{a,k'\omega'}(t) \Big] \label{eq:D_sector_EOM}
\end{align}
\end{widetext}
where $\omega_a$ (and $\omega_b$) represents the transition frequency of the $a$-th (and $b$-th) emitter. 
By explicitly evolving the coupled amplitude equations, the hierarchy retains the non-Markovian retardation effects associated with sequential photon emission within the two-excitation manifold. Consequently, the total probability is conserved, allowing for an evaluation of the reduced atomic observables, including the ground-state population, directly from the retained dynamical sectors. 
The above procedure is summarized in Fig.~\ref{fig:ECM_theory_flow}, and a detailed derivation of norm preservation and bosonic symmetry is provided in Appendix~\ref{app:consistency_check}.

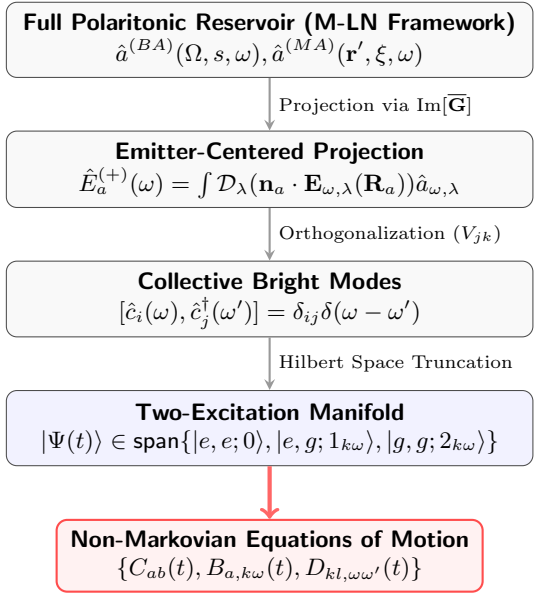
\begin{figure}
\centering
\begin{tikzpicture}[node distance=0.7cm, font=\sffamily\small]
    \tikzset{
        layer_box/.style={rectangle, rounded corners, minimum width=7cm, minimum height=1cm, draw=black!70, fill=gray!5},
        concept_box/.style={rectangle, rounded corners, minimum width=3cm, minimum height=0.7cm, draw=blue!70, fill=blue!2, thick},
        arrow_style/.style={->, >=stealth, thick, gray!80}
    }

    \node (L1) [layer_box] {
        \begin{tabular}{c} 
            \textbf{Full Polaritonic Reservoir (M-LN Framework)} \\ 
            $\hat{a}^{(BA)}(\Omega,s,\omega), \hat{a}^{(MA)}(\mathbf{r}',\xi,\omega)$
        \end{tabular}
    };

    \node (L2) [layer_box, below=of L1] {
        \begin{tabular}{c} 
            \textbf{Emitter-Centered Projection} \\ 
            $\hat{E}_a^{(+)}(\omega) = \int \mathcal{D}_\lambda (\mathbf{n}_a \cdot \mathbf{E}_{\omega,\lambda}(\mathbf{R}_a)) \hat{a}_{\omega,\lambda}$
        \end{tabular}
    };
    \draw [arrow_style] (L1) -- node[right, black, font=\scriptsize] {Projection via $\mathrm{Im}[\overline{\mathbf{G}}]$} (L2);

    \node (L3) [layer_box, below=of L2] {
        \begin{tabular}{c} 
            \textbf{Collective Bright Modes} \\ 
            $[\hat{c}_i(\omega), \hat{c}_j^\dagger(\omega')] = \delta_{ij}\delta(\omega-\omega')$
        \end{tabular}
    };
    \draw [arrow_style] (L2) -- node[right, black, font=\scriptsize] {Orthogonalization ($V_{jk}$)} (L3);

    \node (L4) [layer_box, below=of L3, solid, fill=blue!5] {
        \begin{tabular}{c} 
            \textbf{Two-Excitation Manifold} \\ 
            $|\Psi(t)\rangle \in \text{span} \{ |e,e;0\rangle, |e,g;1_{k\omega}\rangle, |g,g;2_{k\omega}\rangle \}$
        \end{tabular}
    };
    \draw [arrow_style] (L3) -- node[right, black, font=\scriptsize] {Hilbert Space Truncation} (L4);

    \node (Kernel) [below=0.7cm of L4, concept_box, draw=red!70, fill=red!5] {
        \begin{tabular}{c}
             \textbf{Non-Markovian Equations of Motion}\\ 
            $\{ C_{ab}(t), B_{a,k\omega}(t), D_{kl,\omega\omega'}(t) \}$
        \end{tabular}
    };
    \draw [->,  ultra thick, red!60] (L4) -- (Kernel);

\end{tikzpicture}
\caption{Schematic of the Green’s-function-based M-LN/ECM framework and its reduction to the two-excitation manifold.}
\label{fig:ECM_theory_flow}
\end{figure}

\subsection{Observables}

To characterize the atomic correlations and local observables, we construct the reduced atomic density matrix $\rho_{\mathrm{atom}}(t)$ by tracing out the photonic DoFs from the full wavefunction $|\Psi(t)\rangle$. For a two-emitter system ($N=2$), the reduced density matrix in the basis $\{|e_1,e_2\rangle, |e_1,g_2\rangle, |g_1,e_2\rangle, |g_1,g_2\rangle\}$ takes the form
\begin{equation}
    \rho_{\mathrm{atom}}(t) =
    \begin{pmatrix}
        P_{ee}(t) & 0 & 0 & 0 \\
        0 & P_{eg}(t) & Z_{12}(t) & 0 \\
        0 & Z_{12}^*(t) & P_{ge}(t) & 0 \\
        0 & 0 & 0 & P_{gg}(t)
    \end{pmatrix}.
\end{equation}
Within the present two-quantum hierarchy, each matrix element is obtained directly from the sector amplitudes. The population of the doubly excited state is given by
\begin{equation}
    P_{ee}(t) = |C_{12}(t)|^2.
\end{equation}
The single-excitation populations and the atomic coherence are determined by the intermediate one-photon sector,
\begin{align}
    P_{eg}(t) &= \sum_k \int_0^\infty d\omega\, |B_{1,k\omega}(t)|^2, \\
    P_{ge}(t) &= \sum_k \int_0^\infty d\omega\, |B_{2,k\omega}(t)|^2, \\
    Z_{12}(t) &= \sum_k \int_0^\infty d\omega\, B_{1,k\omega}(t) B_{2,k\omega}^*(t).
\end{align}
The ground-state population is obtained from the pure two-photon sector,
\begin{equation}
    P_{gg}(t)
    =
    \frac{1}{2}
    \sum_{k,l}
    \iint_0^\infty d\omega\, d\omega'\,
    |D_{k\omega,l\omega'}(t)|^2 .
\end{equation}
Accordingly, the normalization of the reduced atomic density matrix follows from the norm conservation of the full two-quantum wavefunction,
\begin{equation}
    \mathrm{Tr}\,\rho_{\mathrm{atom}}(t)=1.
\end{equation}

To quantify the entanglement between the two emitters, we use Wootters' concurrence. For the density matrix above, it reduces to
\begin{equation}
    \mathcal{C}(t)
    =
    2 \max\!\left(0, |Z_{12}(t)| - \sqrt{P_{ee}(t)P_{gg}(t)}\right).
    \label{eq:concurrence}
\end{equation}
In addition, the same reduced density matrix provides direct access to other quantities of interest, such as the Bell-state fidelities
\begin{align}
    F_{\pm}(t)
    =
    \frac{P_{eg}(t)+P_{ge}(t)}{2}
    \pm \mathrm{Re}\, Z_{12}(t),
\end{align}

These quantities allow us to characterize the populations, coherences, entanglement, and mixedness of the atomic subsystem within a unified framework.

The formulation can also reconstruct the emitted field in real space using the explicit hierarchy amplitudes and Eq.~\eqref{eq:new_field_reconstruction}. Since the non-Markovian dynamics are initialized and contained within the two-quantum manifold, the most natural photonic observable is the first-order field intensity rather than the field expectation value itself. We therefore introduce the positive-frequency electric field operator expanded in the ECM basis as
\begin{equation}
    \hat{\mathbf{E}}^{(+)}(\mathbf{r}) = \sum_k \int_0^\infty d\omega\, \mathbf{E}_k(\mathbf{r},\omega)\, \hat{c}_{k\omega},
    \label{eq:Eplus_ECM}
\end{equation}
where $\mathbf{E}_k(\mathbf{r},\omega)$ denotes the spatial mode profile associated with the $k$-th bright channel in Eq.~\eqref{eq:new_mode_profile}. Crucially, in the macroscopic QED framework, this spatial profile is rigorously determined by the dyadic Green's tensor $\overline{\mathbf{G}}(\mathbf{r}, \mathbf{R}_k, \omega)$, which dictates the electromagnetic propagation from the $k$-th emitter location $\mathbf{R}_k$ to the arbitrary observation point $\mathbf{r}$.
The corresponding first-order spatial field intensity is defined by 
\begin{equation}
    I(\mathbf{r},t) = \big\langle \Psi(t)\big| \hat{\mathbf{E}}^{(-)}(\mathbf{r}) \cdot \hat{\mathbf{E}}^{(+)}(\mathbf{r}) \big|\Psi(t)\big\rangle.
    \label{eq:intensity_def}
\end{equation}
This quantity is determined by the one-body photonic density kernel Eqs.~\eqref{eq:B_sector_EOM} and~\eqref{eq:D_sector_EOM}, which is analytically derived from the truncated wavefunction as
\begin{align}
    \Gamma_{k\omega, k'\omega'}(t) 
    &= \sum_{a=1}^N B_{a, k\omega}^*(t) B_{a, k'\omega'}(t) \nonumber \\
    &\quad + \sum_{l=1}^N \int_0^\infty d\nu\, D_{kl, \omega\nu}^*(t) D_{k'l, \omega'\nu}(t).
    \label{eq:photonic_kernel}
\end{align}
The first term originates from the intermediate one-photon sector, while the second term captures the exact contribution from the pure two-photon continuum. In terms of this density kernel, the spatiotemporal field intensity is expressed as (See Appendix~\ref{app:field_reconstruction_2}),
\begin{equation}
    I(\mathbf{r},t) = \sum_{k,k'} \iint_0^\infty d\omega d\omega'\, \mathbf{E}_k^*(\mathbf{r},\omega) \cdot \mathbf{E}_{k'}(\mathbf{r},\omega')\, \Gamma_{k\omega, k'\omega'}(t).
    \label{eq:intensity_kernel_form}
\end{equation}

\section{Numerical Simulations}

\subsection{Single-end photonic waveguide with lossy dielectric slab}

\begin{figure}
    \centering
    \includegraphics[width=\linewidth]{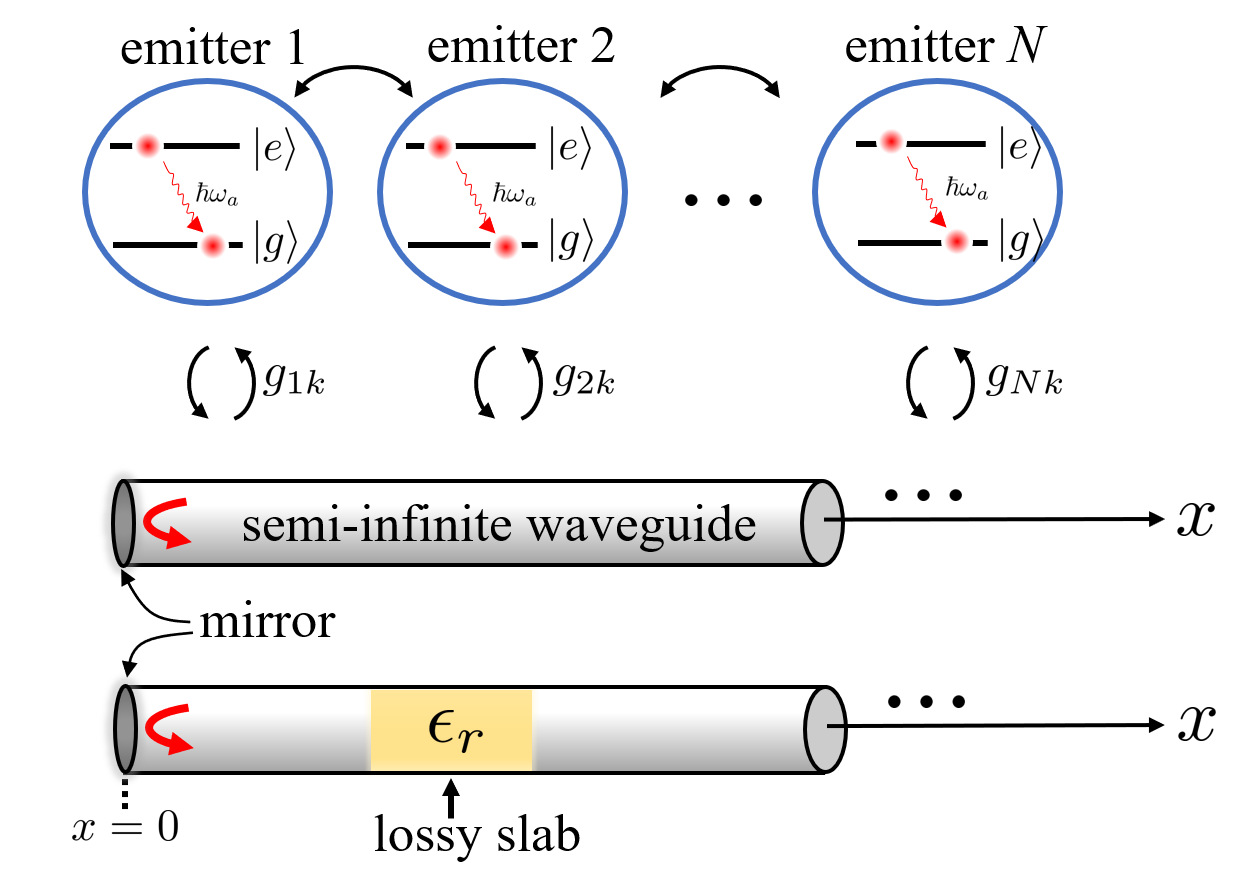}
    \caption{
    Schematic illustration of single-end photonic waveguide configurations. 
    }
    \label{fig:waveguide_slab}
\end{figure}

As shown in Fig.~\ref{fig:waveguide_slab}, we examine both a homogeneous half-space waveguide and a modified geometry in which a finite dielectric slab is embedded.
In both configurations, a perfectly reflecting mirror is placed at $x=0$, enforcing a single-end waveguide geometry. Multiple emitters are positioned at fixed locations $\{x_a\}$ along the waveguide. In the homogeneous case, the emitters interact via the mirror-modified vacuum Green's function. In the slab configuration, an additional dielectric region is introduced over the interval $x \in [x_s^{(1)}, x_s^{(2)}]$, characterized by a generally complex permittivity $\epsilon_r = \epsilon' + i \epsilon''$. 
\begin{figure}[t]
    \centering

    \begin{subfigure}{0.9\columnwidth}
        \centering
        \includegraphics[width=\linewidth]{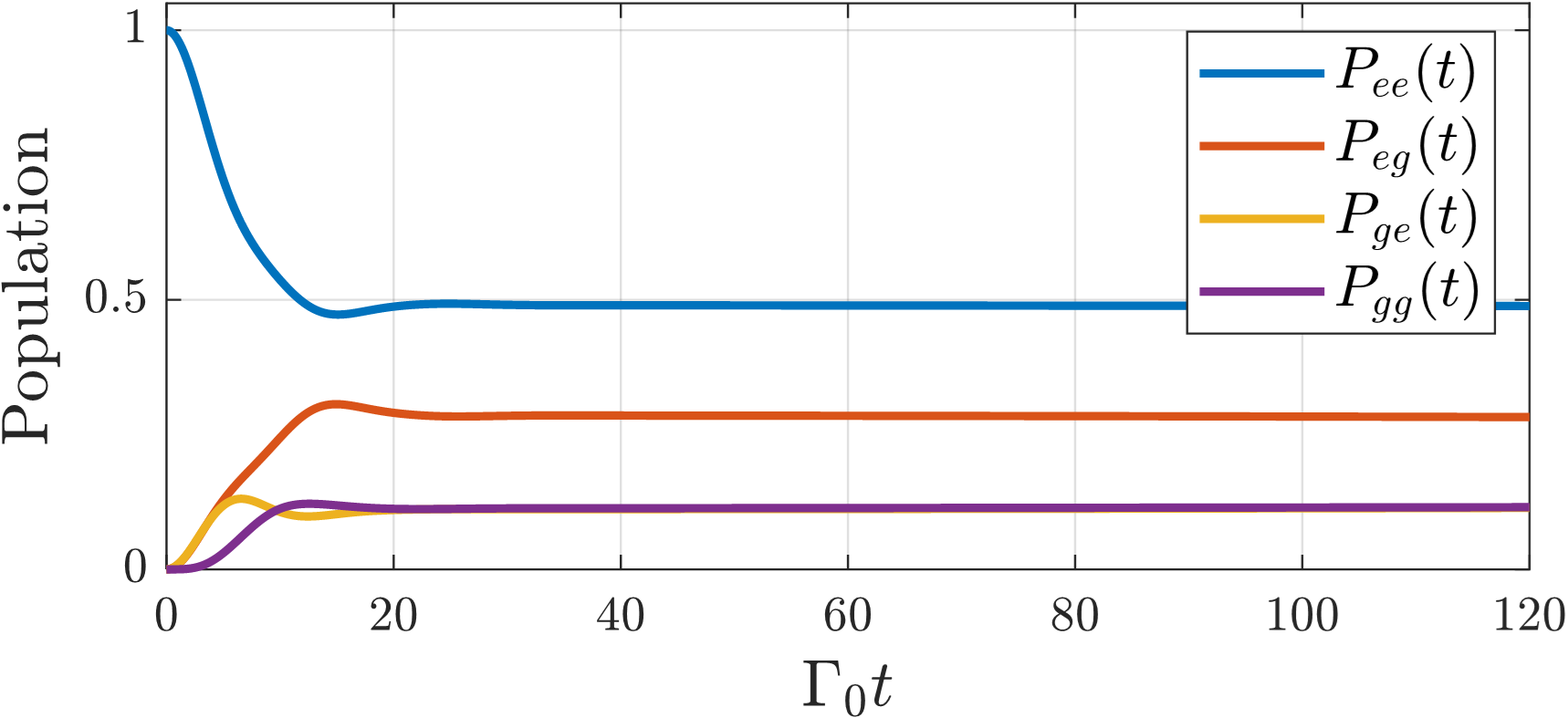}
        \caption{Population}
    \end{subfigure}

    \begin{subfigure}{0.9\columnwidth}
        \centering
        \includegraphics[width=\linewidth]{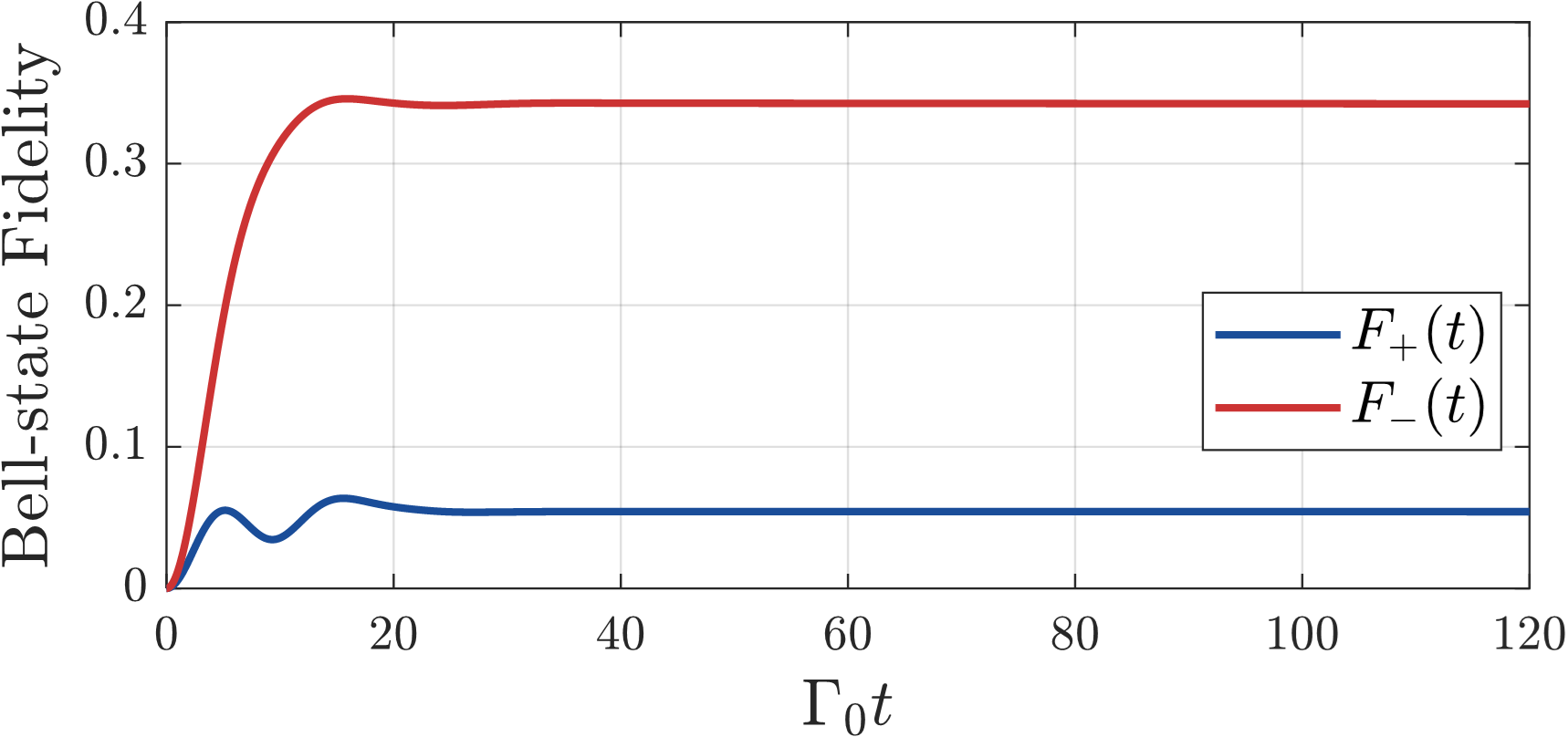}
        \caption{Fidelity}
    \end{subfigure}

    \begin{subfigure}{0.95\columnwidth}
        \centering
        \includegraphics[width=\linewidth]{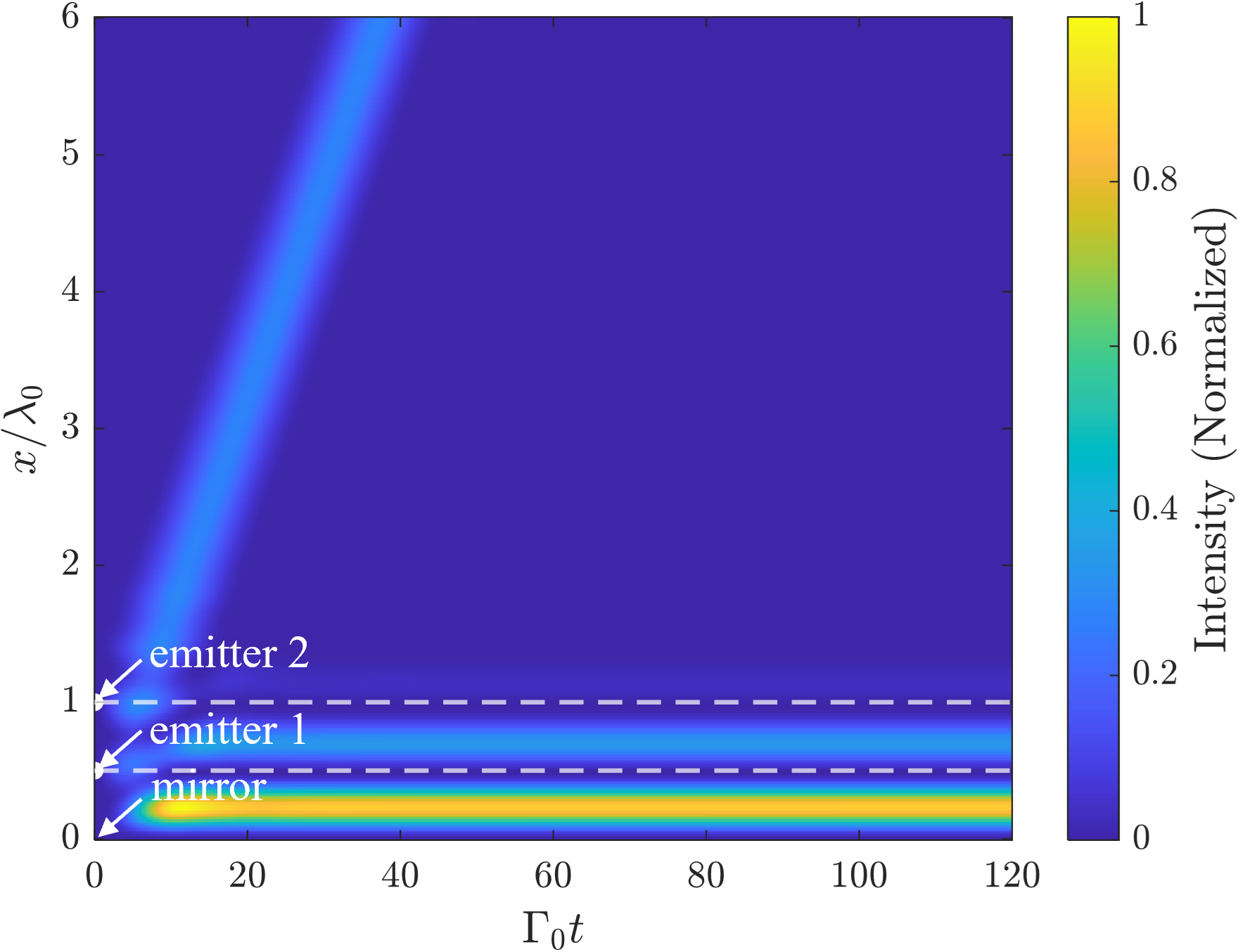}
        \caption{Field intensity}
    \end{subfigure}

    \caption{
    Dynamics in a semi-infinite waveguide. 
    Emitters located near nodes lead to suppressed decay and a bound-state-like component.}
    \label{fig:1D_case1}
\end{figure}

\begin{figure}[t]
    \centering

    \begin{subfigure}{0.9\columnwidth}
        \centering
        \includegraphics[width=\linewidth]{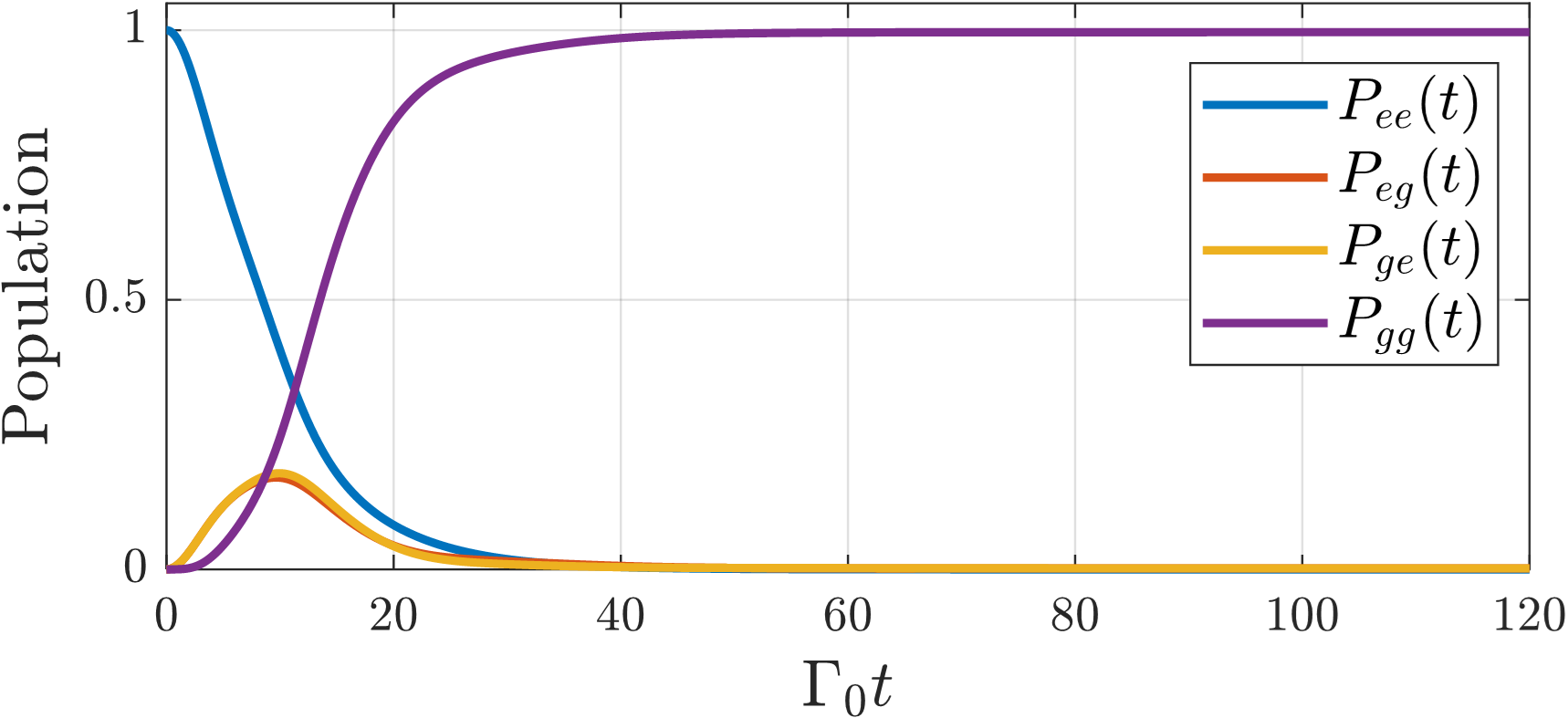}
        \caption{Population}
    \end{subfigure}

    \begin{subfigure}{0.9\columnwidth}
        \centering
        \includegraphics[width=\linewidth]{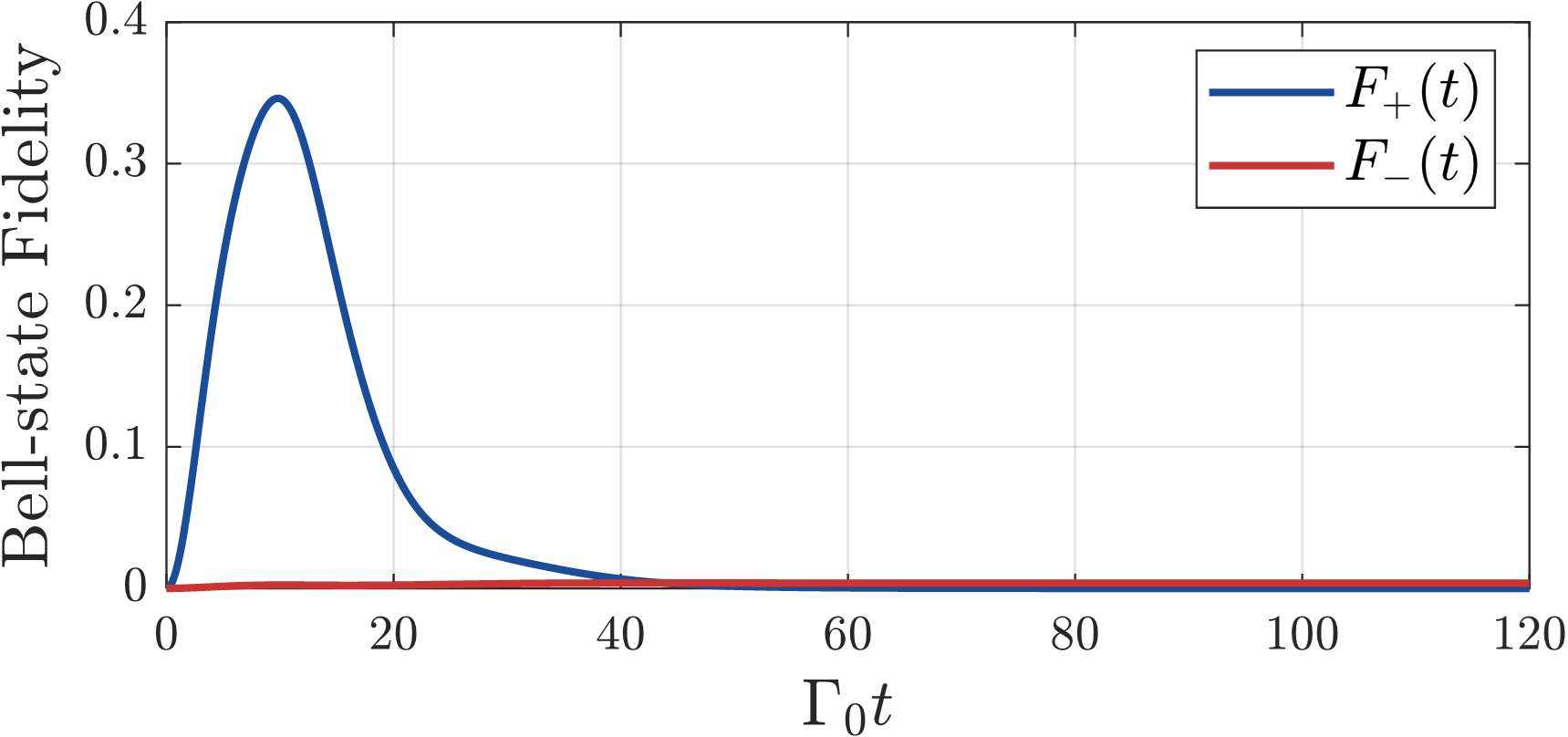}
        \caption{Fidelity}
    \end{subfigure}

    \begin{subfigure}{0.95\columnwidth}
        \centering
        \includegraphics[width=\linewidth]{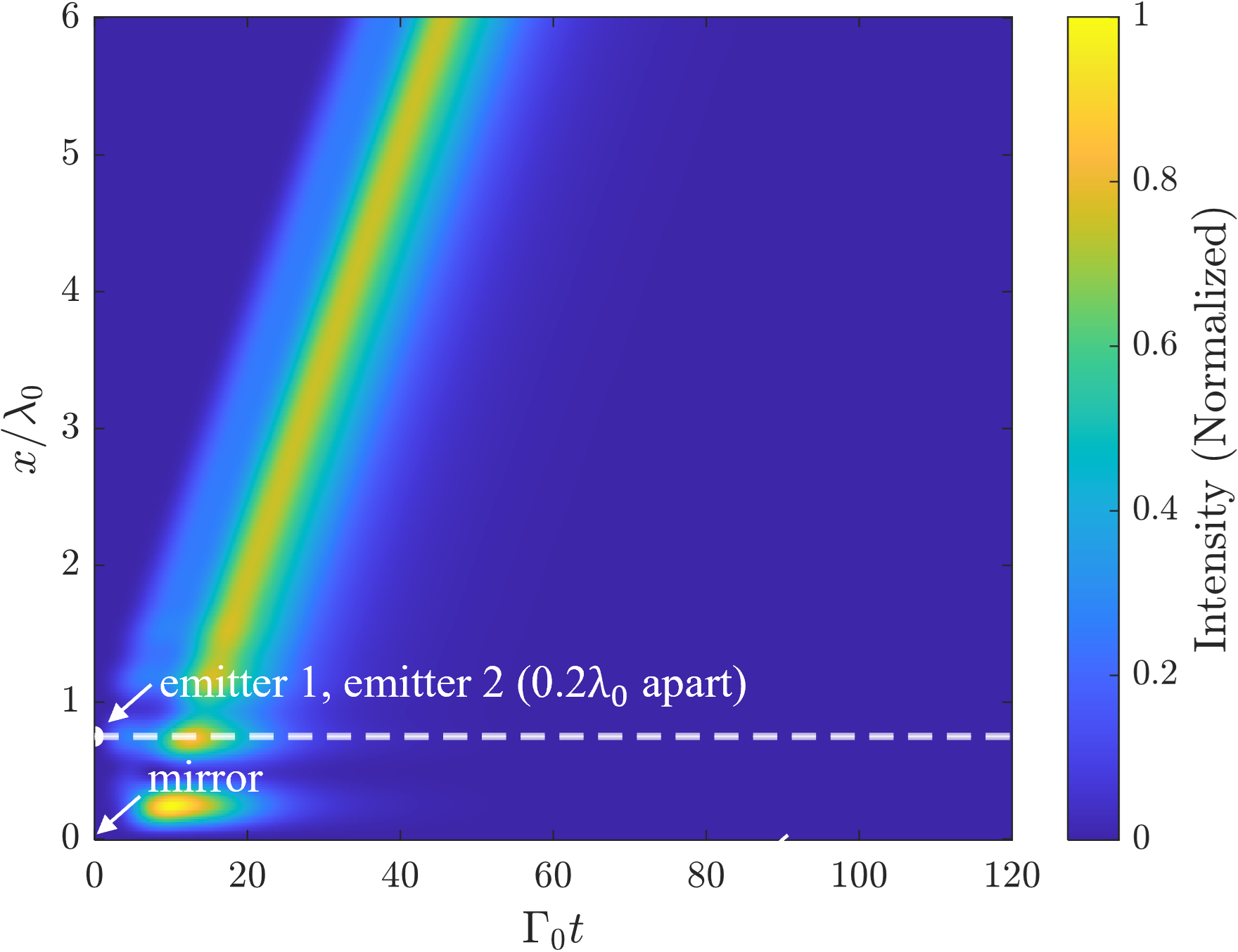}
        \caption{Field intensity}
    \end{subfigure}

    \caption{
    Dynamics in a semi-infinite waveguide. 
    Both emitters near antinodes result in efficient radiative decay.}
    \label{fig:1D_case2}
\end{figure}

We first consider the homogeneous waveguide configuration without the dielectric slab in order to establish a clear baseline for the dynamics in the presence of a single reflecting boundary. In this case, the emitter-emitter interaction is primarily mediated by the mirror-modified vacuum Green's function, and the dynamics are governed by position-dependent interference between the emitters and their mirror images. Here we use normalized units with $\hbar=1$, while the waveguide velocity and electromagnetic prefactors are absorbed into the Green-function normalization and into the definition of the effective dipole parameter. With $\omega_a=10$ and an effective dimensionless dipole moment $d_{\mathrm{eff}} = 0.224$, the mode-resolved effective couplings generated from the Green-function kernel remain much smaller than $\omega_a$, so the rotating-wave approximation is well justified. Fig.~\ref{fig:1D_case1} summarizes the population dynamics, Bell-state fidelities, and field propagation for the first representative configuration, where the emitters are positioned at $\{0.5\lambda_a, 1\lambda_a\}$ where $\lambda_a$ is the wavelength corresponding to $\omega_a$. In this setup, the emitters are located at the nodes of the standing-wave pattern induced by the mirror. As a result, the emitters experience strongly suppressed radiative coupling due to destructive interference with their mirror images, leading to the formation of a long-lived excitation component. This is directly reflected in the population dynamics (Fig.~\ref{fig:1D_case1}a), where a finite excitation probability persists at long times, indicating the emergence of a bound-state-like contribution. The spatiotemporal field intensity as defined in Eq.~\eqref{eq:intensity_def} corroborates this (Fig.~\ref{fig:1D_case1}c), showing that a localized field component remains trapped near the emitter positions. In contrast, Fig.~\ref{fig:1D_case2} illustrates the dynamics for the second configuration, where both emitters are placed near antinodes at $\{0.74\lambda_a, 0.76\lambda_a\}$. In this regime, the coupling to the waveguide modes is enhanced via constructive interference, and both emitters efficiently radiate into the continuum. Consequently, the system relaxes predominantly into the two-photon sector, with the ground-state population approaching unity at long times (Fig.~\ref{fig:1D_case2}a). The emitted photons propagate away from the emitters without significant backaction, forming an outgoing wavefront as seen in the field intensity (Fig.~\ref{fig:1D_case2}c).  The Bell-state fidelities (Figs.~\ref{fig:1D_case1}b and \ref{fig:1D_case2}b) further reveal that the symmetric and antisymmetric collective channels are selectively populated depending on the emitter configuration. In particular, the bound-state regime preferentially stabilizes the antisymmetric Bell state $F_{-}(t)$, whereas the radiative regime is dominated by the symmetric component $F_{+}(t)$, which subsequently decays rapidly.

\begin{figure}[t]
    \centering

    \begin{subfigure}{0.9\columnwidth}
        \centering
        \includegraphics[width=\linewidth]{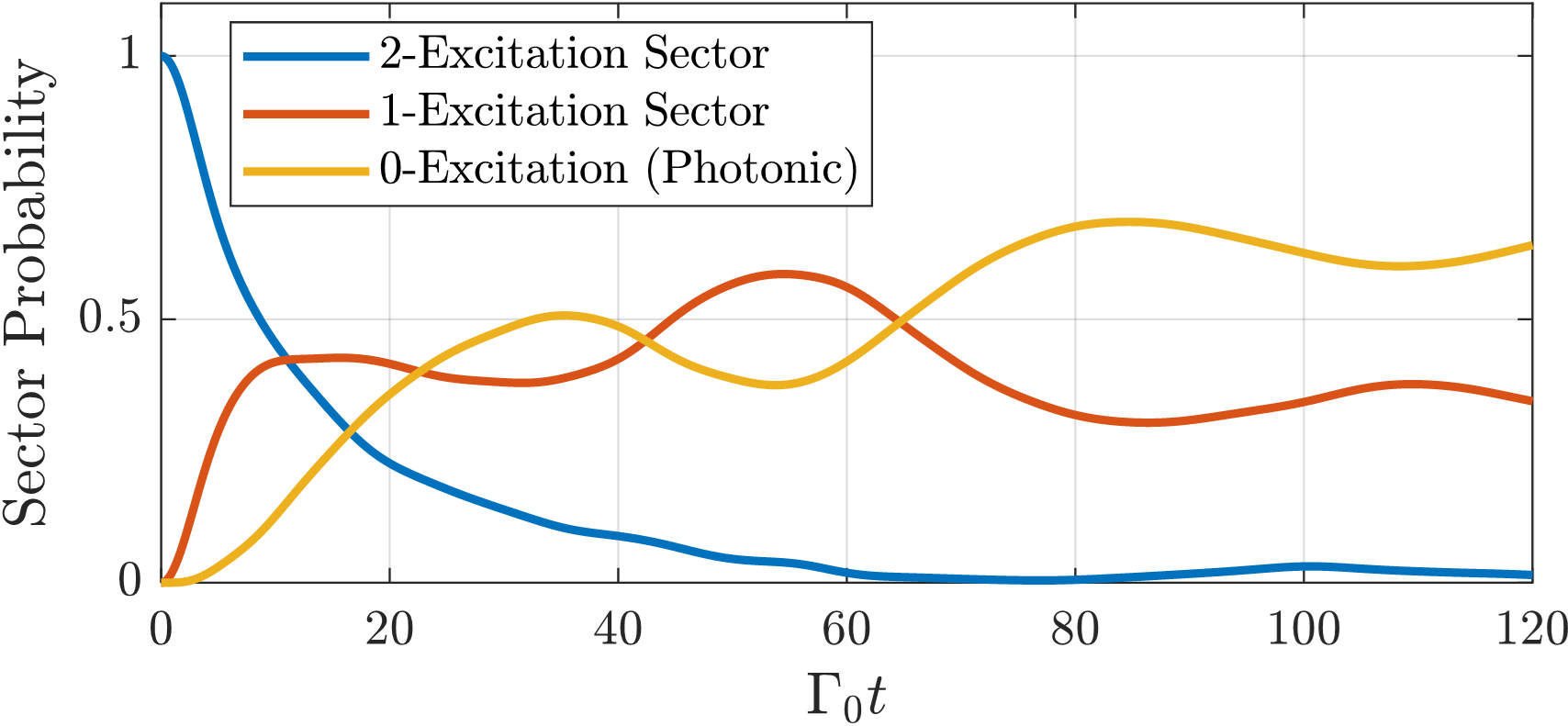}
        \caption{Sector probability}
    \end{subfigure}

    \begin{subfigure}{0.9\columnwidth}
        \centering
        \includegraphics[width=\linewidth]{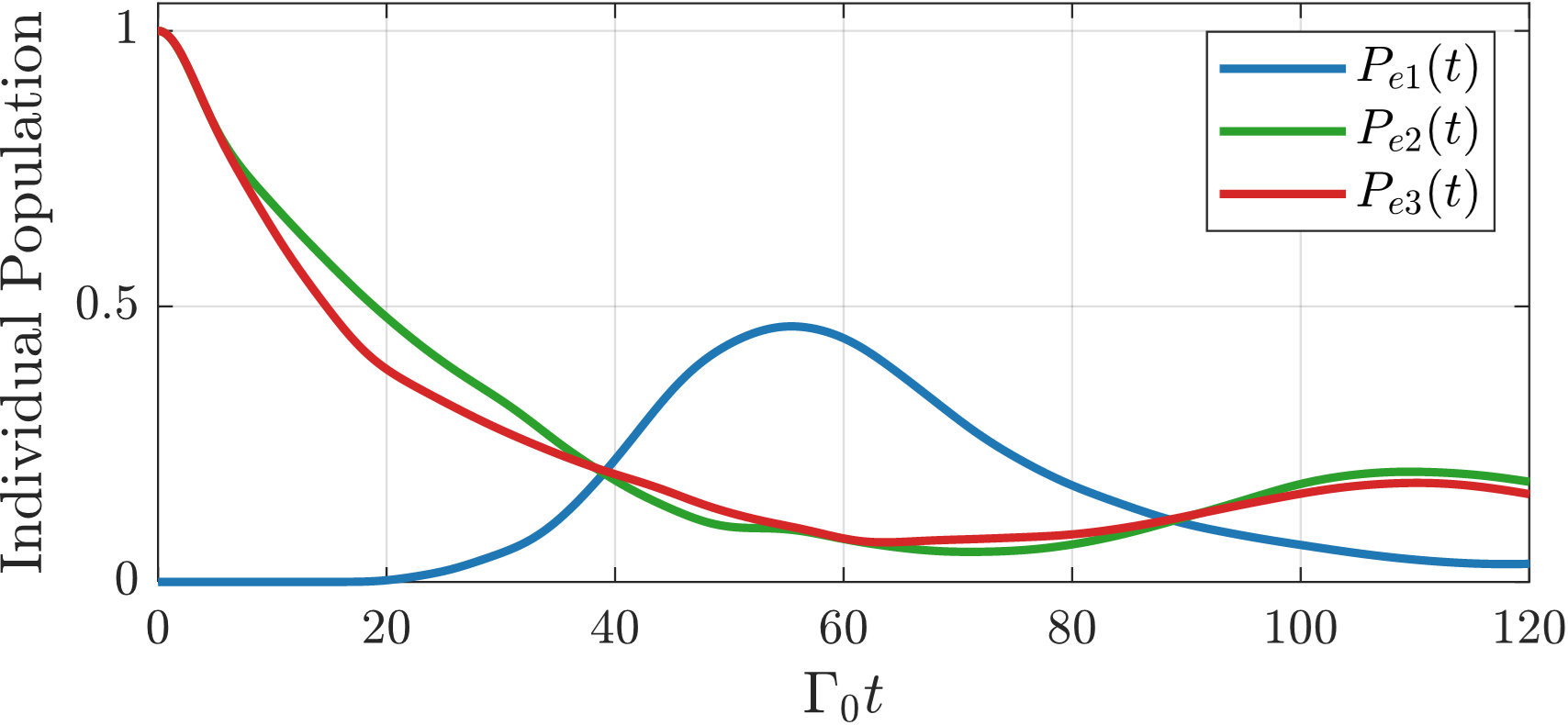}
        \caption{Individual population}
    \end{subfigure}

    \begin{subfigure}{0.95\columnwidth}
        \centering
        \includegraphics[width=\linewidth]{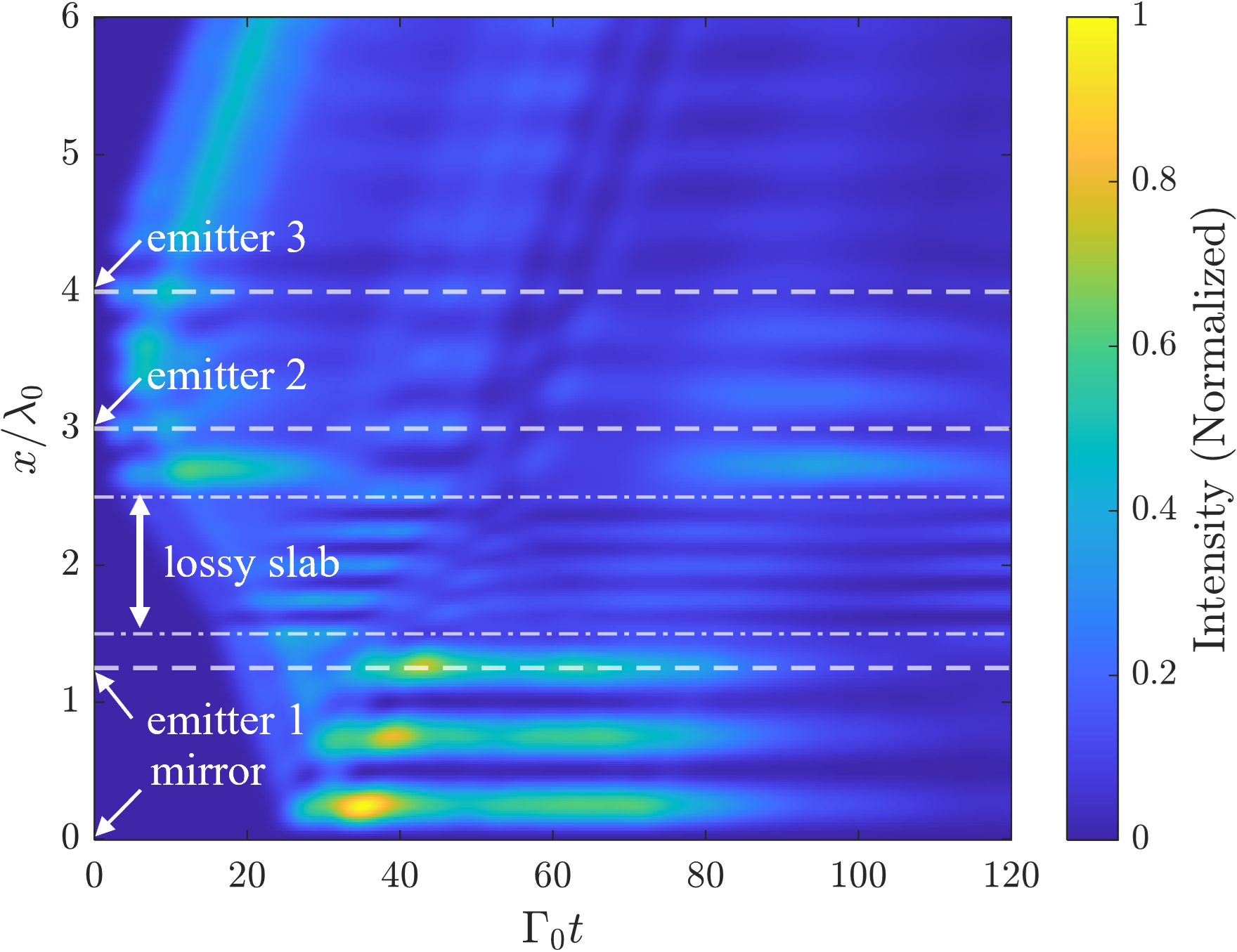}
        \caption{Field intensity}
    \end{subfigure}

    \caption{
    Dynamics of three emitters in a waveguide with a lossy slab. 
    Sector probabilities, individual populations, and field intensity showing delayed propagation and partial confinement.}
    \label{fig:1D_lossy_slab}
\end{figure}

We now turn to the case of three emitters coupled to a structured environment in the presence of a lossy dielectric slab. The transition frequency $\omega_a$ and the effective dipole moment are maintained at the previous values. The emitters are positioned at $\{1.25\lambda_a,\, 3.0\lambda_a,\,4.0\lambda_a\}$, while the slab occupies the region $x \in [1.5\lambda_a,\,2.5\lambda_a]$, introducing both scattering and absorption into the photonic environment.
Fig.~\ref{fig:1D_lossy_slab} shows the resulting dynamics for the initial state $|g,e,e\rangle$, where emitter 1 is in the ground state and emitters 2 and 3 are initially excited. In contrast to the homogeneous waveguide case, the presence of the slab breaks the spatial symmetry of the system and effectively partitions the emitters into distinct photonic environments. In particular, emitter 1 is located to the left of the slab and remains strongly influenced by the mirror, whereas emitters 2 and 3 are located on the far side of the slab and experience filtered and attenuated photonic coupling.
This asymmetry is clearly reflected in the individual emitter populations. At early times, emitters 2 and 3 exhibit rapid decay, transferring excitation into the one-photon sector. In contrast, emitter 1 shows a delayed population buildup, reaching a maximum at intermediate times before gradually decaying. This behavior indicates that the excitation is first emitted by the distant emitters and subsequently reabsorbed or scattered back toward the region near emitter 1, leading to a temporally delayed excitation transfer.
The sector-resolved probabilities further reveal a nontrivial cascade process. The two-excitation manifold decays rapidly at short times, while the one-excitation sector develops a pronounced transient population that persists over an extended time window. Unlike the mirror-only case, the presence of the slab leads to a clear separation of timescales, with a slow relaxation from the one-photon sector to the final two-photon sector. This behavior reflects the frequency-dependent filtering and partial absorption introduced by the slab, which together slow the transfer of population out of the intermediate manifold.
These features are corroborated by the spatiotemporal field intensity. The emitted field exhibits strong localization and multiple scattering within the slab region, with a significant portion of the intensity remaining confined near the slab boundaries for extended times. In addition, delayed wavefronts propagating toward the mirror and back are clearly visible, demonstrating the presence of memory effects induced by the structured environment.

\subsection{Dynamics of symmetric Dicke state in structured environments}

In this subsection, we investigate the emission dynamics starting from a highly correlated many-body state. We focus on the two-excitation symmetric Dicke state, hereafter denoted as
\begin{align}
    |\bar{W}\rangle = \frac{1}{\sqrt{3}}(|e_1 e_2 g_3\rangle + |e_1 g_2 e_3\rangle + |g_1 e_2 e_3\rangle).
\end{align}
 In the context of the Dicke ladder, this state represents a collective excitation where the energy is uniformly shared among the emitters. In the ideal limit, permutation symmetry leads to a purely superradiant cascade, in which the system transitions from the doubly excited $|\bar{W}\rangle$ state exclusively to the single-excitation symmetric Dicke state~\cite{Dicke_1954,PRA_dicke_2000}, commonly known as the $W$ state, given by
 \begin{align}
     |W\rangle = \frac{1}{\sqrt{3}}(|e_1 g_2 g_3\rangle + |g_1 e_2 g_3\rangle + |g_1 g_2 e_3\rangle).
 \end{align}
The exact C-B-D hierarchy allows us to track this process by resolving the probability flow across the distinct excitation manifolds. We apply a collective decomposition to the sector amplitudes: in the atomic sector, the state is projected onto $|\bar{W}\rangle$ and its orthogonal dark pair states as
\begin{align}
|D_1^{(C)}\rangle &= \frac{1}{\sqrt{2}}(|e_1 e_2 g_3\rangle - |e_1 g_2 e_3\rangle), \\
|D_2^{(C)}\rangle &= \frac{1}{\sqrt{6}}(|e_1 e_2 g_3\rangle + |e_1 g_2 e_3\rangle - 2|g_1 e_2 e_3\rangle).
\end{align}
As shown below, any deviation from this ideal cascade can be interpreted in terms of interference between bright and dark decay pathways in the structured reservoir.
In structured reservoirs, the collective decay is not a simple exponential relaxation but arises from interference between two distinct pathways: a broadband radiative continuum associated with direct emission into propagating modes, and a spectrally narrow resonance originating from delayed feedback mediated by the environment. When the geometric delay is compatible with the collective phase relation of the emitters, destructive interference can suppress the radiative channel and favor population transfer into subradiant dark states.

\begin{figure}
    \centering
    \begin{minipage}{0.9\linewidth}
        \centering
        \includegraphics[width=\linewidth]{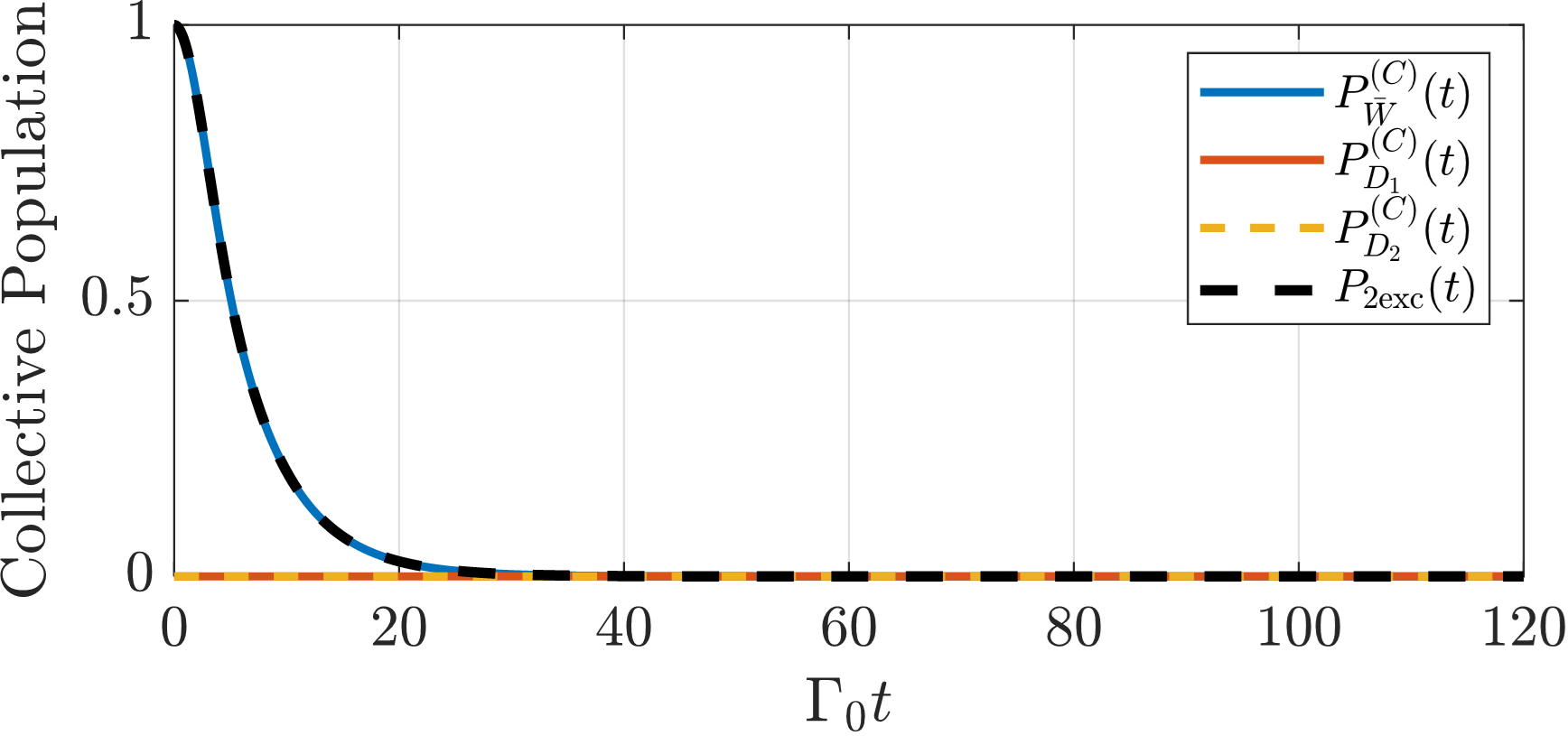}
        \subcaption{Populations in two-excitation}
    \end{minipage}
    \hfill
    \begin{minipage}{0.9\linewidth}
        \centering
        \includegraphics[width=\linewidth]{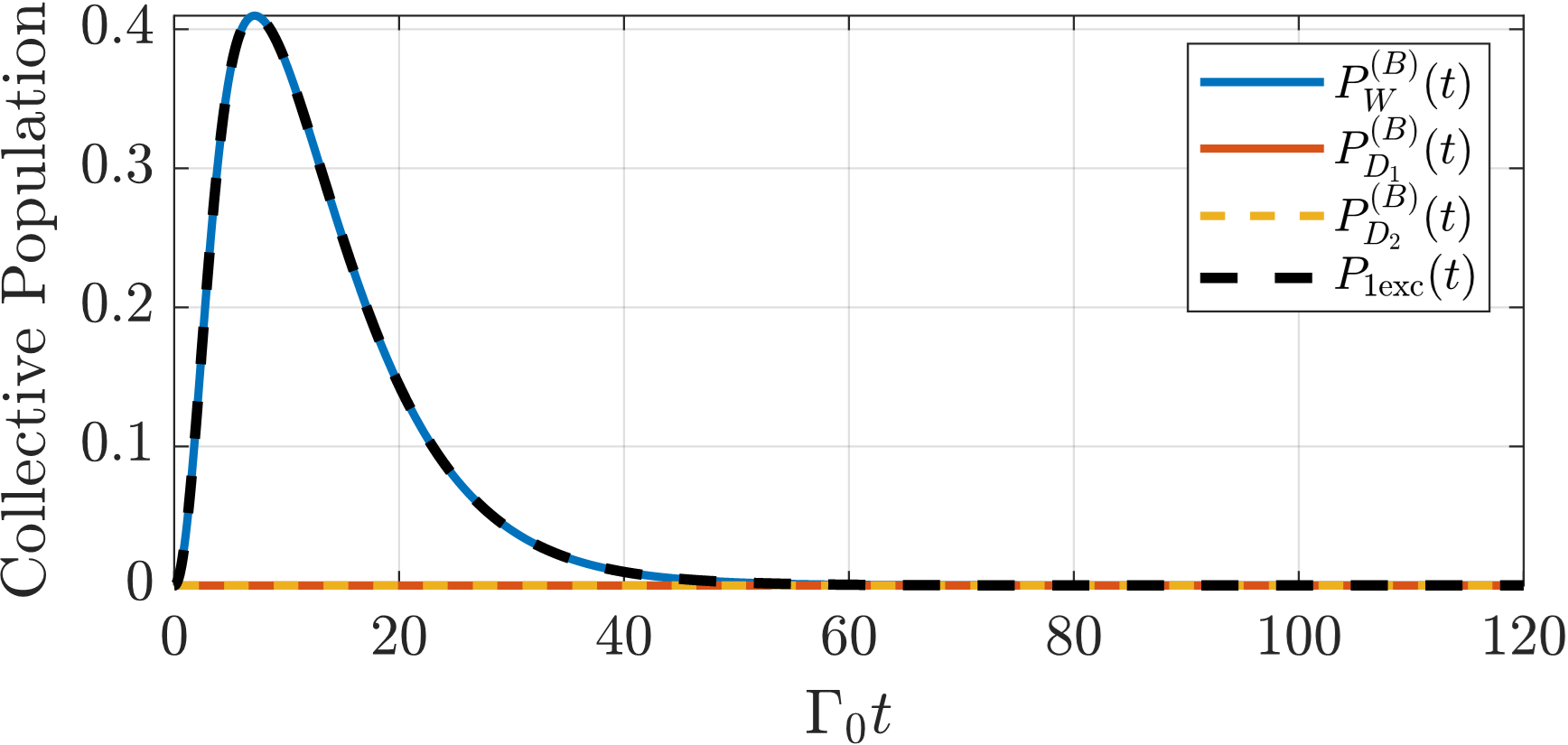}
        \subcaption{Populations in single-excitation}
    \end{minipage}
    \caption{
        Collective decay dynamics of the symmetric Dicke state in free space with $d = 0.01\lambda_a$. 
    }
    \label{fig:Dicke_freespace}
\end{figure}

Fig.~\ref{fig:Dicke_freespace} illustrates the free-space results, where the emitters are separated by $d = 0.01\,\lambda_a$. In this example, we keep the atomic frequency the same as in the previous case, but set $d_{\mathrm{eff}}=0.1$ to access a regime in which collective phase coherence is preserved without being obscured by strong individual decay. In the symmetric limit, the retardation-induced phase mismatch is negligible across the emission bandwidth, and the permutation symmetry of the emitters is effectively preserved. Consequently, the population $P_{\bar W}^{(C)}(t)$ flows entirely into the bright $P_{W}^{(B)}(t)$ channel without any leakage into the dark manifold. This benchmark is consistent with the expected Dicke-like collective decay in the near-symmetric limit and provides a useful reference point for the structured-environment calculations.

In contrast to the free-space benchmark above, where permutation symmetry is preserved, the presence of a structured environment significantly modifies the collective emission pathways. We introduce a high-index dielectric slab ($\epsilon = 12 + 0.05i$) positioned at $x \in [1.25\lambda_a,\,1.5\lambda_a]$. This structure induces strong feedback, enabling an analysis of how structural reflection and dissipation collectively reshape the many-body non-Markovian dynamics. The emitters are positioned asymmetrically with respect to the mirror-induced standing wave, $\{0.25\lambda_a,\, 0.5\lambda_a,\,0.75\lambda_a\}$. This geometry breaks permutation symmetry, as emitter 2 experiences strongly reduced radiative coupling while emitters 1 and 3 remain efficiently coupled to the continuum.

\begin{figure}
    \centering
    \begin{minipage}{0.9\linewidth}
        \centering
        \includegraphics[width=\linewidth]{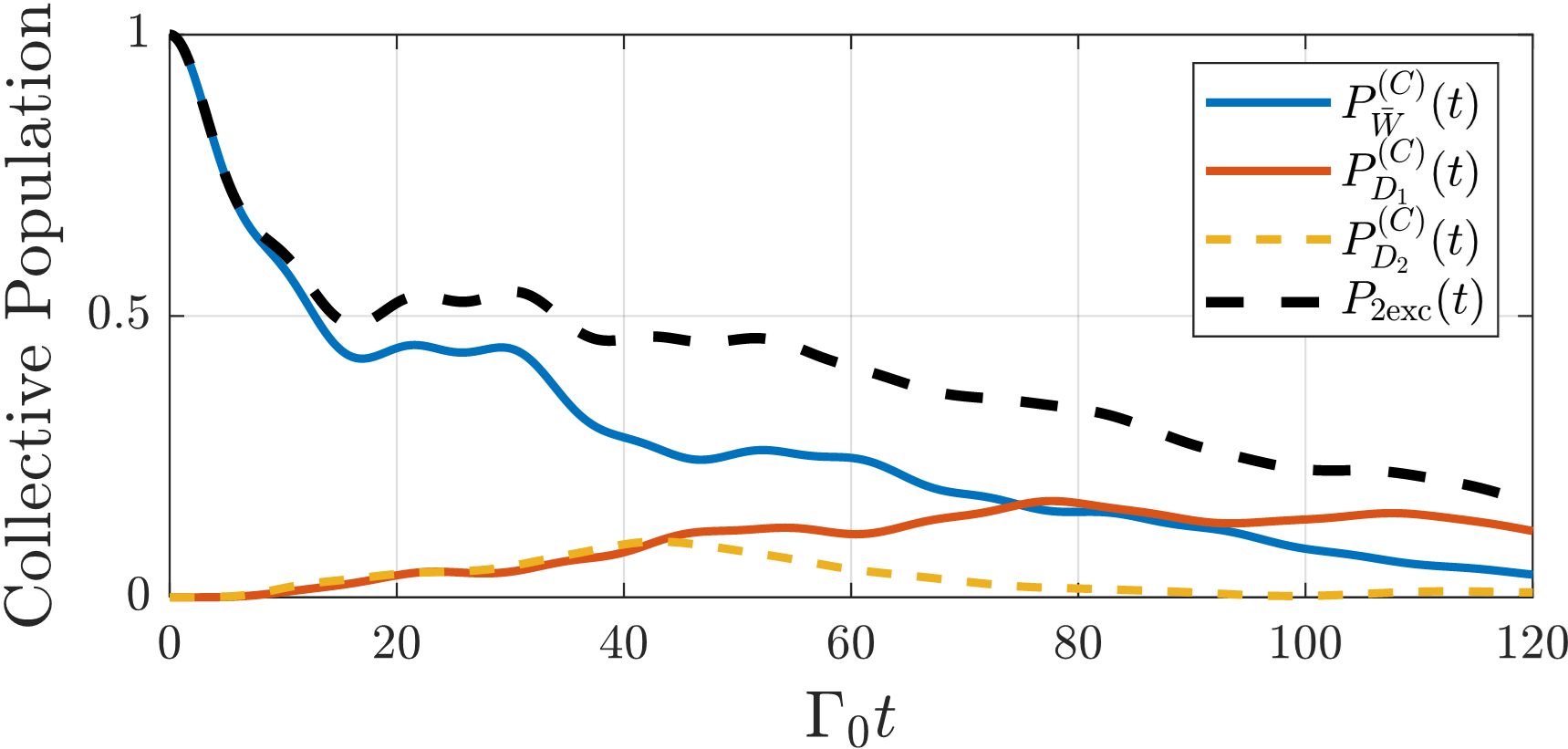}
        \subcaption{Populations in two-excitation}
    \end{minipage}
    \hfill
    \begin{minipage}{0.9\linewidth}
        \centering
        \includegraphics[width=\linewidth]{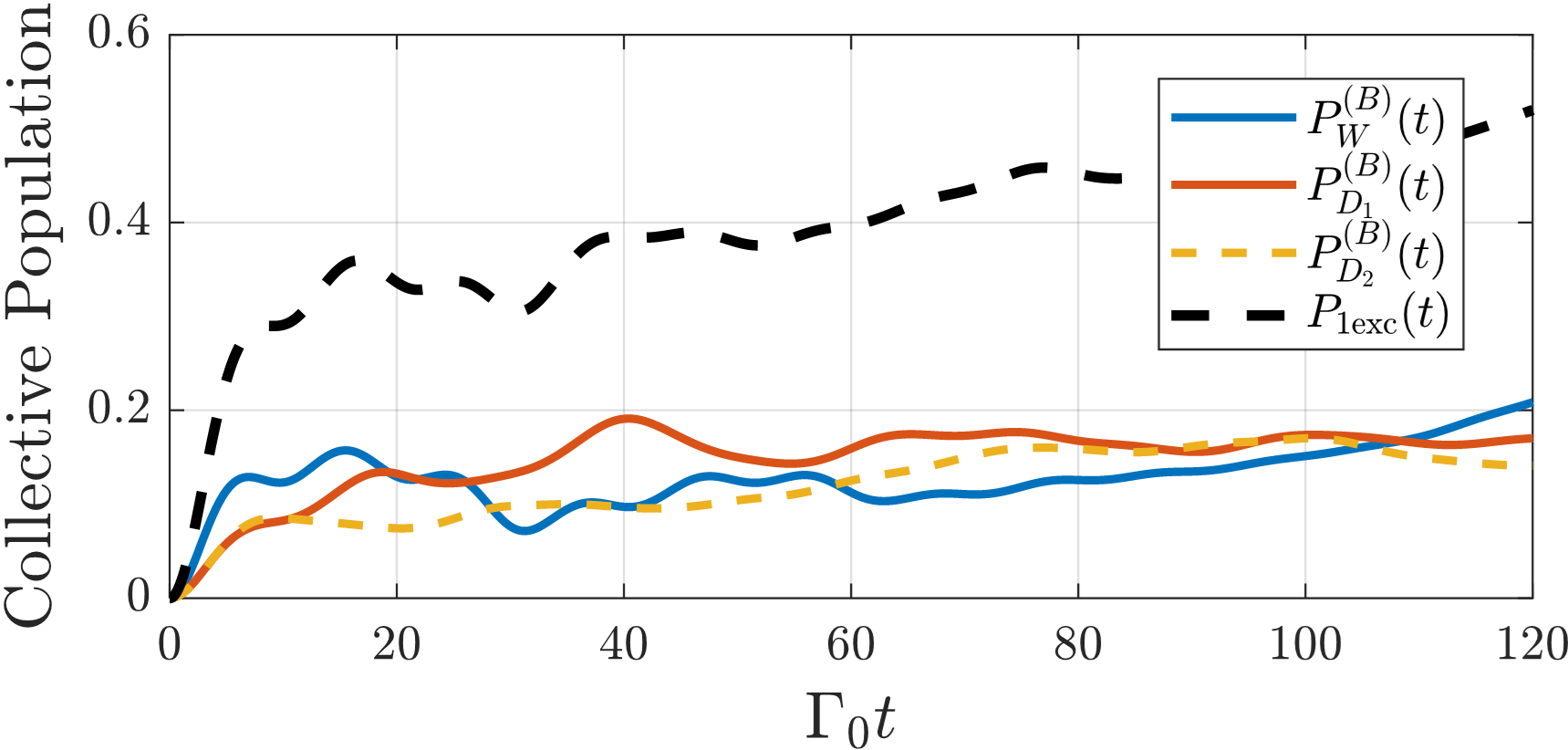}
        \subcaption{Populations in single-excitation}
    \end{minipage}
    \caption{
        Collective decay dynamics of the symmetric Dicke state in structured environments. 
    }
    \label{fig:Dicke_structured}
\end{figure}

Fig.~\ref{fig:Dicke_structured} shows the resulting non-Markovian dynamics. In the doubly-excited $C$ sector (top panel), the initial population $P_{\bar W}^{(C)}(t)$ exhibits pronounced oscillatory behavior due to environmental feedback mediated by the structured reservoir. The geometric asymmetry induces a substantial transfer of population into the orthogonal dark pair states $P_{D_1}^{(C)}(t)$ and $P_{D_2}^{(C)}(t)$, reflecting the breakdown of collective permutation symmetry at the level of the emitter subspace.
This symmetry breaking becomes even more pronounced in the intermediate one-photon $B$ sector. Because emitter 2 is located near a node of the standing wave, radiative decay from this site is strongly inhibited. As a result, the excitation is preferentially redistributed into configurations where emitter 2 remains excited. When expressed in the collective basis, this leads to a significant population transfer into the subradiant dark channels $P_{D_1}^{(B)}(t)$ and $P_{D_2}^{(B)}(t)$.

\begin{figure}
    \centering
    \includegraphics[width=\linewidth]{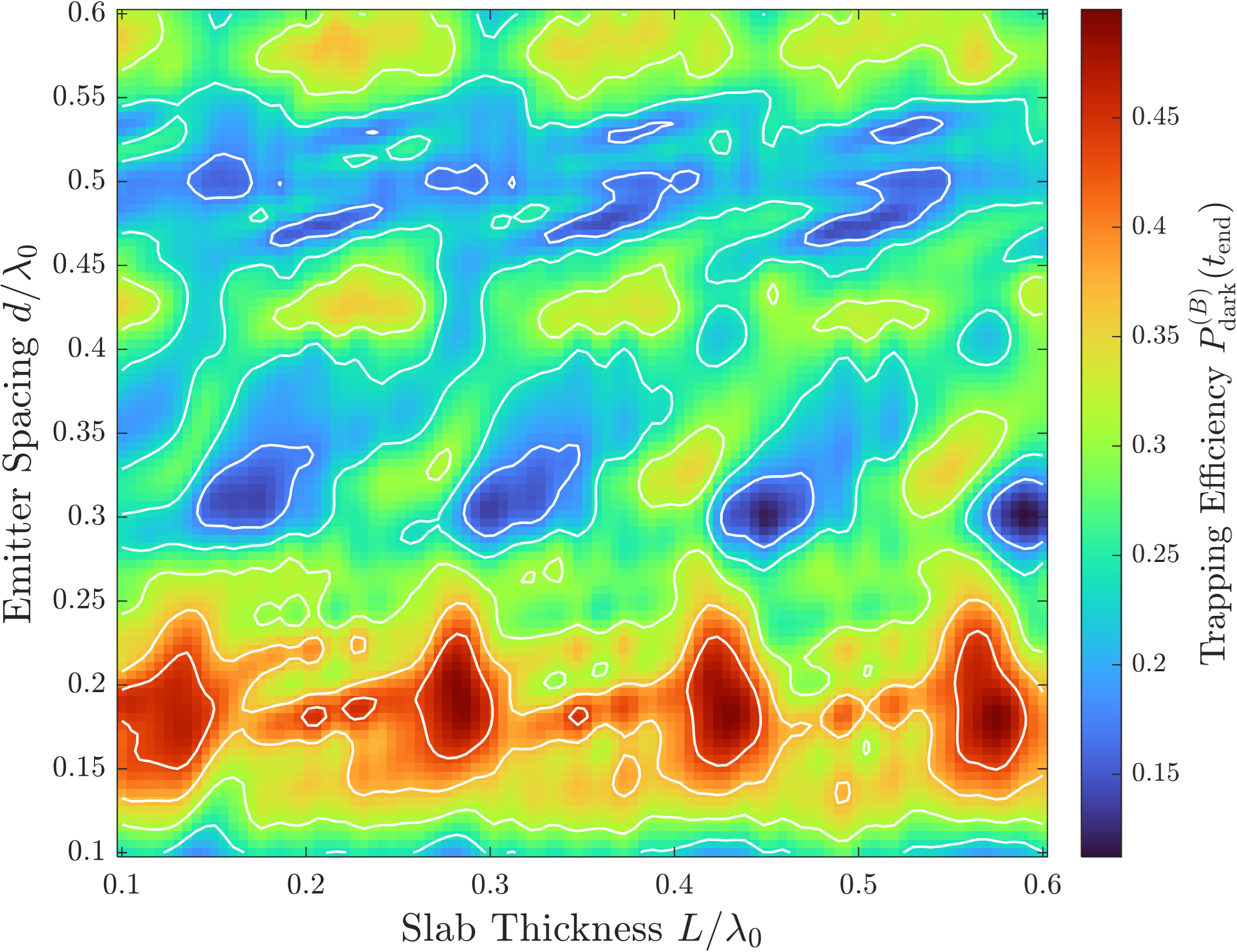}
    \caption{Two-dimensional map of the asymptotic trapping efficiency $P_{\mathrm{dark}}^{(B)}(t_{\mathrm{end}})$ as a function of the slab thickness $L/\lambda_a$ and the inter-emitter spacing $d/\lambda_a$. }
    \label{fig:trapping_phase_map}
\end{figure}

We also map the final-time dark-manifold population across a two-dimensional geometric parameter space. By varying the emitter spacing $d$ and the slab thickness $L$, we evaluate the final-time population accumulated in the single-excitation dark manifold,
\begin{equation}
P_{\mathrm{dark}}^{(B)}(t_{\mathrm{end}})
=
P_{D_1}^{(B)}(t_{\mathrm{end}})
+
P_{D_2}^{(B)}(t_{\mathrm{end}}).
\end{equation}
In this configuration, three emitters are placed sequentially at
$\{1.0\lambda_a,\;1.0\lambda_a+d,\;1.0\lambda_a+2d\}$,
while a high-index slab of thickness $L$ and complex permittivity
$\varepsilon_s = 12 + 0.05 i$
is introduced over the interval
$x\in[3.5\lambda_a,\;3.5\lambda_a+L]$.
The resulting phase map, shown in Fig.~\ref{fig:trapping_phase_map}, exhibits a structured interference landscape. Varying $L$ modifies the phase accumulated through slab-mediated delayed feedback, while varying $d$ changes the collective phase relation among the emitters. As a result, the final dark-manifold population displays pronounced resonance-like regions in which the radiative cascade into bright channels is substantially suppressed and population is preferentially redirected into the dark manifold. These results demonstrate that the present framework enables systematic identification of geometric parameter regimes associated with enhanced dark-manifold trapping in structured photonic environments.

\subsection{Atomic entanglement dynamics}

We now examine the bipartite atomic entanglement dynamics extracted directly from the reduced atomic density matrix. For all simulations in this subsection, the intrinsic atomic parameters are the same as those used in the previous examples. To induce non-Markovian retardation and dissipative feedback, we consider the single-end one-dimensional waveguide configuration with an embedded lossy dielectric slab in the interval $x \in [1.5\lambda_a,\,2.5\lambda_a]$, with complex permittivity $\epsilon_r = 12 + 0.03 i$. The reduced atomic density matrix is reconstructed from the exact hierarchy amplitudes, allowing us to evaluate the concurrence in Eq.~\eqref{eq:concurrence}.

\begin{figure}[t!]
    \centering
    \begin{subfigure}{0.9\columnwidth}
        \centering
        \includegraphics[width=\linewidth]{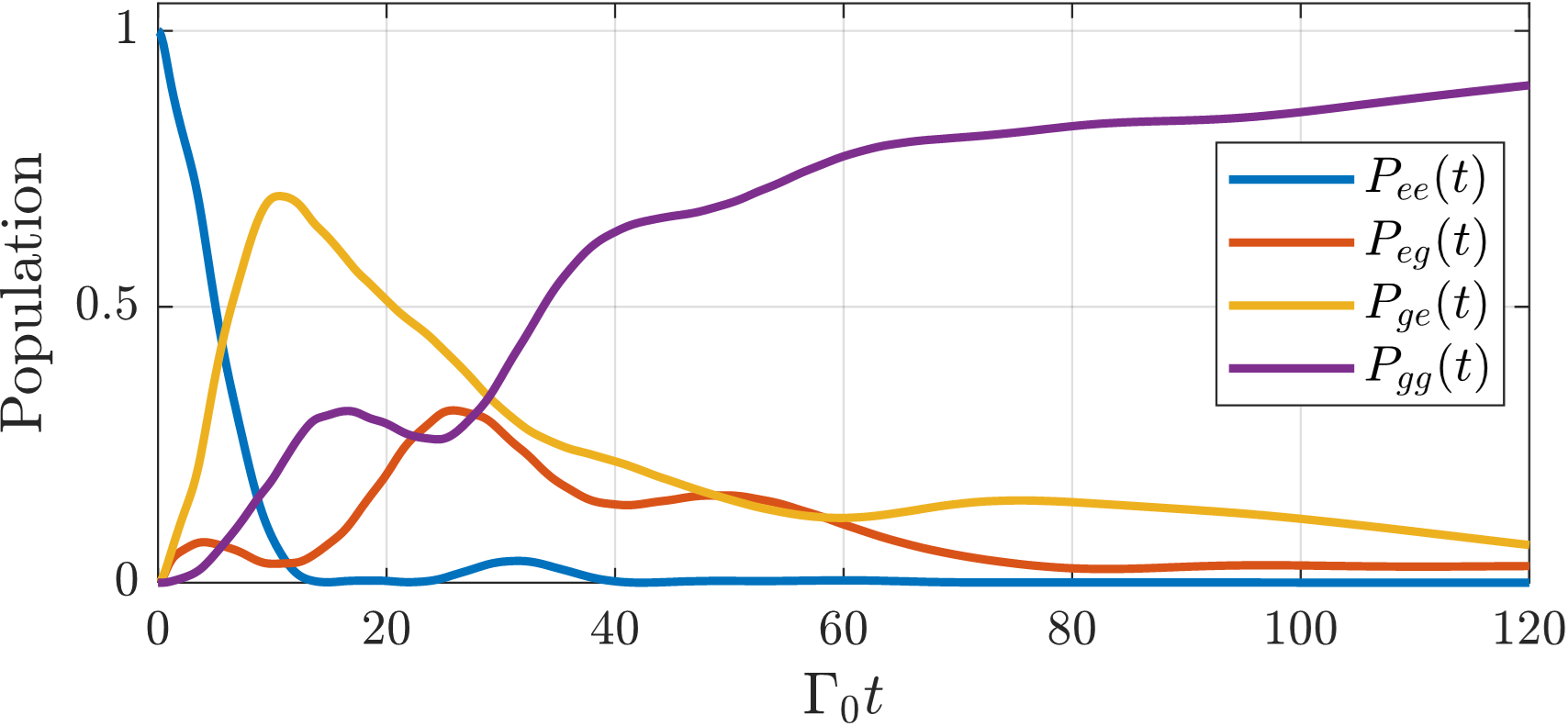}
        \caption{Atomic populations}
        \label{fig:entanglement_ee_pop}
    \end{subfigure}
    
    \begin{subfigure}{0.9\columnwidth}
        \centering
        \includegraphics[width=\linewidth]{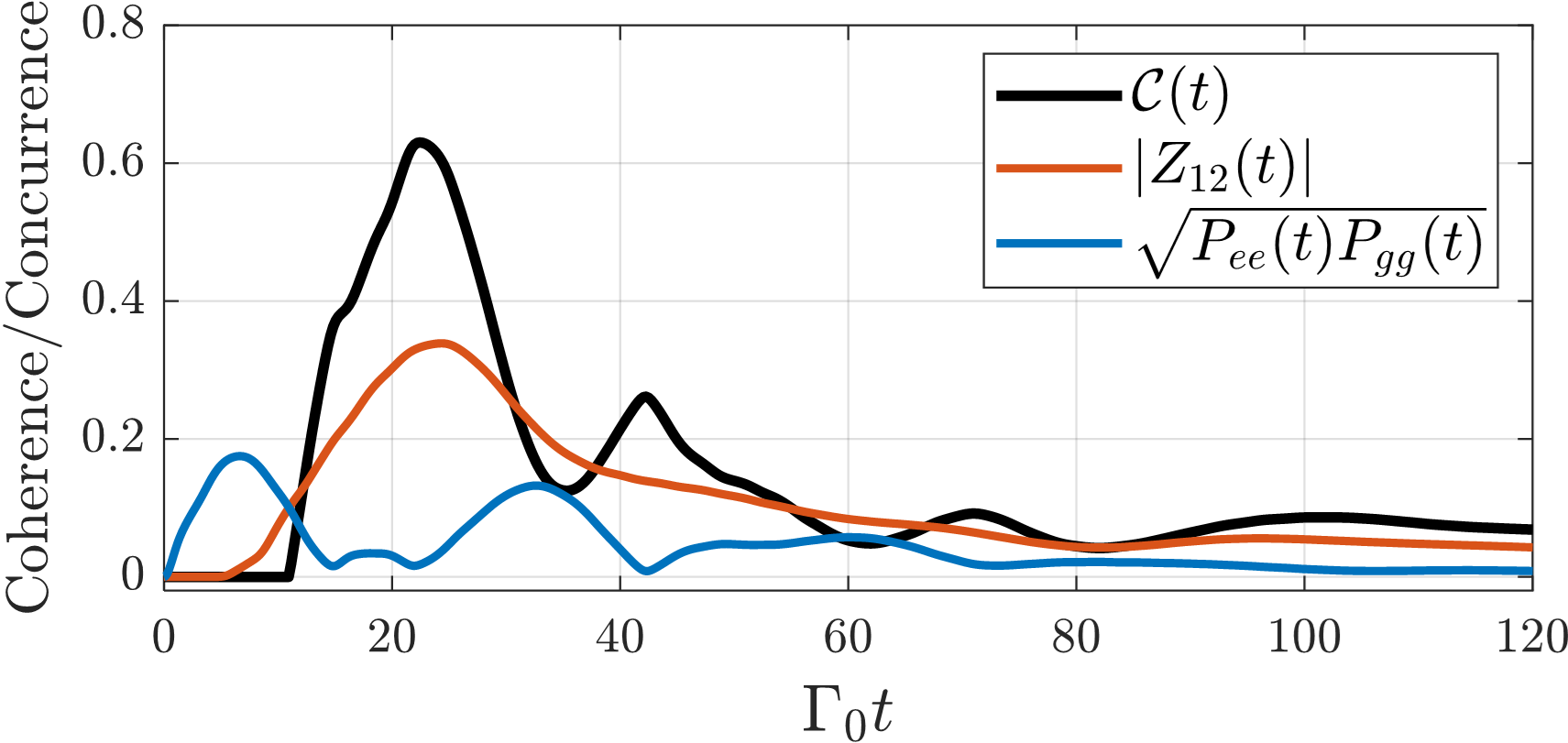}
        \caption{Coherence and concurrence}
        \label{fig:entanglement_ee_conc}
    \end{subfigure}
    
    \caption{
        Bipartite entanglement dynamics for two emitters initialized in the separable state $|e_1, e_2\rangle$. 
    }
    \label{fig:entanglement_ee}
\end{figure}

We first consider the two-emitter configuration initialized in the separable doubly excited state $|e_1,e_2\rangle$. The emitters are located at $\{1.25\lambda_a,\,3\lambda_a\}$
so that emitter 1 lies between the mirror and the slab, whereas emitter 2 is positioned beyond the slab. In this case, the initial reduced density matrix satisfies $P_{ee}(0)=1$ and $P_{eg}(0)=P_{ge}(0)=P_{gg}(0)=Z_{12}(0)=0$, so that the concurrence vanishes at $t=0$. As the dynamics evolve, the doubly excited population is transferred into the single-excitation sectors, while the structured reservoir generates a finite coherence $Z_{12}(t)$ between the two emitters. Fig.~\ref{fig:entanglement_ee} shows that the concurrence remains zero during the early stage of the evolution even though the coherence begins to build up, because the inequality
\begin{equation}
    |Z_{12}(t)| \le \sqrt{P_{ee}(t)P_{gg}(t)}
\end{equation}
is still satisfied. After a finite retardation time, however, the coherence exceeds the population threshold, and $\mathcal{C}(t)$ becomes nonzero. This behavior is consistent with entanglement sudden birth induced by the structured photonic environment. The later oscillatory decay and weak revival of $\mathcal{C}(t)$ are consistent with delayed photon-mediated reinteraction between the emitters, reflecting the non-Markovian nature of the relaxation dynamics.

\begin{figure}[t!]
    \centering
    \begin{subfigure}{0.9\columnwidth}
        \centering
        \includegraphics[width=\linewidth]{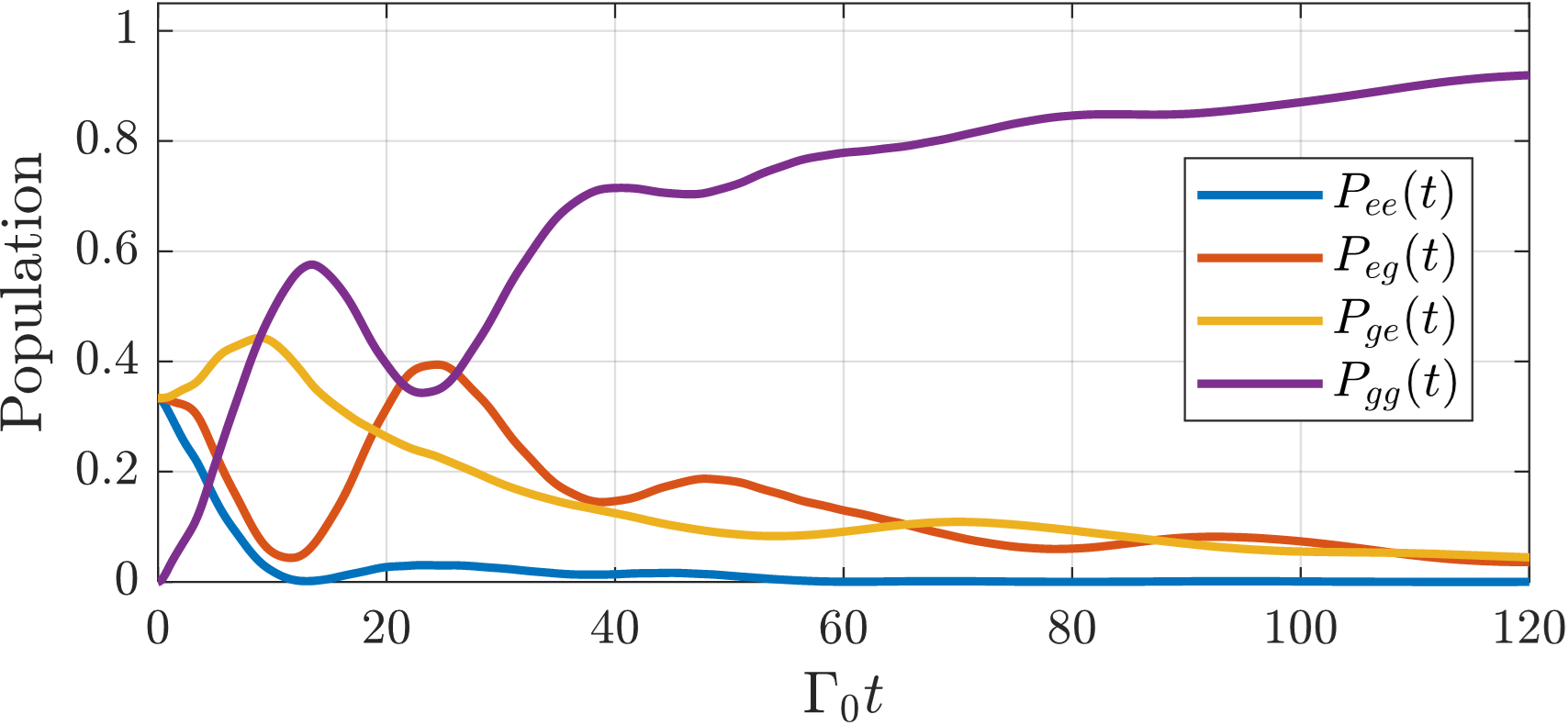}
        \caption{Atomic populations}
        \label{fig:entanglement_dicke_pop}
    \end{subfigure}
        \begin{subfigure}{0.9\columnwidth}
        \centering
        \includegraphics[width=\linewidth]{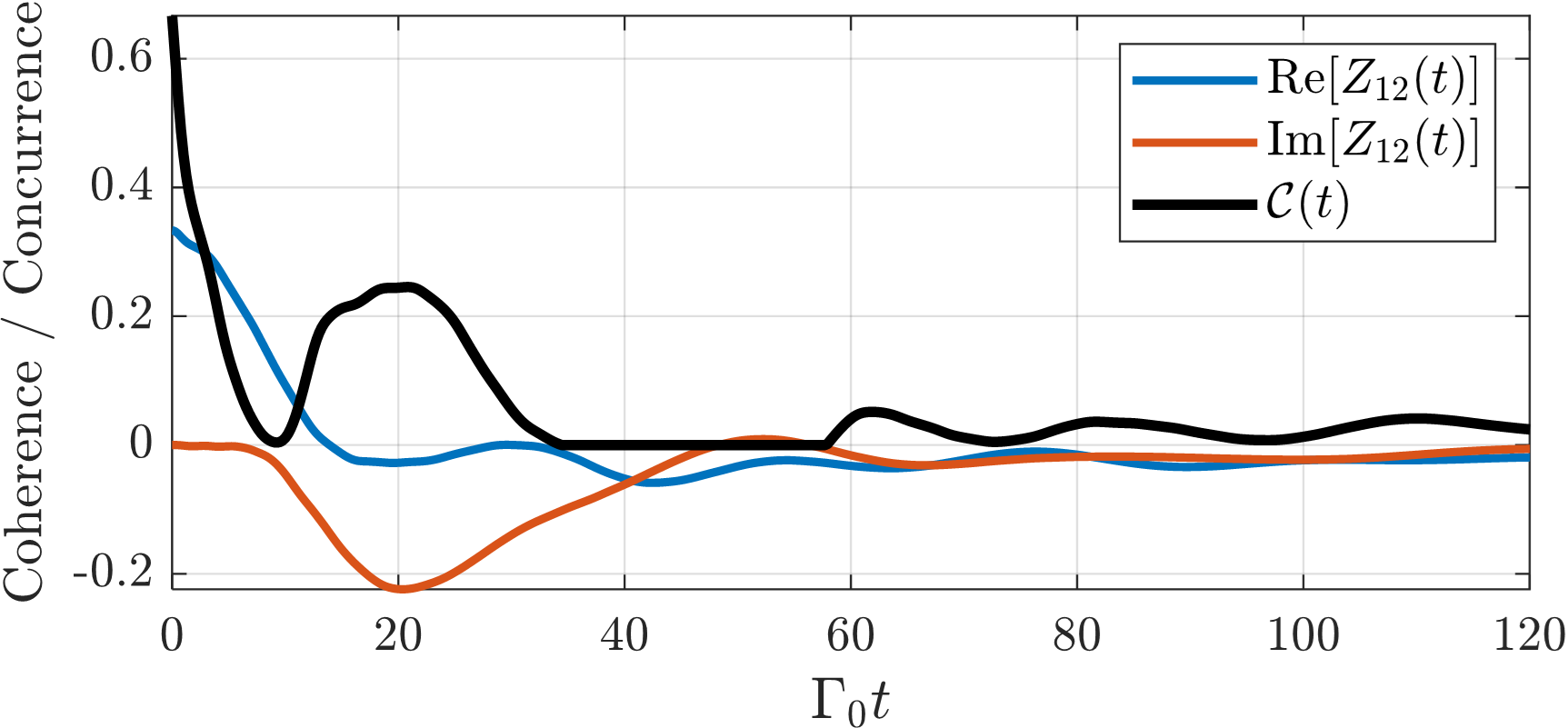}
        \caption{Coherence and concurrence}
        \label{fig:entanglement_dicke_conc}
    \end{subfigure}
    
    \caption{
        Entanglement dynamics for three emitters initialized in the symmetric Dicke state $|\bar{W}\rangle$. 
    }
    \label{fig:entanglement_dicke}
\end{figure}

We next turn to the three-emitter case initialized in the two-excitation symmetric Dicke state $|\bar W\rangle$, introduced above. In this simulation, the emitters are placed at
$x_1 = 1.25\lambda_a$, $x_2 = 3.0\lambda_a$ and $
x_3 = 4.0\lambda_a$. We then construct the reduced bipartite state of emitters 1 and 2 by tracing out emitter 3 together with the photonic DoFs. Since the initial state $|\bar W\rangle$ already contains pairwise quantum coherence, the reduced two-emitter concurrence is finite at $t=0$. The resulting dynamics therefore reflect the decay and partial revival of bipartite entanglement that is already present at $t=0$. Fig.~\ref{fig:entanglement_dicke} shows that the initial concurrence decreases rapidly as the collective excitation is redistributed into asymmetric atomic sectors and the photonic continuum. At intermediate times, the concurrence becomes strongly suppressed when $|Z_{12}(t)|$ falls below $\sqrt{P_{ee}(t)P_{gg}(t)}$. Nevertheless, the entanglement does not decay monotonically to zero. Instead, small but clearly resolved revivals emerge at later times, demonstrating that the structured reservoir transiently stores and feeds back bipartite quantum correlations into the selected atomic pair.

Taken together, these two examples illustrate complementary aspects of atomic entanglement dynamics in the exact two-excitation manifold. For the initially separable state $|e_1,e_2\rangle$, the concurrence exhibits entanglement sudden birth generated by reservoir-mediated coherence. For the initially correlated state $|\bar W\rangle$, the same framework captures the decay, temporary suppression, and partial revival of bipartite entanglement embedded in a larger many-body excitation. In both cases, the dynamics are obtained directly from the exact hierarchy amplitudes and therefore retain the full retardation, dissipation, and multi-photon interference effects of the structured environment.

\section{Conclusion}

In this work, we developed a Green’s-function-based computational framework for non-Markovian multi-emitter quantum electrodynamics in the two-excitation manifold. By combining the modified Langevin noise formalism with the emitter-centered mode construction, the resulting formulation yields a tractable yet explicit hierarchy that retains both the intermediate one-photon sector and the pure two-photon sector.

A central feature of the framework is that the electromagnetic environment enters entirely through the dyadic Green’s function evaluated at the emitter positions, while the subsequent dynamics are propagated within the same emitter-centered hierarchy. This structure provides direct access to both atomic and photonic observables from a unified set of amplitudes. In particular, it enables the evaluation of reduced atomic quantities such as populations, Bell-state fidelities, and concurrence, while also allowing reconstruction of the spatiotemporal field intensity. Because the two-photon sector is retained explicitly, the same framework also provides a systematic starting point for analyzing photonic quantities such as spectral correlations and photonic entanglement, as outlined in Appendix~\ref{app:two_photon}.

The representative semi-infinite waveguide examples demonstrate that the present approach captures a broad range of multi-excitation non-Markovian phenomena, including bound-state-like behavior, delayed excitation transfer, symmetry-breaking redistribution between bright and dark collective channels, and entanglement sudden birth and revival in structured dissipative environments. These results show that the present framework goes beyond reduced single-excitation descriptions by resolving how retardation, loss, and multi-photon interference jointly shape collective quantum dynamics. Although the numerical demonstrations in this work are restricted to representative one-dimensional geometries, the formulation itself is organized around the Green’s function and is therefore naturally suited for future extension to more complex electromagnetic environments and higher excitation manifolds.

\appendix

\section{Derivation of the ECM field commutation relation}
\label{app:Gamma_matrix}

The commutator between the projected positive-frequency operator at $\mathbf{R}_i$ and the negative-frequency operator at $\mathbf{R}_j$ is evaluated by expressing the field in the BA--MA basis as
\begin{align}
    &[\hat{E}^{(+)}_{i}(\omega), \hat{E}^{(-)}_{j}(\omega')] \nonumber \\
    &= \int_{\mathcal{D}_\lambda} d\lambda \int_{\mathcal{D}_\mu} d\mu (\mathbf{n}_i \cdot \mathbf{E}_{\omega, \lambda}(\mathbf{R}_i)) (\mathbf{n}_j \cdot \mathbf{E}_{\omega', \mu}^*(\mathbf{R}_j)) \nonumber \\
    &\quad \times [\hat{a}_{\omega, \lambda}, \hat{a}_{\omega', \mu}^\dagger].
\end{align}
By applying the fundamental bosonic commutation relation, $[\hat{a}_{\omega, \lambda}, \hat{a}_{\omega', \mu}^{\dagger}] = \delta_{\lambda\mu} \delta(\omega - \omega')$, the double integral collapses into a single integral over the degeneracy space as
\begin{align}
    &[\hat{E}^{(+)}_{i}(\omega), \hat{E}^{(-)}_{j}(\omega')] \nonumber \\
    &= \delta(\omega - \omega') \int_{\mathcal{D}_\lambda} d\lambda (\mathbf{n}_i \cdot \mathbf{E}_{\omega, \lambda}(\mathbf{R}_i)) (\mathbf{n}_j \cdot \mathbf{E}_{\omega, \lambda}^*(\mathbf{R}_j)).
\end{align}
Using the dyadic identity $(\mathbf{A} \cdot \mathbf{B})(\mathbf{C} \cdot \mathbf{D}) = \mathbf{A} \cdot (\mathbf{B} \otimes \mathbf{D}) \cdot \mathbf{C}$, the integrand is rearranged to reveal the dyadic product:
\begin{align}
    &[\hat{E}^{(+)}_{i}(\omega), \hat{E}^{(-)}_{j}(\omega')] \nonumber \\
    &= \delta(\omega - \omega') \mathbf{n}_i \cdot \left( \int_{\mathcal{D}_\lambda} d\lambda \, \mathbf{E}_{\omega, \lambda}(\mathbf{R}_i) \otimes \mathbf{E}_{\omega, \lambda}^*(\mathbf{R}_j) \right) \cdot \mathbf{n}_j.
    \label{eq:B3}
\end{align}
Finally, substituting \eqref{eq:BA--MA_completeness_TMC} into \eqref{eq:B3}, the commutation relation is expressed as
\begin{align}
    &[\hat{E}^{(+)}_{i}(\omega), \hat{E}^{(-)}_{j}(\omega')] \nonumber \\
    &= \frac{\hbar \omega^2 \delta(\omega - \omega')}{\pi \epsilon_0 c^2} \left( \mathbf{n}_i \cdot \Im[\overline{\mathbf{G}}(\mathbf{R}_i, \mathbf{R}_j, \omega)] \cdot \mathbf{n}_j \right) .
    \end{align}

\section{Explicit reconstruction of the electric field operator: Extension to M-LN framework}
\label{app:ecm_field_profile}

In this section, we explicitly derive the spatial profile of the ECM and the resulting reconstruction of the total electric field operator at an arbitrary position $\mathbf{r}$. Our goal is to rigorously extend the ECM framework—originally formulated using the macroscopic QED approach—to the M-LN formalism. In the macroscopic QED derivation, the field is expressed as a volume integral over the MA polarization density.
We prove that with the explicit addition of the BA continuum, the effective field reconstruction retains its dependence on the Green's function. We define the effective spatial profile $\mathbf{\Phi}_k(\mathbf{r}, \omega)$ of the $k$-th ECM operator $\hat{c}_k$ via the commutator
\begin{align}
    \mathbf{\Phi}_k(\mathbf{r}, \omega)
    &\equiv \left[ \hat{\mathbf{E}}^{(+)}(\mathbf{r}, \omega), \hat{c}_k^\dagger(\omega) \right] \nonumber \\
    &= \mathcal{A}(\omega) \sqrt{\gamma_k(\omega)} \sum_{j=1}^{N} V_{jk}(\omega) \mathbf{\Psi}_j(\mathbf{r}, \omega),
\end{align}
where $\mathbf{\Psi}_j(\mathbf{r}, \omega)$ is the localized field profile associated with the $j$-th emitter. Using the adjoint of the local field operator $\hat{E}_j^{(-)}$, we have

\begin{align}
    \hat{E}_j^{(-)}(\omega)
    = \int_{\mathcal{D}_\lambda} d\lambda \left( \mathbf{n}_j \cdot \mathbf{E}_{\omega, \lambda}(\mathbf{R}_j) \right)^* \hat{a}_{\omega, \lambda}^\dagger.
    \label{eq:law_ECM}
\end{align}
Expanding the positive-frequency field operator over the complete set of BA--MA modes as
\begin{align}
    \hat{\mathbf{E}}^{(+)}(\mathbf{r}, \omega)
    = \int_{\mathcal{D}_\lambda} d\lambda \, \mathbf{E}_{\omega, \lambda}(\mathbf{r}) \, \hat{a}_{\omega, \lambda},
\end{align}
We define the field profile as
\begin{align}
    \mathbf{\Psi}_j(\mathbf{r}, \omega) \equiv \frac{[\hat{\mathbf{E}}^{(+)}(\mathbf{r}, \omega), \hat{E}_j^{(-)}(\omega)]} { \mathcal{A}(\omega)\Gamma_{jj}(\omega)}
    \label{eq:field profile}
\end{align}
substitute Eq.~\eqref{eq:law_ECM} into Eq.~\eqref{eq:field profile}
\begin{align}
    \mathbf{\Psi}_j(\mathbf{r}, \omega)
    &= \frac{1}{\mathcal{A}(\omega)^2\Gamma_{jj}(\omega)} \int_{\mathcal{D}_\lambda} d\lambda \int_{\mathcal{D}_{\lambda'}} d\lambda' \mathbf{E}_{\omega, \lambda}(\mathbf{r}) \nonumber \\
    &\quad \times \left( \mathbf{E}_{\omega, \lambda'}^*(\mathbf{R}_j) \cdot \mathbf{n}_j \right) [\hat{a}_{\omega, \lambda}, \hat{a}_{\omega, \lambda'}^\dagger] \nonumber \\
    &= \frac{ \left( \int_{\mathcal{D}_\lambda} d\lambda \, \mathbf{E}_{\omega, \lambda}(\mathbf{r}) \otimes \mathbf{E}_{\omega, \lambda}^*(\mathbf{R}_j) \right) \cdot \mathbf{n}_j}{\mathcal{A}(\omega)^2\Gamma_{jj}(\omega)}.
\end{align}
Invoking Eq.~\eqref{eq:BA--MA_completeness_TMC}
we arrive at the localized field profile,
\begin{align}
    \mathbf{\Psi}_{j}(\mathbf{r}, \omega)
    = \frac{ \Im[\overline{\mathbf{G}}(\mathbf{r}, \mathbf{R}_j, \omega)] \cdot \mathbf{n}_j}{\Gamma_{jj}(\omega)}.
\end{align}
Finally, the reconstructed field operator in the orthonormal ECM basis is given by
\begin{align}
    \hat{\mathbf{E}}^{(+)}(\mathbf{r})
    &= \sum_{k=1}^{N} \int_{0}^{\infty} d\omega \left( \sum_{j=1}^{N} V_{jk}(\omega) \mathcal{A}(\omega) \sqrt{\gamma_k} \mathbf{\Psi}_j(\mathbf{r}, \omega) \right) \nonumber \\
    &\quad \times \hat{c}_k(\omega).
    \label{eq:new_field_reconstruction_final}
\end{align}

\section{Derivation of norm preservation and bosonic symmetry}
\label{app:consistency_check}
In this appendix, we demonstrate that the derived non-Markovian hierarchy exactly preserves two fundamental physical properties: the conservation of total probability and the bosonic exchange symmetry of the two-photon continuum.

The total probability is given by the norm of the exact two-excitation state in Eq.~\eqref{eq:Main_Ansatz_Revised}
\begin{equation}
\begin{aligned}
P_{\mathrm{tot}}(t) = &\sum_{a<b} |C_{ab}(t)|^2 + \sum_{a,k} \int_0^\infty d\omega \, |B_{a,k\omega}(t)|^2 \\
&+ \frac{1}{2}\sum_{k,l}\iint_0^\infty d\omega \, d\omega' \, |D_{k\omega,l\omega'}(t)|^2.
\end{aligned}
\label{eq:F1}
\end{equation}
Differentiating with respect to time gives
\begin{widetext}
\begin{equation}
\frac{d}{dt}P_{\mathrm{tot}}(t) = 2\,\mathrm{Re} \left[ \sum_{a<b} C_{ab}^*(t)\dot{C}_{ab}(t) + \sum_{a,k}\int_0^\infty d\omega \, B_{a,k\omega}^*(t)\dot{B}_{a,k\omega}(t) + \frac{1}{2}\sum_{k,l}\iint_0^\infty d\omega \, d\omega' \, D_{k\omega,l\omega'}^*(t)\dot{D}_{k\omega,l\omega'}(t) \right].
\label{eq:F2}
\end{equation}
\end{widetext}
Substituting Eqs.~\eqref{eq:C_sector_EOM}, \eqref{eq:B_sector_EOM} and \eqref{eq:D_sector_EOM} into Eq.~\eqref{eq:F2}, all free-evolution terms proportional to $(\omega_a+\omega_b)$, $(\omega_a+\omega)$, and $(\omega+\omega')$ drop out identically, since they are purely imaginary inside the real part. The remaining interaction contributions can be grouped into probability fluxes between adjacent sectors
\begin{widetext}
\begin{align}
\frac{d}{dt}P_{\mathrm{tot}}(t)
&= i\sum_{a<b}\sum_k\int_0^\infty d\omega
\Big[
g_{bk}^*(\omega)C_{ab}(t)B_{a,k\omega}^*(t)
+
g_{ak}^*(\omega)C_{ab}(t)B_{b,k\omega}^*(t)
- \mathrm{c.c.}
\Big]
\notag\\
&\quad
+ i\sum_{a<b}\sum_k\int_0^\infty d\omega
\Big[
g_{bk}(\omega)C_{ab}^*(t)B_{a,k\omega}(t)
+
g_{ak}(\omega)C_{ab}^*(t)B_{b,k\omega}(t)
- \mathrm{c.c.}
\Big]
\notag\\
&\quad
+ i\sum_{a,k,l}\iint_0^\infty d\omega\,d\omega'
\Big[
g_{al}^*(\omega')B_{a,k\omega}(t)D_{k\omega,l\omega'}^*(t)
- \mathrm{c.c.}
\Big]
\notag\\
&\quad
+ i\sum_{a,k,l}\iint_0^\infty d\omega\,d\omega'
\Big[
g_{al}(\omega')B_{a,k\omega}^*(t)D_{k\omega,l\omega'}(t)
- \mathrm{c.c.}
\Big].
\label{eq:F3}
\end{align}
\end{widetext}
where $\text{c.c.}$ means complex conjugate.
The first two lines in Eq.~\ref{eq:F3} cancel pairwise and represent the probability flux exchanged between the doubly excited sector and the intermediate one-photon sector. Likewise, the last two lines cancel pairwise and represent the flux exchanged between the one-photon sector and the pure two-photon sector. Therefore, all interaction-induced probability currents vanish exactly, and one obtains
\begin{equation}
\frac{d}{dt}P_{\mathrm{tot}}(t)=0.
\label{eq:F4}
\end{equation}
This establishes that the exact non-Markovian hierarchy preserves the norm of the state and is therefore fully consistent with the unitary dynamics generated by the Hermitian Hamiltonian within the two-excitation manifold.

We next examine the bosonic exchange symmetry of the two-photon amplitude. Because the photons are indistinguishable bosons, the pure two-photon sector must satisfy
\begin{equation}
D_{k\omega,l\omega'}(t)=D_{l\omega',k\omega}(t).
\label{eq:F5}
\end{equation}
To verify that this symmetry is dynamically preserved, we define the antisymmetric component
\begin{equation}
\Delta_{k\omega,l\omega'}(t)
\equiv
D_{k\omega,l\omega'}(t)-D_{l\omega',k\omega}(t).
\label{eq:F6}
\end{equation}
Using Eq.~\eqref{eq:D_sector_EOM}, we obtain
\begin{align}
\begin{aligned}
i\dot{D}_{k\omega,l\omega'}(t) &= (\omega+\omega')D_{k\omega,l\omega'}(t) \\
&+ \sum_a \Big[ g_{al}^*(\omega')B_{a,k\omega}(t) + g_{ak}^*(\omega)B_{a,l\omega'}(t) \Big],
\end{aligned}
\label{eq:F7}
\end{align}
and, after exchanging $(k,\omega)$$(l,\omega')$,
\begin{align}
\begin{aligned}
i\dot{D}_{l\omega',k\omega}(t) &= (\omega+\omega')D_{l\omega',k\omega}(t) \\
&+ \sum_a \Big[ g_{ak}^*(\omega)B_{a,l\omega'}(t) + g_{al}^*(\omega')B_{a,k\omega}(t) \Big].
\end{aligned}
\label{eq:F8}
\end{align}
The source terms on the right-hand sides of Equations.~(\ref{eq:F7}) and (\ref{eq:F8}) are identical. Subtracting the two equations therefore gives
\begin{equation}
i\dot{\Delta}_{k\omega,l\omega'}(t)
=
(\omega+\omega')\Delta_{k\omega,l\omega'}(t).
\label{eq:F9}
\end{equation}
Its solution is
\begin{equation}
\Delta_{k\omega,l\omega'}(t)
=
e^{-i(\omega+\omega')t}\Delta_{k\omega,l\omega'}(0).
\label{eq:F10}
\end{equation}
Hence, if the initial state is symmetric, for example
\begin{equation}
D_{k\omega,l\omega'}(0)=D_{l\omega',k\omega}(0),
\label{eq:F11}
\end{equation}
or in particular $D_{k\omega,l\omega'}(0)=0$, then Eq.~\eqref{eq:F10} implies
\begin{equation}
\Delta_{k\omega,l\omega'}(t)=0
\label{eq:F12}
\end{equation}
Therefore,
\begin{equation}
D_{k\omega,l\omega'}(t)=D_{l\omega',k\omega}(t)
\label{eq:F13}
\end{equation}

showing that the exact non-Markovian hierarchy remains strictly confined to the symmetric two-photon subspace. The bosonic indistinguishability of the emitted photons is thus preserved automatically by the time evolution, without requiring any additional symmetrization procedure.
\begin{figure}[htbp]
    \centering
    \includegraphics[width=\linewidth]{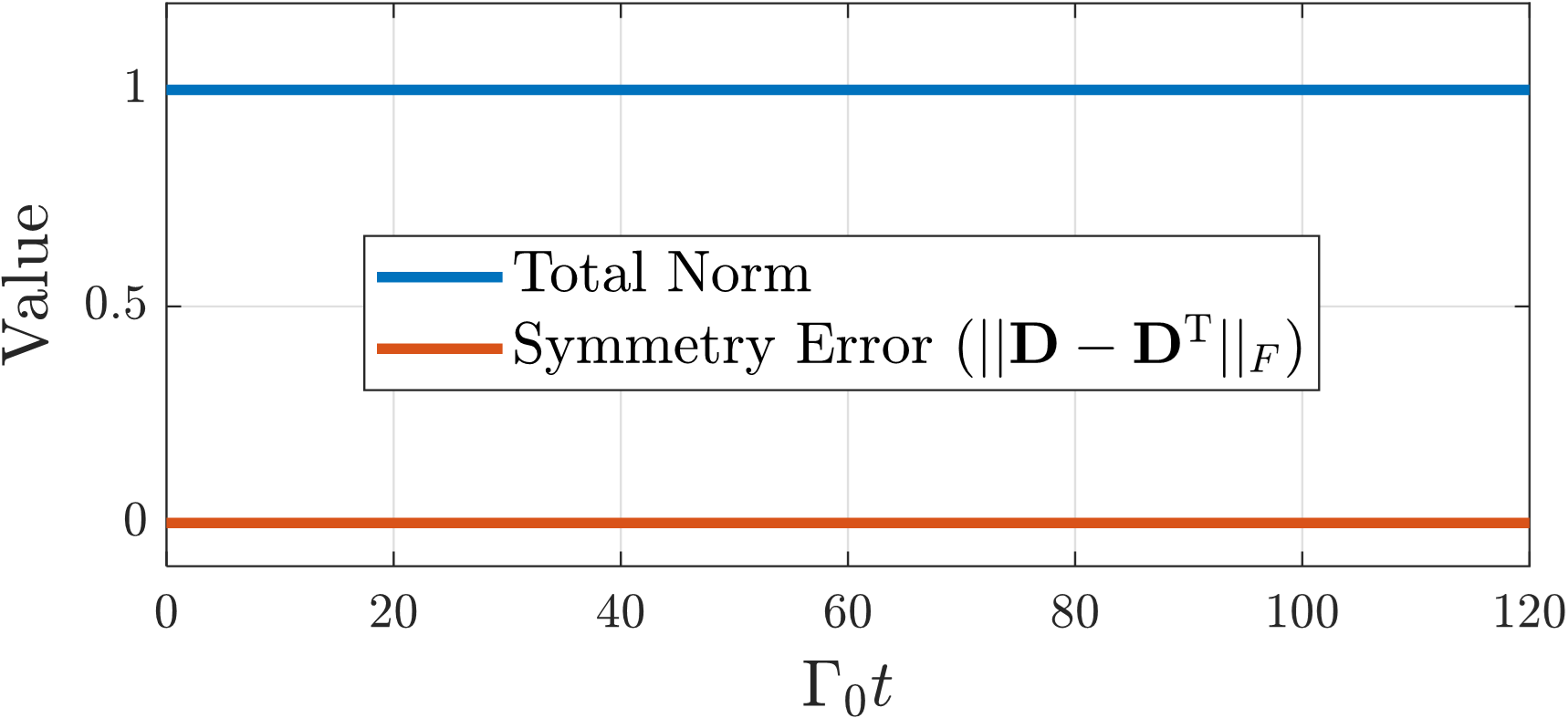}
    \caption{Consistency analysis for the non-Markovian hierarchy in Simulation of Fig.~\ref{fig:1D_lossy_slab}}
    \label{fig:consistency_check}
\end{figure}

\section{Reconstruction of the Spatiotemporal Field Intensity}
\label{app:field_reconstruction_2}

In this appendix, we detail the derivation of the one-body photonic density kernel $\Gamma_{k\omega, k'\omega'}(t)$ and the spatiotemporal field intensity $I(\mathbf{r},t)$ from the exact two-quantum state vector $|\Psi(t)\rangle$. 
The first-order spatiotemporal field intensity is defined as the expectation value of the photon number density at position $\mathbf{r}$ and time $t$,
\begin{equation}
    I(\mathbf{r},t) = \big\langle \Psi(t)\big| \hat{\mathbf{E}}^{(-)}(\mathbf{r}) \cdot \hat{\mathbf{E}}^{(+)}(\mathbf{r}) \big|\Psi(t)\big\rangle.
    \label{eq:app_intensity_norm}
\end{equation}
The positive-frequency electric field operator, expanded in the complete set of Emitter-Centered Modes (ECMs), is given by
\begin{equation}
    \hat{\mathbf{E}}^{(+)}(\mathbf{r}) = \sum_p \int_0^\infty d\Omega\, \mathbf{E}_p(\mathbf{r},\Omega)\, \hat{c}_{p\Omega},
    \label{eq:app_E_operator}
\end{equation}
where $\hat{c}_{p\Omega}$ is the bosonic annihilation operator for the $p$-th bright mode at frequency $\Omega$.

To evaluate Eq.~\eqref{eq:app_intensity_norm}, we apply the annihilation operator $\hat{c}_{p\Omega}$ to the truncated ansatz $|\Psi(t)\rangle$ defined in the main text. The application of $\hat{c}_{p\Omega}$ lowers the photon number of each sector: the pure atomic sector vanishes, the intermediate one-photon sector is reduced to the zero-photon state, and the pure two-photon continuum is reduced to a single-photon state. Mathematically, this yields
\begin{align}
    \hat{c}_{p\Omega} |\Psi(t)\rangle 
    &= \sum_{a=1}^N B_{a, p\Omega}(t) |e_a; \{0\}\rangle \nonumber \\
    &\quad + \sum_{l=1}^N \int_0^\infty d\nu\, D_{pl, \Omega\nu}(t) |\{g\}; 1_{l\nu}\rangle,
    \label{eq:app_annihilated_state}
\end{align}
where we have explicitly used the bosonic symmetry of the pure two-photon amplitude, $D_{kl, \omega\nu}(t) = D_{lk, \nu\omega}(t)$, to combine identical terms arising from the destruction of either the first or the second photon.

Substituting Eq.~\eqref{eq:app_E_operator} into Eq.~\eqref{eq:app_intensity_norm}, the field intensity is expressed as
\begin{align}
    I(\mathbf{r},t) &= \sum_{p,p'} \iint_0^\infty d\Omega d\Omega'\, \mathbf{E}_p^*(\mathbf{r},\Omega) \cdot \mathbf{E}_{p'}(\mathbf{r},\Omega') \nonumber \\
    &\quad \times \big\langle \Psi(t)\big| \hat{c}_{p\Omega}^\dagger \hat{c}_{p'\Omega'} \big|\Psi(t)\big\rangle.
\end{align}
The expectation value $\langle \Psi(t)| \hat{c}_{p\Omega}^\dagger \hat{c}_{p'\Omega'} |\Psi(t)\rangle$ precisely defines the one-body photonic density kernel $\Gamma_{p\Omega, p'\Omega'}(t)$. Taking the inner product of the state in Eq.~\eqref{eq:app_annihilated_state} with its dual, and recognizing that the atomic basis states $|e_a\rangle$ and $|\{g\}\rangle$ are strictly orthogonal ($\langle e_a | \{g\} \rangle = 0$), all cross terms between the intermediate and pure two-photon sectors identically vanish. The kernel thus separates into two distinct contributions:
\begin{align}
    \Gamma_{p\Omega, p'\Omega'}(t) &= \sum_{a=1}^N B_{a, p\Omega}^*(t) B_{a, p'\Omega'}(t) \nonumber \\
    &\quad + \sum_{l=1}^N \int_0^\infty d\nu\, D_{pl, \Omega\nu}^*(t) D_{p'l, \Omega'\nu}(t).
    \label{eq:app_final_kernel}
\end{align}
Inserting Eq.~\eqref{eq:app_final_kernel} back into the spatial mode expansion yields the final expression for $I(\mathbf{r},t)$ as
\begin{equation}
    I(\mathbf{r},t) = \sum_{k,k'} \iint_0^\infty d\omega d\omega'\, \mathbf{E}_k^*(\mathbf{r},\omega) \cdot \mathbf{E}_{k'}(\mathbf{r},\omega')\, \Gamma_{k\omega, k'\omega'}(t).
\end{equation}

\section{Reduction to the single-excitation limit}

We demonstrate that the theoretical framework developed for the two-excitation manifold is fundamentally consistent with the standard single-excitation Wigner-Weisskopf theory. By restricting the global Hilbert space to the single-excitation manifold, the state vector of the system at time $t$ is exactly expressed as
\begin{align}
\begin{aligned}
|\psi^{(1)}(t)\rangle = &\sum_{a=1}^N C_a(t) |e_a, 0\rangle \\
&+ \sum_{k} \int_0^\infty d\omega\, B_{k\omega}(t) |g, 1_{k\omega}\rangle.
\end{aligned}
\end{align}
where $|e_a, 0\rangle$ denotes the state where the $a$-th emitter is excited while all other emitters are in the ground state with zero photons in the reservoir. The state $|g, 1_{k\omega}\rangle$ represents the configuration where all emitters are in the ground state and a single photon with frequency $\omega$ is present in the reservoir mode $k$. The amplitudes $C_a(t)$ and $B_{k\omega}(t)$ are the corresponding probability amplitudes for these states.

\begin{figure}[htbp]
    \centering
    \begin{subfigure}{0.95\linewidth}
        \centering
        \includegraphics[width=\linewidth]{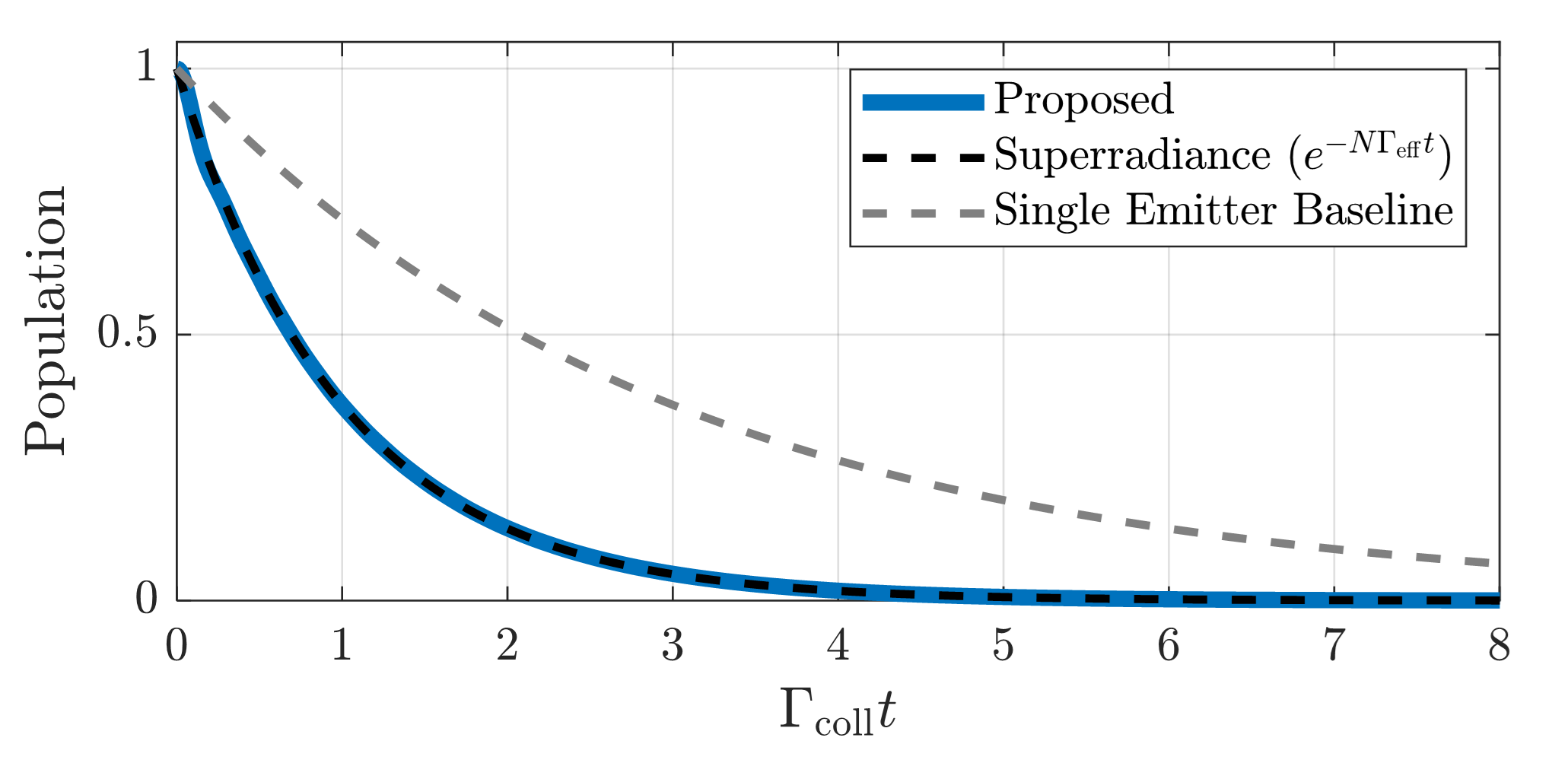}
        \caption{Population}
        \label{fig:single_population}
    \end{subfigure}
    \begin{subfigure}{0.95\linewidth}
        \centering
        \includegraphics[width=\linewidth]{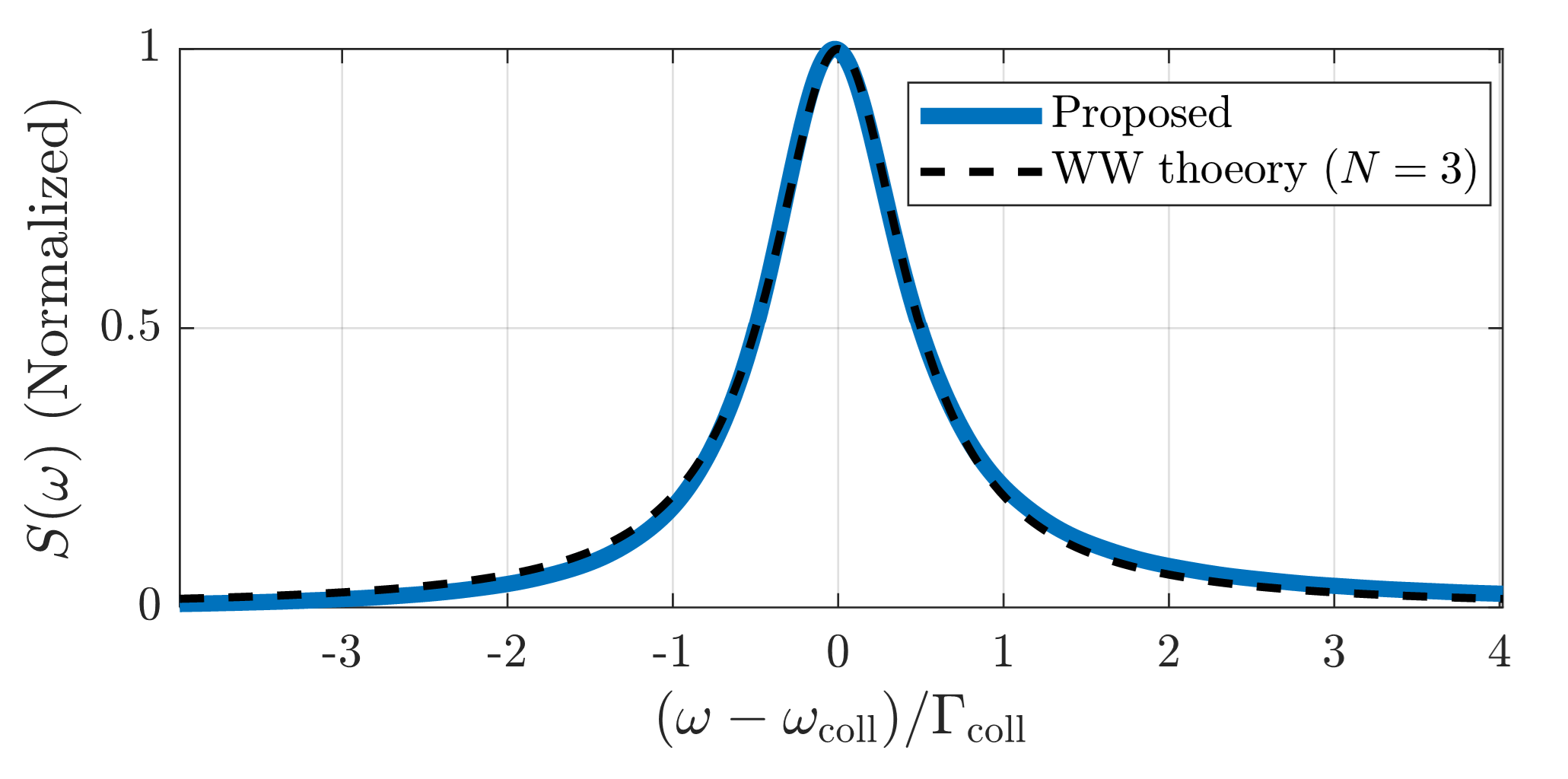}
        \caption{Spectral response}
        \label{fig:single_spectral}
    \end{subfigure}
    \caption{The dynamics of single-excitation and free space limit for three emitters compared with proposed framework and the analytical theory.}
    \label{fig:single_quanta_total}
\end{figure}

Applying the Schrödinger equation $i\partial_t |\psi^{(1)}(t)\rangle = H |\psi^{(1)}(t)\rangle$ using the total Hamiltonian yields the exact equations of motion for the single-excitation amplitudes:
\begin{align}
    i\dot{c}_a(t) &= \omega_a c_a(t) + \sum_{k} \int_0^\infty d\omega\, g_{ak}(\omega) b_{k\omega}(t) \label{eq:single_c} \\
    i\dot{b}_{k\omega}(t) &= \omega b_{k\omega}(t) + \sum_{a=1}^N g_{ak}^*(\omega) c_a(t) \label{eq:single_b}
\end{align}

This coupled system perfectly mirrors the fundamental interaction sub-structures observed in the two-excitation hierarchy. Specifically, the dynamic energy exchange between $C_a(t)$ and $B_{k\omega}(t)$ is governed by the identical mode-dependent coupling coefficient $g_{ak}(\omega)$. Consequently, the numerical discretization scheme and the macroscopic Green's function formalism utilized to construct the interaction matrices for the two-excitation dynamics are proven to be universally valid. The matrices and operators derived for the multi-excitation manifold can be directly applied to evaluate the single-excitation dynamics without any mathematical modification or loss of generality. 

In free space, as validated in Figure \ref{fig:single_population} and \ref{fig:single_spectral}, the numerical superradiance and single-photon emission spectrum from our formulation (with the correction of finite truncation effect) reproduces the analytical results.

\section{Numerical Implementation}

To solve the exact hierarchical equations and extract the physical observables, we transform the continuous frequency domain into a discrete grid. The frequency continuum is truncated at a sufficiently large bandwidth $\omega_c$ and discretized into $Q$ points $\{\omega_q\}_{q=1}^Q$ with corresponding quadrature weights $\{w_q\}_{q=1}^Q$. This discretization converts the analytical integro-differential hierarchy into a closed finite set of ordinary differential equations (ODEs), which are integrated using the standard fourth-order Runge-Kutta method.

During the numerical integration, the continuous field amplitudes are transformed into pre-weighted discrete vectors and matrices. Specifically, the intermediate one-photon amplitude is discretized as a vector $\mathbf{B}_a$ with elements $[\mathbf{B}_a]_q = \sqrt{w_q} B_{a, \omega_q}(t)$, and the pure two-photon amplitude is discretized as a symmetric matrix $\mathbf{D}$ with elements $[\mathbf{D}]_{qq'} = \sqrt{w_q w_{q'}} D_{\omega_q, \omega_{q'}}(t)$. 

With this weighted discrete basis, the atomic populations at any time step $t_n$ are directly evaluated using standard vector and matrix norms. For the two-emitter configuration, the fully excited population is simply given by the pair sector amplitude:
\begin{equation}
    P_{ee}(t_n) = |C_{12}(t_n)|^2.
\end{equation}
The single-excitation populations are strictly determined by the squared Euclidean norm of the weighted one-photon vectors,
\begin{equation}
    P_{e_a, g_b}(t_n) = \left\| \mathbf{B}_a(t_n) \right\|^2,
\end{equation}
where $a \neq b$. Crucially, the ground-state population is calculated exactly from the Frobenius norm of the two-photon matrix,
\begin{equation}
    P_{gg}(t_n) = \frac{1}{2} \left\| \mathbf{D}(t_n) \right\|_F^2.
\end{equation}
By explicitly tracking the $\mathbf{D}$ matrix, the numerical integration inherently preserves the total probability $\mathrm{Tr}[\rho_{\mathrm{atom}}(t)] = 1$ up to the precision of the ODE solver, entirely circumventing the need for artificial trace corrections.

To complement the atomic observables, we reconstruct the emitted field in real space. For the general numerical implementation, the spatial field profiles, including the Green's tensor dependence and the frequency quadrature weights, are absorbed into a generalized reconstruction vector $\mathbf{F}(\mathbf{r})$, with its elements defined as $F_q(\mathbf{r}) = \sqrt{w_q} \mathbf{E}(\mathbf{r}, \omega_q)$.

By exploiting the bosonic symmetry of the pure two-photon amplitude matrix ($\mathbf{D}^T = \mathbf{D}$), the numerical evaluation of the field intensity at a given spatial grid point $\mathbf{r}_j$ and time step $t_n$ simplifies to the exact matrix operation:
\begin{equation}
    I(\mathbf{r}_j,t_n) = \sum_{a=1}^N \left| \mathbf{F}^\dagger(\mathbf{r}_j)\mathbf{B}_a(t_n) \right|^2 + \left\| \mathbf{D}(t_n)\mathbf{F}(\mathbf{r}_j) \right\|^2,
    \label{eq:intensity_discrete}
\end{equation}
where $\mathbf{B}_a$ are the discretized one-photon amplitude vectors, and $\mathbf{D}$ is the two-photon amplitude matrix defined above. Eq.~\eqref{eq:intensity_discrete} provides a direct, exact spatiotemporal reconstruction of the emitted photonic intensity from the dynamical hierarchy, fully preserving the non-Markovian retardation and multi-photon interference effects dictated by the arbitrary structured environment.
\section{Analytic solution of Green's function for Configuration in Fig.~\ref{fig:waveguide_slab}}

To obtain the one-dimensional Green's function for the semi-infinite waveguide with a perfect mirror at $x=0$ and a lossy dielectric slab in the interval $x\in[x_{s1},x_{s2}]$, we consider the scalar Helmholtz equation
\begin{equation}
\left[\partial_x^2 + k^2(x,\omega)\right]G(x,x',\omega) = -\delta(x-x'),
\end{equation}
where
\begin{equation}
k(x,\omega)=
\begin{cases}
k_0=\omega/v_g, & x<x_{s1}\ \text{or}\ x>x_{s2},\\
k_s=\omega\sqrt{\epsilon_s(\omega)}/v_g, & x_{s1}\le x\le x_{s2}.
\end{cases}
\end{equation}
The Green's function is constructed from two linearly independent homogeneous solutions, $\phi_L(x,\omega)$ and $\phi_R(x,\omega)$, satisfying the left and right boundary conditions, respectively as 
\begin{equation}
G(x, x', \omega) = -\frac{\phi_L(x_<, \omega) \phi_R(x_>, \omega)}{W(\omega)}
\end{equation}
where $x_< = \min(x, x')$, $x_> = \max(x, x')$, and $W(\omega)$ is the spatially invariant Wronskian. 

The left-satisfying solution $\phi_L(x, \omega)$ meets the Dirichlet boundary condition at the PEC mirror ($\phi_L(0, \omega) = 0$). In the vacuum region ($x < x_{s1}$), it is given by $\phi_L(x, \omega) = \sin(k_0 x)/k_0$, where $k_0 = \omega/v_g$. We propagate the state vector $\mathbf{u}(x) = [\phi(x), \partial_x \phi(x)]^T$ across different media using the transfer matrix $T(k, d)$ as

\begin{equation}
T(k, d) = 
\begin{bmatrix}
\cos(kd) & \frac{1}{k}\sin(kd) \\
-k\sin(kd) & \cos(kd)
\end{bmatrix}
\end{equation}

For an arbitrary configuration of quantum emitters located at coordinates $x_a$ and $x_b$, the environmental coupling spectrum utilized in the modified Langevin noise formalism is directly proportional to the imaginary part of the evaluated Green's function:

\begin{equation}
\text{Im}[G(x_a, x_b, \omega)] = \text{Im} \left[ -\frac{\phi_L(x_<, \omega) \phi_R(x_>, \omega)}{W(\omega)} \right]
\end{equation}

\section{Further discussion of modeling two-photon joint spectra and photonic entanglement}
\label{app:two_photon}

In the main text, we focused on atomic observables and real-space field dynamics obtained from the exact two-excitation hierarchy. Since the full two-photon sector is retained explicitly in the present formulation, the same framework also gives direct access to spectral correlations of the emitted radiation and to photonic entanglement diagnostics. Here, we briefly summarize how these quantities can be extracted from the asymptotic two-photon amplitude.

Let $D_{\mu\nu}(\omega_1,\omega_2;t)$ denote the two-photon amplitude in the channel-frequency representation, where $\mu,\nu$ label the photonic propagation channels. At sufficiently long times, when the residual atomic excitation is negligible, this amplitude can be interpreted as an asymptotic biphoton wavefunction. If a small non-photonic population remains at the final simulation time $t_f$, the corresponding spectral observables are understood conditionally, i.e., post-selected on the two-photon sector. We therefore introduce the normalized conditional two-photon amplitude
\begin{equation}
D^{(\mathrm{cond})}_{\mu\nu}(\omega_1,\omega_2)
=
\frac{D_{\mu\nu}(\omega_1,\omega_2;t_f)}{\sqrt{P_{gg}(t_f)}},
\end{equation}
where $P_{gg}(t_f)$ is the total two-photon-sector population at $t_f$.
\begin{figure}
    \centering
    \includegraphics[width=0.9\linewidth]{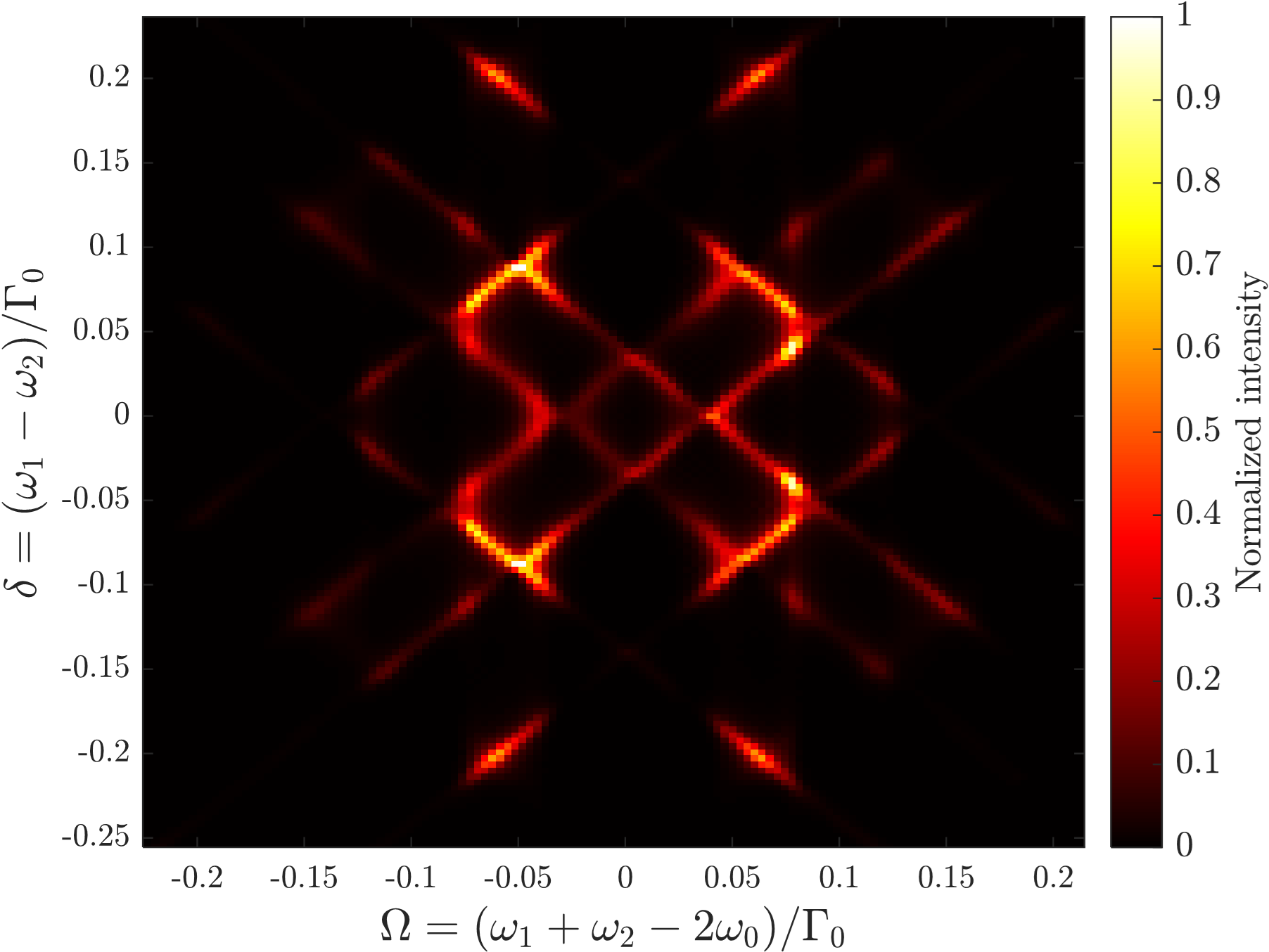}
    \caption{Post-selected conditional joint spectral density (JSD) $J(\omega_1, \omega_2)$ of the emitted photon pair, represented in the rotated coordinates $\Omega = (\omega_1 + \omega_2 - 2\omega_0)/\Gamma_0$ and $\delta = (\omega_1 - \omega_2)/\Gamma_0$.}    \label{fig:conditional_jsd_appendix}
\end{figure}
From this quantity, the conditional joint spectral density (JSD) is defined as
\begin{equation}
J(\omega_1,\omega_2)
=
\sum_{\mu,\nu}
\bigl|
D^{(\mathrm{cond})}_{\mu\nu}(\omega_1,\omega_2)
\bigr|^2,
\end{equation}
where,
\begin{equation}
\int d\omega_1 d\omega_2\, J(\omega_1,\omega_2)=1.
\end{equation}
This object visualizes frequency-frequency correlations in the emitted two-photon state. It is convenient to represent the JSD in the rotated coordinates
\begin{equation}
\Omega=\frac{\omega_1+\omega_2-2\omega_0}{\Gamma_0},
\qquad
\delta=\frac{\omega_1-\omega_2}{\Gamma_0},
\end{equation}
which separate collective energy shifts from relative-frequency correlations. A representative example is shown in Fig.~\ref{fig:conditional_jsd_appendix}, modeled with two emitters at $\{0.75\lambda_a, 3.0\lambda_a\}$ and a lossy dielectric slab ($\epsilon = 16 + 0.05i$) over the interval $x \in [1.5\lambda_a, 2.5\lambda_a]$. The structured cavity-waveguide environment produces several correlated spectral branches rather than a single dominant ridge, indicating nontrivial multimode frequency correlations in the emitted radiation. The structured cavity-waveguide environment produces several correlated spectral branches rather than a single dominant ridge, indicating nontrivial multimode frequency correlations in the emitted radiation.
\begin{equation}
\mathrm{Tr}\,\rho_1^2,
\end{equation}
the effective Schmidt number
\begin{equation}
K_{\mathrm{eff}}=\frac{1}{\mathrm{Tr}\,\rho_1^2},
\end{equation}
and the von Neumann entropy
\begin{equation}
S_{\mathrm{vN}}=-\mathrm{Tr}(\rho_1\ln\rho_1).
\end{equation}

To quantify the corresponding photonic entanglement, we construct the conditional one-photon reduced density matrix
\begin{equation}
\rho_1(\alpha,\alpha')
=
\frac{1}{2}
\sum_{\beta}
D^{(\mathrm{cond})}_{\alpha\beta}
\bigl(D^{(\mathrm{cond})}_{\alpha'\beta}\bigr)^*,
\end{equation}
where $\alpha$ and $\beta$ denote the combined channel-frequency indices. The eigenvalue distribution of $\rho_1$ determines the Schmidt-mode content of the biphoton state: a nearly separable state would yield one dominant eigenvalue, whereas a broader distribution signals multimode photonic entanglement. For compact characterization, we use the purity

\begin{table}
\centering
\caption{Summary of conditional photonic correlation and entanglement diagnostics.}
\label{tab:photonic_summary_appendix}
\begin{tabular}{lc}
\hline\hline
Quantity & Value \\
\hline
Two-photon population $P_{gg}$ & $0.9996$ \\
Photonic purity $\mathrm{Tr}(\rho_1^2)$ & $0.2474$ \\
Effective Schmidt number $K_{\mathrm{eff}}$ & $4.0418$ \\
von Neumann entropy $S_{\mathrm{vN}}$ (nats) & $1.6891$ \\
von Neumann entropy $S_{\mathrm{vN}}$ (bits) & $2.4368$ \\
\hline\hline
\end{tabular}
\end{table}

For the representative parameter set used in Fig.~\ref{fig:conditional_jsd_appendix}, the extracted diagnostics are summarized in Table~\ref{tab:photonic_summary_appendix}. The relatively small purity and the corresponding value $K_{\mathrm{eff}}\simeq 4$ indicate that the conditional two-photon state is not dominated by a single spectral mode but instead contains appreciable multimode entanglement. The nonzero entropy likewise supports this interpretation. In this sense, the JSD and the reduced-density-matrix diagnostics play complementary roles: the former shows the geometry of the spectral correlations, while the latter quantifies their effective multimode entanglement content.

\bibliography{sorsamp}

\begin{thebibliography}{59}%
\makeatletter
\providecommand \@ifxundefined [1]{%
 \@ifx{#1\undefined}
}%
\providecommand \@ifnum [1]{%
 \ifnum #1\expandafter \@firstoftwo
 \else \expandafter \@secondoftwo
 \fi
}%
\providecommand \@ifx [1]{%
 \ifx #1\expandafter \@firstoftwo
 \else \expandafter \@secondoftwo
 \fi
}%
\providecommand \natexlab [1]{#1}%
\providecommand \enquote  [1]{``#1''}%
\providecommand \bibnamefont  [1]{#1}%
\providecommand \bibfnamefont [1]{#1}%
\providecommand \citenamefont [1]{#1}%
\providecommand \href@noop [0]{\@secondoftwo}%
\providecommand \href [0]{\begingroup \@sanitize@url \@href}%
\providecommand \@href[1]{\@@startlink{#1}\@@href}%
\providecommand \@@href[1]{\endgroup#1\@@endlink}%
\providecommand \@sanitize@url [0]{\catcode `\\12\catcode `\$12\catcode
  `\&12\catcode `\#12\catcode `\^12\catcode `\_12\catcode `\%12\relax}%
\providecommand \@@startlink[1]{}%
\providecommand \@@endlink[0]{}%
\providecommand \url  [0]{\begingroup\@sanitize@url \@url }%
\providecommand \@url [1]{\endgroup\@href {#1}{\urlprefix }}%
\providecommand \urlprefix  [0]{URL }%
\providecommand \Eprint [0]{\href }%
\providecommand \doibase [0]{https://doi.org/}%
\providecommand \selectlanguage [0]{\@gobble}%
\providecommand \bibinfo  [0]{\@secondoftwo}%
\providecommand \bibfield  [0]{\@secondoftwo}%
\providecommand \translation [1]{[#1]}%
\providecommand \BibitemOpen [0]{}%
\providecommand \bibitemStop [0]{}%
\providecommand \bibitemNoStop [0]{.\EOS\space}%
\providecommand \EOS [0]{\spacefactor3000\relax}%
\providecommand \BibitemShut  [1]{\csname bibitem#1\endcsname}%
\let\auto@bib@innerbib\@empty
\bibitem [{\citenamefont {Lodahl}\ \emph {et~al.}(2015)\citenamefont {Lodahl},
  \citenamefont {Mahmoodian},\ and\ \citenamefont
  {Stobbe}}]{Lodahl_RevModPhy_nanostructure_2015}%
  \BibitemOpen
  \bibfield  {author} {\bibinfo {author} {\bibfnamefont {P.}~\bibnamefont
  {Lodahl}}, \bibinfo {author} {\bibfnamefont {S.}~\bibnamefont {Mahmoodian}},\
  and\ \bibinfo {author} {\bibfnamefont {S.}~\bibnamefont {Stobbe}},\
  }\bibfield  {title} {\bibinfo {title} {Interfacing single photons and single
  quantum dots with photonic nanostructures},\ }\href
  {https://doi.org/10.1103/RevModPhys.87.347} {\bibfield  {journal} {\bibinfo
  {journal} {Rev. Mod. Phys.}\ }\textbf {\bibinfo {volume} {87}},\ \bibinfo
  {pages} {347} (\bibinfo {year} {2015})}\BibitemShut {NoStop}%
\bibitem [{\citenamefont {Tame}\ \emph {et~al.}(2013)\citenamefont {Tame},
  \citenamefont {McEnery}, \citenamefont {\"{O}zdemir}, \citenamefont {Lee},
  \citenamefont {Maier},\ and\ \citenamefont {Kim}}]{Tame2013}%
  \BibitemOpen
  \bibfield  {author} {\bibinfo {author} {\bibfnamefont {M.~S.}\ \bibnamefont
  {Tame}}, \bibinfo {author} {\bibfnamefont {K.~R.}\ \bibnamefont {McEnery}},
  \bibinfo {author} {\bibfnamefont {Å.~K.}\ \bibnamefont {\"{O}zdemir}},
  \bibinfo {author} {\bibfnamefont {J.}~\bibnamefont {Lee}}, \bibinfo {author}
  {\bibfnamefont {S.~A.}\ \bibnamefont {Maier}},\ and\ \bibinfo {author}
  {\bibfnamefont {M.~S.}\ \bibnamefont {Kim}},\ }\bibfield  {title} {\bibinfo
  {title} {Quantum plasmonics},\ }\href {https://doi.org/10.1038/nphys2615}
  {\bibfield  {journal} {\bibinfo  {journal} {Nature Physics}\ }\textbf
  {\bibinfo {volume} {9}},\ \bibinfo {pages} {329–340} (\bibinfo {year}
  {2013})}\BibitemShut {NoStop}%
\bibitem [{\citenamefont {Dowran}\ \emph {et~al.}(2025)\citenamefont {Dowran},
  \citenamefont {Kilic}, \citenamefont {Lamichhane}, \citenamefont {Erickson},
  \citenamefont {Barker}, \citenamefont {Schubert}, \citenamefont {Liou},
  \citenamefont {Argyropoulos},\ and\ \citenamefont
  {Laraoui}}]{Dowran_Laserphotonics_nanocavity_2025}%
  \BibitemOpen
  \bibfield  {author} {\bibinfo {author} {\bibfnamefont {M.}~\bibnamefont
  {Dowran}}, \bibinfo {author} {\bibfnamefont {U.}~\bibnamefont {Kilic}},
  \bibinfo {author} {\bibfnamefont {S.}~\bibnamefont {Lamichhane}}, \bibinfo
  {author} {\bibfnamefont {A.}~\bibnamefont {Erickson}}, \bibinfo {author}
  {\bibfnamefont {J.}~\bibnamefont {Barker}}, \bibinfo {author} {\bibfnamefont
  {M.}~\bibnamefont {Schubert}}, \bibinfo {author} {\bibfnamefont {S.-H.}\
  \bibnamefont {Liou}}, \bibinfo {author} {\bibfnamefont {C.}~\bibnamefont
  {Argyropoulos}},\ and\ \bibinfo {author} {\bibfnamefont {A.}~\bibnamefont
  {Laraoui}},\ }\bibfield  {title} {\bibinfo {title} {Plasmonic nanocavity to
  boost single photon emission from defects in thin hexagonal boron nitride},\
  }\href {https://doi.org/https://doi.org/10.1002/lpor.202400705} {\bibfield
  {journal} {\bibinfo  {journal} {Laser \& Photonics Reviews}\ }\textbf
  {\bibinfo {volume} {19}},\ \bibinfo {pages} {2400705} (\bibinfo {year}
  {2025})}\BibitemShut {NoStop}%
\bibitem [{\citenamefont {Ye}\ \emph {et~al.}(2023)\citenamefont {Ye},
  \citenamefont {Tian}, \citenamefont {Lin}, \citenamefont {Luo}, \citenamefont
  {You}, \citenamefont {Hu}, \citenamefont {Zhang}, \citenamefont {Chen},\ and\
  \citenamefont {Li}}]{Ye_PRL_Coldatom_2023}%
  \BibitemOpen
  \bibfield  {author} {\bibinfo {author} {\bibfnamefont {M.}~\bibnamefont
  {Ye}}, \bibinfo {author} {\bibfnamefont {Y.}~\bibnamefont {Tian}}, \bibinfo
  {author} {\bibfnamefont {J.}~\bibnamefont {Lin}}, \bibinfo {author}
  {\bibfnamefont {Y.}~\bibnamefont {Luo}}, \bibinfo {author} {\bibfnamefont
  {J.}~\bibnamefont {You}}, \bibinfo {author} {\bibfnamefont {J.}~\bibnamefont
  {Hu}}, \bibinfo {author} {\bibfnamefont {W.}~\bibnamefont {Zhang}}, \bibinfo
  {author} {\bibfnamefont {W.}~\bibnamefont {Chen}},\ and\ \bibinfo {author}
  {\bibfnamefont {X.}~\bibnamefont {Li}},\ }\bibfield  {title} {\bibinfo
  {title} {Universal quantum optimization with cold atoms in an optical
  cavity},\ }\href {https://doi.org/10.1103/PhysRevLett.131.103601} {\bibfield
  {journal} {\bibinfo  {journal} {Phys. Rev. Lett.}\ }\textbf {\bibinfo
  {volume} {131}},\ \bibinfo {pages} {103601} (\bibinfo {year}
  {2023})}\BibitemShut {NoStop}%
\bibitem [{\citenamefont {Sheremet}\ \emph {et~al.}(2023)\citenamefont
  {Sheremet}, \citenamefont {Petrov}, \citenamefont {Iorsh}, \citenamefont
  {Poshakinskiy},\ and\ \citenamefont
  {Poddubny}}]{Sheremet_RMP_waveguideQED_2023}%
  \BibitemOpen
  \bibfield  {author} {\bibinfo {author} {\bibfnamefont {A.~S.}\ \bibnamefont
  {Sheremet}}, \bibinfo {author} {\bibfnamefont {M.~I.}\ \bibnamefont
  {Petrov}}, \bibinfo {author} {\bibfnamefont {I.~V.}\ \bibnamefont {Iorsh}},
  \bibinfo {author} {\bibfnamefont {A.~V.}\ \bibnamefont {Poshakinskiy}},\ and\
  \bibinfo {author} {\bibfnamefont {A.~N.}\ \bibnamefont {Poddubny}},\
  }\bibfield  {title} {\bibinfo {title} {Waveguide quantum electrodynamics:
  Collective radiance and photon-photon correlations},\ }\href
  {https://doi.org/10.1103/RevModPhys.95.015002} {\bibfield  {journal}
  {\bibinfo  {journal} {Rev. Mod. Phys.}\ }\textbf {\bibinfo {volume} {95}},\
  \bibinfo {pages} {015002} (\bibinfo {year} {2023})}\BibitemShut {NoStop}%
\bibitem [{\citenamefont {Mahmoodian}\ \emph {et~al.}(2018)\citenamefont
  {Mahmoodian}, \citenamefont {\ifmmode~\check{C}\else \v{C}\fi{}epulkovskis},
  \citenamefont {Das}, \citenamefont {Lodahl}, \citenamefont {Hammerer},\ and\
  \citenamefont {S\o{}rensen}}]{Mahmoodian_PRL_waveguide_2018}%
  \BibitemOpen
  \bibfield  {author} {\bibinfo {author} {\bibfnamefont {S.}~\bibnamefont
  {Mahmoodian}}, \bibinfo {author} {\bibfnamefont {M.}~\bibnamefont
  {\ifmmode~\check{C}\else \v{C}\fi{}epulkovskis}}, \bibinfo {author}
  {\bibfnamefont {S.}~\bibnamefont {Das}}, \bibinfo {author} {\bibfnamefont
  {P.}~\bibnamefont {Lodahl}}, \bibinfo {author} {\bibfnamefont
  {K.}~\bibnamefont {Hammerer}},\ and\ \bibinfo {author} {\bibfnamefont
  {A.~S.}\ \bibnamefont {S\o{}rensen}},\ }\bibfield  {title} {\bibinfo {title}
  {Strongly correlated photon transport in waveguide quantum electrodynamics
  with weakly coupled emitters},\ }\href
  {https://doi.org/10.1103/PhysRevLett.121.143601} {\bibfield  {journal}
  {\bibinfo  {journal} {Phys. Rev. Lett.}\ }\textbf {\bibinfo {volume} {121}},\
  \bibinfo {pages} {143601} (\bibinfo {year} {2018})}\BibitemShut {NoStop}%
\bibitem [{\citenamefont {Slussarenko}\ and\ \citenamefont
  {Pryde}(2019)}]{Slussarenko_APL_Photonic_2019}%
  \BibitemOpen
  \bibfield  {author} {\bibinfo {author} {\bibfnamefont {S.}~\bibnamefont
  {Slussarenko}}\ and\ \bibinfo {author} {\bibfnamefont {G.~J.}\ \bibnamefont
  {Pryde}},\ }\bibfield  {title} {\bibinfo {title} {Photonic quantum
  information processing: A concise review},\ }\href@noop {} {\bibfield
  {journal} {\bibinfo  {journal} {Applied Physics Reviews}\ }\textbf {\bibinfo
  {volume} {6}},\ \bibinfo {pages} {041303} (\bibinfo {year}
  {2019})}\BibitemShut {NoStop}%
\bibitem [{\citenamefont {Thomas}\ \emph {et~al.}(2022)\citenamefont {Thomas},
  \citenamefont {Ruscio}, \citenamefont {Morin},\ and\ \citenamefont
  {Rempe}}]{Thomas2022}%
  \BibitemOpen
  \bibfield  {author} {\bibinfo {author} {\bibfnamefont {P.}~\bibnamefont
  {Thomas}}, \bibinfo {author} {\bibfnamefont {L.}~\bibnamefont {Ruscio}},
  \bibinfo {author} {\bibfnamefont {O.}~\bibnamefont {Morin}},\ and\ \bibinfo
  {author} {\bibfnamefont {G.}~\bibnamefont {Rempe}},\ }\bibfield  {title}
  {\bibinfo {title} {Efficient generation of entangled multiphoton graph states
  from a single atom},\ }\href {https://doi.org/10.1038/s41586-022-04987-5}
  {\bibfield  {journal} {\bibinfo  {journal} {Nature}\ }\textbf {\bibinfo
  {volume} {608}},\ \bibinfo {pages} {677–681} (\bibinfo {year}
  {2022})}\BibitemShut {NoStop}%
\bibitem [{psi()}]{psiquantum_2025}%
  \BibitemOpen
  \bibfield  {title} {\bibinfo {title} {A manufacturable platform for photonic
  quantum computing},\ }\href {https://doi.org/10.1038/s41586-025-08820-7}
  {\bibfield  {journal} {\bibinfo  {journal} {Nature}\ }\textbf {\bibinfo
  {volume} {641}},\ \bibinfo {pages} {876–883}}\BibitemShut {NoStop}%
\bibitem [{\citenamefont {Gonz\'alez-Tudela}\ and\ \citenamefont
  {Cirac}(2017)}]{Gonz_PRA_nonMarkovian_2017}%
  \BibitemOpen
  \bibfield  {author} {\bibinfo {author} {\bibfnamefont {A.}~\bibnamefont
  {Gonz\'alez-Tudela}}\ and\ \bibinfo {author} {\bibfnamefont {J.~I.}\
  \bibnamefont {Cirac}},\ }\bibfield  {title} {\bibinfo {title} {Markovian and
  non-markovian dynamics of quantum emitters coupled to two-dimensional
  structured reservoirs},\ }\href {https://doi.org/10.1103/PhysRevA.96.043811}
  {\bibfield  {journal} {\bibinfo  {journal} {Phys. Rev. A}\ }\textbf {\bibinfo
  {volume} {96}},\ \bibinfo {pages} {043811} (\bibinfo {year}
  {2017})}\BibitemShut {NoStop}%
\bibitem [{\citenamefont {de~Vega}\ and\ \citenamefont
  {Alonso}(2017)}]{de_RMP_nonMarkovian_2018}%
  \BibitemOpen
  \bibfield  {author} {\bibinfo {author} {\bibfnamefont {I.}~\bibnamefont
  {de~Vega}}\ and\ \bibinfo {author} {\bibfnamefont {D.}~\bibnamefont
  {Alonso}},\ }\bibfield  {title} {\bibinfo {title} {Dynamics of non-markovian
  open quantum systems},\ }\href {https://doi.org/10.1103/RevModPhys.89.015001}
  {\bibfield  {journal} {\bibinfo  {journal} {Rev. Mod. Phys.}\ }\textbf
  {\bibinfo {volume} {89}},\ \bibinfo {pages} {015001} (\bibinfo {year}
  {2017})}\BibitemShut {NoStop}%
\bibitem [{\citenamefont {Scully}\ and\ \citenamefont
  {Zubairy}(1997)}]{Scully1997quantum}%
  \BibitemOpen
  \bibfield  {author} {\bibinfo {author} {\bibfnamefont {M.~O.}\ \bibnamefont
  {Scully}}\ and\ \bibinfo {author} {\bibfnamefont {M.~S.}\ \bibnamefont
  {Zubairy}},\ }\href@noop {} {\emph {\bibinfo {title} {Quantum Optics}}}\
  (\bibinfo  {publisher} {Cambridge University Press},\ \bibinfo {year}
  {1997})\BibitemShut {NoStop}%
\bibitem [{\citenamefont {Mandel}\ and\ \citenamefont
  {Wolf}(1995)}]{Mandel1995optical}%
  \BibitemOpen
  \bibfield  {author} {\bibinfo {author} {\bibfnamefont {L.}~\bibnamefont
  {Mandel}}\ and\ \bibinfo {author} {\bibfnamefont {E.}~\bibnamefont {Wolf}},\
  }\href@noop {} {\emph {\bibinfo {title} {Optical Coherence and Quantum
  Optics}}}\ (\bibinfo  {publisher} {Cambridge University Press},\ \bibinfo
  {year} {1995})\BibitemShut {NoStop}%
\bibitem [{\citenamefont {Medina}\ \emph {et~al.}(2021)\citenamefont {Medina},
  \citenamefont {Garc\'{\i}a-Vidal}, \citenamefont
  {Fern\'andez-Dom\'{\i}nguez},\ and\ \citenamefont
  {Feist}}]{Vidal_PRL_fewmode_2021}%
  \BibitemOpen
  \bibfield  {author} {\bibinfo {author} {\bibfnamefont {I.}~\bibnamefont
  {Medina}}, \bibinfo {author} {\bibfnamefont {F.~J.}\ \bibnamefont
  {Garc\'{\i}a-Vidal}}, \bibinfo {author} {\bibfnamefont {A.~I.}\ \bibnamefont
  {Fern\'andez-Dom\'{\i}nguez}},\ and\ \bibinfo {author} {\bibfnamefont
  {J.}~\bibnamefont {Feist}},\ }\bibfield  {title} {\bibinfo {title} {Few-mode
  field quantization of arbitrary electromagnetic spectral densities},\ }\href
  {https://doi.org/10.1103/PhysRevLett.126.093601} {\bibfield  {journal}
  {\bibinfo  {journal} {Phys. Rev. Lett.}\ }\textbf {\bibinfo {volume} {126}},\
  \bibinfo {pages} {093601} (\bibinfo {year} {2021})}\BibitemShut {NoStop}%
\bibitem [{\citenamefont {Franke}\ \emph {et~al.}(2019)\citenamefont {Franke},
  \citenamefont {Hughes}, \citenamefont {Dezfouli}, \citenamefont {Kristensen},
  \citenamefont {Busch}, \citenamefont {Knorr},\ and\ \citenamefont
  {Richter}}]{Franke_PRL_QNMQuantization_2019}%
  \BibitemOpen
  \bibfield  {author} {\bibinfo {author} {\bibfnamefont {S.}~\bibnamefont
  {Franke}}, \bibinfo {author} {\bibfnamefont {S.}~\bibnamefont {Hughes}},
  \bibinfo {author} {\bibfnamefont {M.~K.}\ \bibnamefont {Dezfouli}}, \bibinfo
  {author} {\bibfnamefont {P.~T.}\ \bibnamefont {Kristensen}}, \bibinfo
  {author} {\bibfnamefont {K.}~\bibnamefont {Busch}}, \bibinfo {author}
  {\bibfnamefont {A.}~\bibnamefont {Knorr}},\ and\ \bibinfo {author}
  {\bibfnamefont {M.}~\bibnamefont {Richter}},\ }\bibfield  {title} {\bibinfo
  {title} {Quantization of quasinormal modes for open cavities and plasmonic
  cavity quantum electrodynamics},\ }\href
  {https://doi.org/10.1103/PhysRevLett.122.213901} {\bibfield  {journal}
  {\bibinfo  {journal} {Phys. Rev. Lett.}\ }\textbf {\bibinfo {volume} {122}},\
  \bibinfo {pages} {213901} (\bibinfo {year} {2019})}\BibitemShut {NoStop}%
\bibitem [{\citenamefont {Bouten}\ \emph {et~al.}(2004)\citenamefont {Bouten},
  \citenamefont {Guta},\ and\ \citenamefont {Maassen}}]{Bouten2004}%
  \BibitemOpen
  \bibfield  {author} {\bibinfo {author} {\bibfnamefont {L.}~\bibnamefont
  {Bouten}}, \bibinfo {author} {\bibfnamefont {M.}~\bibnamefont {Guta}},\ and\
  \bibinfo {author} {\bibfnamefont {H.}~\bibnamefont {Maassen}},\ }\bibfield
  {title} {\bibinfo {title} {Stochastic schr\"{o}dinger equations},\ }\href
  {https://doi.org/10.1088/0305-4470/37/9/010} {\bibfield  {journal} {\bibinfo
  {journal} {Journal of Physics A: Mathematical and General}\ }\textbf
  {\bibinfo {volume} {37}},\ \bibinfo {pages} {3189–3209} (\bibinfo {year}
  {2004})}\BibitemShut {NoStop}%
\bibitem [{\citenamefont {Jin}(2014)}]{Jin2014FEM}%
  \BibitemOpen
  \bibfield  {author} {\bibinfo {author} {\bibfnamefont {J.-M.}\ \bibnamefont
  {Jin}},\ }\href@noop {} {\emph {\bibinfo {title} {The Finite Element Method
  in Electromagnetics}}},\ \bibinfo {edition} {3rd}\ ed.\ (\bibinfo
  {publisher} {Wiley-IEEE Press},\ \bibinfo {address} {Hoboken, NJ},\ \bibinfo
  {year} {2014})\BibitemShut {NoStop}%
\bibitem [{\citenamefont {Taflove}\ and\ \citenamefont
  {Hagness}(2000)}]{taflove_fdtd}%
  \BibitemOpen
  \bibfield  {author} {\bibinfo {author} {\bibfnamefont {A.}~\bibnamefont
  {Taflove}}\ and\ \bibinfo {author} {\bibfnamefont {S.}~\bibnamefont
  {Hagness}},\ }\href@noop {} {\emph {\bibinfo {title} {Computational
  electrodynamics: the finite-difference time-domain method. 2nd ed}}},\ Vol.\
  \bibinfo {volume} {67–106}\ (\bibinfo {year} {2000})\BibitemShut {NoStop}%
\bibitem [{\citenamefont {Chew}\ \emph {et~al.}(2024)\citenamefont {Chew},
  \citenamefont {Zhang},\ and\ \citenamefont {Zhu}}]{Chew2024}%
  \BibitemOpen
  \bibfield  {author} {\bibinfo {author} {\bibfnamefont {W.~C.}\ \bibnamefont
  {Chew}}, \bibinfo {author} {\bibfnamefont {B.}~\bibnamefont {Zhang}},\ and\
  \bibinfo {author} {\bibfnamefont {J.}~\bibnamefont {Zhu}},\ }\bibfield
  {title} {\bibinfo {title} {Some selected unsolved problems in classical and
  quantum electromagnetics},\ }\href {https://doi.org/10.2528/pier24072910}
  {\bibfield  {journal} {\bibinfo  {journal} {Progress In Electromagnetics
  Research}\ }\textbf {\bibinfo {volume} {180}},\ \bibinfo {pages} {79–87}
  (\bibinfo {year} {2024})}\BibitemShut {NoStop}%
\bibitem [{\citenamefont {Moon}\ \emph {et~al.}(2025)\citenamefont {Moon},
  \citenamefont {Khan}, \citenamefont {Elkin}, \citenamefont {Ryu},
  \citenamefont {Na},\ and\ \citenamefont {Roth}}]{Moon2025CQEM}%
  \BibitemOpen
  \bibfield  {author} {\bibinfo {author} {\bibfnamefont {S.}~\bibnamefont
  {Moon}}, \bibinfo {author} {\bibfnamefont {G.}~\bibnamefont {Khan}}, \bibinfo
  {author} {\bibfnamefont {S.~T.}\ \bibnamefont {Elkin}}, \bibinfo {author}
  {\bibfnamefont {C.~J.}\ \bibnamefont {Ryu}}, \bibinfo {author} {\bibfnamefont
  {D.-Y.}\ \bibnamefont {Na}},\ and\ \bibinfo {author} {\bibfnamefont {T.~E.}\
  \bibnamefont {Roth}},\ }\bibfield  {title} {\bibinfo {title} {Computational
  quantum electromagnetics: Basic concepts and emerging trends},\ }\href
  {https://doi.org/10.1109/MAP.2025.3610069} {\bibfield  {journal} {\bibinfo
  {journal} {IEEE Antennas and Propagation Magazine}\ ,\ \bibinfo {pages} {2}}
  (\bibinfo {year} {2025})}\BibitemShut {NoStop}%
\bibitem [{\citenamefont {Elkin}\ \emph {et~al.}(2025)\citenamefont {Elkin},
  \citenamefont {Khan}, \citenamefont {Forati}, \citenamefont {Langley},
  \citenamefont {Timucin}, \citenamefont {Molavi}, \citenamefont {Sussman},\
  and\ \citenamefont {Roth}}]{elkin_CQEMreview_2025}%
  \BibitemOpen
  \bibfield  {author} {\bibinfo {author} {\bibfnamefont {S.~T.}\ \bibnamefont
  {Elkin}}, \bibinfo {author} {\bibfnamefont {G.}~\bibnamefont {Khan}},
  \bibinfo {author} {\bibfnamefont {E.}~\bibnamefont {Forati}}, \bibinfo
  {author} {\bibfnamefont {B.~W.}\ \bibnamefont {Langley}}, \bibinfo {author}
  {\bibfnamefont {D.}~\bibnamefont {Timucin}}, \bibinfo {author} {\bibfnamefont
  {R.}~\bibnamefont {Molavi}}, \bibinfo {author} {\bibfnamefont
  {S.}~\bibnamefont {Sussman}},\ and\ \bibinfo {author} {\bibfnamefont {T.~E.}\
  \bibnamefont {Roth}},\ }\href {https://arxiv.org/abs/2511.20774} {\bibinfo
  {title} {Opportunities and challenges of computational electromagnetics
  methods for superconducting circuit quantum device modeling: A practical
  review}} (\bibinfo {year} {2025}),\ \Eprint
  {https://arxiv.org/abs/2511.20774} {arXiv:2511.20774 [quant-ph]} \BibitemShut
  {NoStop}%
\bibitem [{\citenamefont {Roth}\ and\ \citenamefont
  {Chew}(2021)}]{Roth2021JMMCT}%
  \BibitemOpen
  \bibfield  {author} {\bibinfo {author} {\bibfnamefont {T.~E.}\ \bibnamefont
  {Roth}}\ and\ \bibinfo {author} {\bibfnamefont {W.~C.}\ \bibnamefont
  {Chew}},\ }\bibfield  {title} {\bibinfo {title} {Macroscopic circuit quantum
  electrodynamics: A new look toward developing full-wave numerical models},\
  }\href {https://doi.org/10.1109/JMMCT.2021.3112808} {\bibfield  {journal}
  {\bibinfo  {journal} {IEEE Journal on Multiscale and Multiphysics
  Computational Techniques}\ }\textbf {\bibinfo {volume} {6}},\ \bibinfo
  {pages} {109} (\bibinfo {year} {2021})}\BibitemShut {NoStop}%
\bibitem [{\citenamefont {Roth}\ and\ \citenamefont
  {Chew}(2022)}]{2022RothJMMCT}%
  \BibitemOpen
  \bibfield  {author} {\bibinfo {author} {\bibfnamefont {T.~E.}\ \bibnamefont
  {Roth}}\ and\ \bibinfo {author} {\bibfnamefont {W.~C.}\ \bibnamefont
  {Chew}},\ }\bibfield  {title} {\bibinfo {title} {Full-wave methodology to
  compute the spontaneous emission rate of a transmon qubit},\ }\href
  {https://doi.org/10.1109/JMMCT.2022.3169460} {\bibfield  {journal} {\bibinfo
  {journal} {IEEE Journal on Multiscale and Multiphysics Computational
  Techniques}\ }\textbf {\bibinfo {volume} {7}},\ \bibinfo {pages} {92}
  (\bibinfo {year} {2022})}\BibitemShut {NoStop}%
\bibitem [{\citenamefont {Na}\ \emph {et~al.}(2021)\citenamefont {Na},
  \citenamefont {Zhu},\ and\ \citenamefont {Chew}}]{Na2021cqNMD}%
  \BibitemOpen
  \bibfield  {author} {\bibinfo {author} {\bibfnamefont {D.-Y.}\ \bibnamefont
  {Na}}, \bibinfo {author} {\bibfnamefont {J.}~\bibnamefont {Zhu}},\ and\
  \bibinfo {author} {\bibfnamefont {W.~C.}\ \bibnamefont {Chew}},\ }\bibfield
  {title} {\bibinfo {title} {Diagonalization of the hamiltonian for
  finite-sized dispersive media: Canonical quantization with numerical mode
  decomposition},\ }\href {https://doi.org/10.1103/PhysRevA.103.063707}
  {\bibfield  {journal} {\bibinfo  {journal} {Phys. Rev. A}\ }\textbf {\bibinfo
  {volume} {103}},\ \bibinfo {pages} {063707} (\bibinfo {year}
  {2021})}\BibitemShut {NoStop}%
\bibitem [{\citenamefont {Na}\ \emph {et~al.}(2023)\citenamefont {Na},
  \citenamefont {Roth}, \citenamefont {Zhu}, \citenamefont {Chew},\ and\
  \citenamefont {Ryu}}]{Na2023quantumEMLossy}%
  \BibitemOpen
  \bibfield  {author} {\bibinfo {author} {\bibfnamefont {D.-Y.}\ \bibnamefont
  {Na}}, \bibinfo {author} {\bibfnamefont {T.~E.}\ \bibnamefont {Roth}},
  \bibinfo {author} {\bibfnamefont {J.}~\bibnamefont {Zhu}}, \bibinfo {author}
  {\bibfnamefont {W.~C.}\ \bibnamefont {Chew}},\ and\ \bibinfo {author}
  {\bibfnamefont {C.~J.}\ \bibnamefont {Ryu}},\ }\bibfield  {title} {\bibinfo
  {title} {Numerical framework for modeling quantum electromagnetic systems
  involving finite-sized lossy dielectric objects in free space},\ }\href
  {https://doi.org/10.1103/PhysRevA.107.063702} {\bibfield  {journal} {\bibinfo
   {journal} {Phys. Rev. A}\ }\textbf {\bibinfo {volume} {107}},\ \bibinfo
  {pages} {063702} (\bibinfo {year} {2023})}\BibitemShut {NoStop}%
\bibitem [{\citenamefont {Moon}\ \emph {et~al.}(2024)\citenamefont {Moon},
  \citenamefont {Na},\ and\ \citenamefont {Roth}}]{Moon2024TAP}%
  \BibitemOpen
  \bibfield  {author} {\bibinfo {author} {\bibfnamefont {S.}~\bibnamefont
  {Moon}}, \bibinfo {author} {\bibfnamefont {D.-Y.}\ \bibnamefont {Na}},\ and\
  \bibinfo {author} {\bibfnamefont {T.~E.}\ \bibnamefont {Roth}},\ }\bibfield
  {title} {\bibinfo {title} {Analytical quantum full-wave solutions for a 3-d
  circuit quantum electrodynamics system},\ }\href
  {https://doi.org/10.1109/TAP.2024.3427362} {\bibfield  {journal} {\bibinfo
  {journal} {IEEE Transactions on Antennas and Propagation}\ }\textbf {\bibinfo
  {volume} {72}},\ \bibinfo {pages} {6702} (\bibinfo {year}
  {2024})}\BibitemShut {NoStop}%
\bibitem [{\citenamefont {Forestiere}\ and\ \citenamefont
  {Miano}(2022)}]{Forestiere2022PRA}%
  \BibitemOpen
  \bibfield  {author} {\bibinfo {author} {\bibfnamefont {C.}~\bibnamefont
  {Forestiere}}\ and\ \bibinfo {author} {\bibfnamefont {G.}~\bibnamefont
  {Miano}},\ }\bibfield  {title} {\bibinfo {title} {Operative approach to
  quantum electrodynamics in dispersive dielectric objects based on a
  polarization-mode expansion},\ }\href
  {https://doi.org/10.1103/PhysRevA.106.033701} {\bibfield  {journal} {\bibinfo
   {journal} {Phys. Rev. A}\ }\textbf {\bibinfo {volume} {106}},\ \bibinfo
  {pages} {033701} (\bibinfo {year} {2022})}\BibitemShut {NoStop}%
\bibitem [{\citenamefont {Forestiere}\ and\ \citenamefont
  {Miano}(2023)}]{Forestiere2023PRA}%
  \BibitemOpen
  \bibfield  {author} {\bibinfo {author} {\bibfnamefont {C.}~\bibnamefont
  {Forestiere}}\ and\ \bibinfo {author} {\bibfnamefont {G.}~\bibnamefont
  {Miano}},\ }\bibfield  {title} {\bibinfo {title} {Integral formulation of the
  macroscopic quantum electrodynamics in dispersive dielectric objects},\
  }\href {https://doi.org/10.1103/PhysRevA.107.063705} {\bibfield  {journal}
  {\bibinfo  {journal} {Phys. Rev. A}\ }\textbf {\bibinfo {volume} {107}},\
  \bibinfo {pages} {063705} (\bibinfo {year} {2023})}\BibitemShut {NoStop}%
\bibitem [{\citenamefont {Ryu}\ \emph {et~al.}(2023{\natexlab{a}})\citenamefont
  {Ryu}, \citenamefont {Na}, \citenamefont {Chew},\ and\ \citenamefont
  {Kudeki}}]{Ryu2023Dicke}%
  \BibitemOpen
  \bibfield  {author} {\bibinfo {author} {\bibfnamefont {C.~J.}\ \bibnamefont
  {Ryu}}, \bibinfo {author} {\bibfnamefont {D.-Y.}\ \bibnamefont {Na}},
  \bibinfo {author} {\bibfnamefont {W.~C.}\ \bibnamefont {Chew}},\ and\
  \bibinfo {author} {\bibfnamefont {E.}~\bibnamefont {Kudeki}},\ }\bibfield
  {title} {\bibinfo {title} {Efficient tensor-network simulation for the
  few-atom multimode dicke model via coupling-matrix transformation},\ }\href
  {https://doi.org/10.1103/PhysRevA.108.043707} {\bibfield  {journal} {\bibinfo
   {journal} {Phys. Rev. A}\ }\textbf {\bibinfo {volume} {108}},\ \bibinfo
  {pages} {043707} (\bibinfo {year} {2023}{\natexlab{a}})}\BibitemShut
  {NoStop}%
\bibitem [{\citenamefont {Ryu}\ \emph {et~al.}(2023{\natexlab{b}})\citenamefont
  {Ryu}, \citenamefont {Na},\ and\ \citenamefont {Chew}}]{Ryu2023MPS}%
  \BibitemOpen
  \bibfield  {author} {\bibinfo {author} {\bibfnamefont {C.~J.}\ \bibnamefont
  {Ryu}}, \bibinfo {author} {\bibfnamefont {D.-Y.}\ \bibnamefont {Na}},\ and\
  \bibinfo {author} {\bibfnamefont {W.~C.}\ \bibnamefont {Chew}},\ }\bibfield
  {title} {\bibinfo {title} {Matrix product states and numerical mode
  decomposition for the analysis of gauge-invariant cavity quantum
  electrodynamics},\ }\href {https://doi.org/10.1103/PhysRevA.107.063707}
  {\bibfield  {journal} {\bibinfo  {journal} {Phys. Rev. A}\ }\textbf {\bibinfo
  {volume} {107}},\ \bibinfo {pages} {063707} (\bibinfo {year}
  {2023}{\natexlab{b}})}\BibitemShut {NoStop}%
\bibitem [{\citenamefont {Huang}\ \emph {et~al.}(2025)\citenamefont {Huang},
  \citenamefont {Ge}, \citenamefont {Cheng}, \citenamefont {He}, \citenamefont
  {Liu},\ and\ \citenamefont {Sha}}]{huang2025VIE}%
  \BibitemOpen
  \bibfield  {author} {\bibinfo {author} {\bibfnamefont {C.}~\bibnamefont
  {Huang}}, \bibinfo {author} {\bibfnamefont {H.}~\bibnamefont {Ge}}, \bibinfo
  {author} {\bibfnamefont {Y.}~\bibnamefont {Cheng}}, \bibinfo {author}
  {\bibfnamefont {Z.}~\bibnamefont {He}}, \bibinfo {author} {\bibfnamefont
  {F.}~\bibnamefont {Liu}},\ and\ \bibinfo {author} {\bibfnamefont {W.~E.~I.}\
  \bibnamefont {Sha}},\ }\href {https://arxiv.org/abs/2508.16471} {\bibinfo
  {title} {Modeling of far-field quantum coherence by dielectric bodies based
  on the volume integral equation method}} (\bibinfo {year} {2025}),\ \Eprint
  {https://arxiv.org/abs/2508.16471} {arXiv:2508.16471 [quant-ph]} \BibitemShut
  {NoStop}%
\bibitem [{\citenamefont {Seo}\ \emph {et~al.}(2025)\citenamefont {Seo},
  \citenamefont {Choi}, \citenamefont {Roth}, \citenamefont {Zhu},
  \citenamefont {Chew},\ and\ \citenamefont {Na}}]{seo_nano_2025}%
  \BibitemOpen
  \bibfield  {author} {\bibinfo {author} {\bibfnamefont {J.}~\bibnamefont
  {Seo}}, \bibinfo {author} {\bibfnamefont {H.}~\bibnamefont {Choi}}, \bibinfo
  {author} {\bibfnamefont {T.~E.}\ \bibnamefont {Roth}}, \bibinfo {author}
  {\bibfnamefont {J.}~\bibnamefont {Zhu}}, \bibinfo {author} {\bibfnamefont
  {W.~C.}\ \bibnamefont {Chew}},\ and\ \bibinfo {author} {\bibfnamefont
  {D.-Y.}\ \bibnamefont {Na}},\ }\href {https://arxiv.org/abs/2205.03388}
  {\bibinfo {title} {Quantum-plasmonic dynamics modeled via a modified langevin
  noise formalism: Numerical studies of single-photon emission and two-photon
  interference}} (\bibinfo {year} {2025}),\ \Eprint
  {https://arxiv.org/abs/2205.03388} {arXiv:2205.03388 [physics.optics]}
  \BibitemShut {NoStop}%
\bibitem [{\citenamefont {Choi}\ \emph
  {et~al.}(2025{\natexlab{a}})\citenamefont {Choi}, \citenamefont {Roth},
  \citenamefont {Chew},\ and\ \citenamefont {Na}}]{Choi_PRApplied}%
  \BibitemOpen
  \bibfield  {author} {\bibinfo {author} {\bibfnamefont {H.}~\bibnamefont
  {Choi}}, \bibinfo {author} {\bibfnamefont {T.~E.}\ \bibnamefont {Roth}},
  \bibinfo {author} {\bibfnamefont {W.~C.}\ \bibnamefont {Chew}},\ and\
  \bibinfo {author} {\bibfnamefont {D.-Y.}\ \bibnamefont {Na}},\ }\bibfield
  {title} {\bibinfo {title} {Non-markovian analysis of atom-field interactions
  in dissipative electromagnetic environments},\ }\href
  {https://doi.org/10.1103/kvxk-sp94} {\bibfield  {journal} {\bibinfo
  {journal} {Phys. Rev. Appl.}\ }\textbf {\bibinfo {volume} {24}},\ \bibinfo
  {pages} {044056} (\bibinfo {year} {2025}{\natexlab{a}})}\BibitemShut
  {NoStop}%
\bibitem [{\citenamefont {Choi}\ \emph
  {et~al.}(2025{\natexlab{b}})\citenamefont {Choi}, \citenamefont {Seo},
  \citenamefont {Chew},\ and\ \citenamefont {Na}}]{Choi_FDTD_QE}%
  \BibitemOpen
  \bibfield  {author} {\bibinfo {author} {\bibfnamefont {H.}~\bibnamefont
  {Choi}}, \bibinfo {author} {\bibfnamefont {J.}~\bibnamefont {Seo}}, \bibinfo
  {author} {\bibfnamefont {W.~C.}\ \bibnamefont {Chew}},\ and\ \bibinfo
  {author} {\bibfnamefont {D.-Y.}\ \bibnamefont {Na}},\ }\href
  {https://doi.org/10.48550/ARXIV.2511.03561} {\bibinfo {title} {Atom-field
  non-markovian dynamics in open and dissipative systems: An efficient
  memory-kernel approach linked to dyadic greens function and cem treatments}}
  (\bibinfo {year} {2025}{\natexlab{b}})\BibitemShut {NoStop}%
\bibitem [{\citenamefont {Xu}\ and\ \citenamefont
  {Li}(2014)}]{PRA_twophoton_2014}%
  \BibitemOpen
  \bibfield  {author} {\bibinfo {author} {\bibfnamefont {X.-W.}\ \bibnamefont
  {Xu}}\ and\ \bibinfo {author} {\bibfnamefont {Y.}~\bibnamefont {Li}},\
  }\bibfield  {title} {\bibinfo {title} {Strongly correlated two-photon
  transport in a one-dimensional waveguide coupled to a weakly nonlinear
  cavity},\ }\href {https://doi.org/10.1103/PhysRevA.90.033832} {\bibfield
  {journal} {\bibinfo  {journal} {Phys. Rev. A}\ }\textbf {\bibinfo {volume}
  {90}},\ \bibinfo {pages} {033832} (\bibinfo {year} {2014})}\BibitemShut
  {NoStop}%
\bibitem [{\citenamefont {Yudson}\ and\ \citenamefont
  {Reineker}(2008)}]{PRA_multiphoton_2008}%
  \BibitemOpen
  \bibfield  {author} {\bibinfo {author} {\bibfnamefont {V.~I.}\ \bibnamefont
  {Yudson}}\ and\ \bibinfo {author} {\bibfnamefont {P.}~\bibnamefont
  {Reineker}},\ }\bibfield  {title} {\bibinfo {title} {Multiphoton scattering
  in a one-dimensional waveguide with resonant atoms},\ }\href
  {https://doi.org/10.1103/PhysRevA.78.052713} {\bibfield  {journal} {\bibinfo
  {journal} {Phys. Rev. A}\ }\textbf {\bibinfo {volume} {78}},\ \bibinfo
  {pages} {052713} (\bibinfo {year} {2008})}\BibitemShut {NoStop}%
\bibitem [{\citenamefont {Reiserer}\ \emph {et~al.}(2014)\citenamefont
  {Reiserer}, \citenamefont {Kalb}, \citenamefont {Rempe},\ and\ \citenamefont
  {Ritter}}]{Reiserer_gate_nature_2014}%
  \BibitemOpen
  \bibfield  {author} {\bibinfo {author} {\bibfnamefont {A.}~\bibnamefont
  {Reiserer}}, \bibinfo {author} {\bibfnamefont {N.}~\bibnamefont {Kalb}},
  \bibinfo {author} {\bibfnamefont {G.}~\bibnamefont {Rempe}},\ and\ \bibinfo
  {author} {\bibfnamefont {S.}~\bibnamefont {Ritter}},\ }\bibfield  {title}
  {\bibinfo {title} {A quantum gate between a flying optical photon and a
  single trapped atom},\ }\href {https://doi.org/10.1038/nature13177}
  {\bibfield  {journal} {\bibinfo  {journal} {Nature}\ }\textbf {\bibinfo
  {volume} {508}},\ \bibinfo {pages} {237–240} (\bibinfo {year}
  {2014})}\BibitemShut {NoStop}%
\bibitem [{\citenamefont {Shen}\ and\ \citenamefont
  {Fan}(2007)}]{Shen_PRL_twophoton_2007}%
  \BibitemOpen
  \bibfield  {author} {\bibinfo {author} {\bibfnamefont {J.-T.}\ \bibnamefont
  {Shen}}\ and\ \bibinfo {author} {\bibfnamefont {S.}~\bibnamefont {Fan}},\
  }\bibfield  {title} {\bibinfo {title} {Strongly correlated two-photon
  transport in a one-dimensional waveguide coupled to a two-level system},\
  }\href {https://doi.org/10.1103/PhysRevLett.98.153003} {\bibfield  {journal}
  {\bibinfo  {journal} {Phys. Rev. Lett.}\ }\textbf {\bibinfo {volume} {98}},\
  \bibinfo {pages} {153003} (\bibinfo {year} {2007})}\BibitemShut {NoStop}%
\bibitem [{\citenamefont {Roy}(2011)}]{Roy_PRL_twophoton_2011}%
  \BibitemOpen
  \bibfield  {author} {\bibinfo {author} {\bibfnamefont {D.}~\bibnamefont
  {Roy}},\ }\bibfield  {title} {\bibinfo {title} {Two-photon scattering by a
  driven three-level emitter in a one-dimensional waveguide and
  electromagnetically induced transparency},\ }\href
  {https://doi.org/10.1103/PhysRevLett.106.053601} {\bibfield  {journal}
  {\bibinfo  {journal} {Phys. Rev. Lett.}\ }\textbf {\bibinfo {volume} {106}},\
  \bibinfo {pages} {053601} (\bibinfo {year} {2011})}\BibitemShut {NoStop}%
\bibitem [{\citenamefont {Gu}\ \emph {et~al.}(2024)\citenamefont {Gu},
  \citenamefont {Li}, \citenamefont {Tian}, \citenamefont {Yi},\ and\
  \citenamefont {Li}}]{Gu_PRA_twophoton_2024}%
  \BibitemOpen
  \bibfield  {author} {\bibinfo {author} {\bibfnamefont {W.}~\bibnamefont
  {Gu}}, \bibinfo {author} {\bibfnamefont {T.}~\bibnamefont {Li}}, \bibinfo
  {author} {\bibfnamefont {Y.}~\bibnamefont {Tian}}, \bibinfo {author}
  {\bibfnamefont {Z.}~\bibnamefont {Yi}},\ and\ \bibinfo {author}
  {\bibfnamefont {G.-x.}\ \bibnamefont {Li}},\ }\bibfield  {title} {\bibinfo
  {title} {Two-photon dynamics in non-markovian waveguide qed with a giant
  atom},\ }\href {https://doi.org/10.1103/PhysRevA.110.033707} {\bibfield
  {journal} {\bibinfo  {journal} {Phys. Rev. A}\ }\textbf {\bibinfo {volume}
  {110}},\ \bibinfo {pages} {033707} (\bibinfo {year} {2024})}\BibitemShut
  {NoStop}%
\bibitem [{\citenamefont {Xie}\ \emph {et~al.}(2025)\citenamefont {Xie},
  \citenamefont {Mirza},\ and\ \citenamefont {Schotland}}]{PRA_twophoton_2025}%
  \BibitemOpen
  \bibfield  {author} {\bibinfo {author} {\bibfnamefont {W.}~\bibnamefont
  {Xie}}, \bibinfo {author} {\bibfnamefont {I.~M.}\ \bibnamefont {Mirza}},\
  and\ \bibinfo {author} {\bibfnamefont {J.~C.}\ \bibnamefont {Schotland}},\
  }\bibfield  {title} {\bibinfo {title} {Two-photon superradiance and
  subradiance},\ }\href {https://doi.org/10.1103/dbny-pjqc} {\bibfield
  {journal} {\bibinfo  {journal} {Phys. Rev. A}\ }\textbf {\bibinfo {volume}
  {112}},\ \bibinfo {pages} {043712} (\bibinfo {year} {2025})}\BibitemShut
  {NoStop}%
\bibitem [{\citenamefont {Stefano}\ \emph {et~al.}(2001)\citenamefont
  {Stefano}, \citenamefont {Savasta},\ and\ \citenamefont
  {Girlanda}}]{DiStefano2001ModeExpansion}%
  \BibitemOpen
  \bibfield  {author} {\bibinfo {author} {\bibfnamefont {O.~D.}\ \bibnamefont
  {Stefano}}, \bibinfo {author} {\bibfnamefont {S.}~\bibnamefont {Savasta}},\
  and\ \bibinfo {author} {\bibfnamefont {R.}~\bibnamefont {Girlanda}},\
  }\bibfield  {title} {\bibinfo {title} {Mode expansion and photon operators in
  dispersive and absorbing dielectrics},\ }\href
  {https://doi.org/10.1080/09500340108235155} {\bibfield  {journal} {\bibinfo
  {journal} {Journal of Modern Optics}\ }\textbf {\bibinfo {volume} {48}},\
  \bibinfo {pages} {67} (\bibinfo {year} {2001})}\BibitemShut {NoStop}%
\bibitem [{\citenamefont {Drezet}(2017)}]{Drezet2017QuantizingPolaritons}%
  \BibitemOpen
  \bibfield  {author} {\bibinfo {author} {\bibfnamefont {A.}~\bibnamefont
  {Drezet}},\ }\bibfield  {title} {\bibinfo {title} {Quantizing polaritons in
  inhomogeneous dissipative systems},\ }\href
  {https://doi.org/10.1103/PhysRevA.95.023831} {\bibfield  {journal} {\bibinfo
  {journal} {Phys. Rev. A}\ }\textbf {\bibinfo {volume} {95}},\ \bibinfo
  {pages} {023831} (\bibinfo {year} {2017})}\BibitemShut {NoStop}%
\bibitem [{\citenamefont {Ciattoni}(2024)}]{ciatonni_MLN_2024}%
  \BibitemOpen
  \bibfield  {author} {\bibinfo {author} {\bibfnamefont {A.}~\bibnamefont
  {Ciattoni}},\ }\bibfield  {title} {\bibinfo {title} {Quantum electrodynamics
  of lossy magnetodielectric samples in vacuum: Modified langevin noise
  formalism},\ }\href {https://doi.org/10.1103/PhysRevA.110.013707} {\bibfield
  {journal} {\bibinfo  {journal} {Phys. Rev. A}\ }\textbf {\bibinfo {volume}
  {110}},\ \bibinfo {pages} {013707} (\bibinfo {year} {2024})}\BibitemShut
  {NoStop}%
\bibitem [{\citenamefont {Ciattoni}(2026)}]{ciattoni_MLN_2026}%
  \BibitemOpen
  \bibfield  {author} {\bibinfo {author} {\bibfnamefont {A.}~\bibnamefont
  {Ciattoni}},\ }\href {https://arxiv.org/abs/2603.04336} {\bibinfo {title}
  {Direct derivation of the modified langevin noise formalism from the
  canonical quantization of macroscopic electromagnetism}} (\bibinfo {year}
  {2026}),\ \Eprint {https://arxiv.org/abs/2603.04336} {arXiv:2603.04336
  [quant-ph]} \BibitemShut {NoStop}%
\bibitem [{\citenamefont {Gruner}\ and\ \citenamefont
  {Welsch}(1996)}]{Gruner1996QEDEvanescent}%
  \BibitemOpen
  \bibfield  {author} {\bibinfo {author} {\bibfnamefont {T.}~\bibnamefont
  {Gruner}}\ and\ \bibinfo {author} {\bibfnamefont {D.-G.}\ \bibnamefont
  {Welsch}},\ }\bibfield  {title} {\bibinfo {title} {Green-function approach to
  the radiation-field quantization for homogeneous and inhomogeneous
  kramers-kronig dielectrics},\ }\href
  {https://doi.org/10.1103/PhysRevA.53.1818} {\bibfield  {journal} {\bibinfo
  {journal} {Phys. Rev. A}\ }\textbf {\bibinfo {volume} {53}},\ \bibinfo
  {pages} {1818} (\bibinfo {year} {1996})}\BibitemShut {NoStop}%
\bibitem [{\citenamefont {Dung}\ \emph {et~al.}(2000)\citenamefont {Dung},
  \citenamefont {Kn\"oll},\ and\ \citenamefont
  {Welsch}}]{Dung2000QEDLocalized}%
  \BibitemOpen
  \bibfield  {author} {\bibinfo {author} {\bibfnamefont {H.~T.}\ \bibnamefont
  {Dung}}, \bibinfo {author} {\bibfnamefont {L.}~\bibnamefont {Kn\"oll}},\ and\
  \bibinfo {author} {\bibfnamefont {D.-G.}\ \bibnamefont {Welsch}},\ }\bibfield
   {title} {\bibinfo {title} {Spontaneous decay in the presence of dispersing
  and absorbing bodies: General theory and application to a spherical cavity},\
  }\href {https://doi.org/10.1103/PhysRevA.62.053804} {\bibfield  {journal}
  {\bibinfo  {journal} {Phys. Rev. A}\ }\textbf {\bibinfo {volume} {62}},\
  \bibinfo {pages} {053804} (\bibinfo {year} {2000})}\BibitemShut {NoStop}%
\bibitem [{\citenamefont {Scheel}\ and\ \citenamefont
  {Buhmann}(2008)}]{Scheel2008MacroscopicQED}%
  \BibitemOpen
  \bibfield  {author} {\bibinfo {author} {\bibfnamefont {S.}~\bibnamefont
  {Scheel}}\ and\ \bibinfo {author} {\bibfnamefont {S.~Y.}\ \bibnamefont
  {Buhmann}},\ }\bibfield  {title} {\bibinfo {title} {Macroscopic quantum
  electrodynamics – concepts and applications},\ }\href@noop {} {\bibfield
  {journal} {\bibinfo  {journal} {Acta Physica Slovaca}\ }\textbf {\bibinfo
  {volume} {58}},\ \bibinfo {pages} {675} (\bibinfo {year} {2008})}\BibitemShut
  {NoStop}%
\bibitem [{\citenamefont {Feist}\ \emph {et~al.}(2021)\citenamefont {Feist},
  \citenamefont {Fernández-Domínguez},\ and\ \citenamefont
  {García-Vidal}}]{ECM_nanophotonics_2021}%
  \BibitemOpen
  \bibfield  {author} {\bibinfo {author} {\bibfnamefont {J.}~\bibnamefont
  {Feist}}, \bibinfo {author} {\bibfnamefont {A.~I.}\ \bibnamefont
  {Fernández-Domínguez}},\ and\ \bibinfo {author} {\bibfnamefont {F.~J.}\
  \bibnamefont {García-Vidal}},\ }\bibfield  {title} {\bibinfo {title}
  {Macroscopic qed for quantum nanophotonics: emitter-centered modes as a
  minimal basis for multiemitter problems},\ }\href
  {https://doi.org/doi:10.1515/nanoph-2020-0451} {\bibfield  {journal}
  {\bibinfo  {journal} {Nanophotonics}\ }\textbf {\bibinfo {volume} {10}},\
  \bibinfo {pages} {477} (\bibinfo {year} {2021})}\BibitemShut {NoStop}%
\bibitem [{\citenamefont {Sánchez-Barquilla}\ \emph
  {et~al.}(2022)\citenamefont {Sánchez-Barquilla}, \citenamefont
  {García-Vidal}, \citenamefont {Fernández-Domínguez},\ and\ \citenamefont
  {Feist}}]{Monica2022ECM}%
  \BibitemOpen
  \bibfield  {author} {\bibinfo {author} {\bibfnamefont {M.}~\bibnamefont
  {Sánchez-Barquilla}}, \bibinfo {author} {\bibfnamefont {F.~J.}\ \bibnamefont
  {García-Vidal}}, \bibinfo {author} {\bibfnamefont {A.~I.}\ \bibnamefont
  {Fernández-Domínguez}},\ and\ \bibinfo {author} {\bibfnamefont
  {J.}~\bibnamefont {Feist}},\ }\bibfield  {title} {\bibinfo {title} {Few-mode
  field quantization for multiple emitters},\ }\href
  {https://doi.org/doi:10.1515/nanoph-2021-0795} {\bibfield  {journal}
  {\bibinfo  {journal} {Nanophotonics}\ }\textbf {\bibinfo {volume} {11}},\
  \bibinfo {pages} {4363} (\bibinfo {year} {2022})}\BibitemShut {NoStop}%
\bibitem [{\citenamefont {Miano}\ \emph
  {et~al.}(2025{\natexlab{a}})\citenamefont {Miano}, \citenamefont {Cangemi},\
  and\ \citenamefont {Forestiere}}]{Forestire_nanophotonics_2025}%
  \BibitemOpen
  \bibfield  {author} {\bibinfo {author} {\bibfnamefont {G.}~\bibnamefont
  {Miano}}, \bibinfo {author} {\bibfnamefont {L.~M.}\ \bibnamefont {Cangemi}},\
  and\ \bibinfo {author} {\bibfnamefont {C.}~\bibnamefont {Forestiere}},\
  }\bibfield  {title} {\bibinfo {title} {Quantum emitter interacting with a
  dispersive dielectric object: a model based on the modified langevin noise
  formalism},\ }\href {https://doi.org/doi:10.1515/nanoph-2024-0703} {\bibfield
   {journal} {\bibinfo  {journal} {Nanophotonics}\ }\textbf {\bibinfo {volume}
  {14}},\ \bibinfo {pages} {4019} (\bibinfo {year}
  {2025}{\natexlab{a}})}\BibitemShut {NoStop}%
\bibitem [{\citenamefont {Miano}\ \emph
  {et~al.}(2025{\natexlab{b}})\citenamefont {Miano}, \citenamefont {Cangemi},\
  and\ \citenamefont {Forestiere}}]{Forestire_PRA_2025}%
  \BibitemOpen
  \bibfield  {author} {\bibinfo {author} {\bibfnamefont {G.}~\bibnamefont
  {Miano}}, \bibinfo {author} {\bibfnamefont {L.~M.}\ \bibnamefont {Cangemi}},\
  and\ \bibinfo {author} {\bibfnamefont {C.}~\bibnamefont {Forestiere}},\
  }\bibfield  {title} {\bibinfo {title} {Spectral densities of a dispersive
  dielectric sphere in the modified langevin noise formalism},\ }\href
  {https://doi.org/10.1103/g71y-hglg} {\bibfield  {journal} {\bibinfo
  {journal} {Phys. Rev. A}\ }\textbf {\bibinfo {volume} {112}},\ \bibinfo
  {pages} {033712} (\bibinfo {year} {2025}{\natexlab{b}})}\BibitemShut
  {NoStop}%
\bibitem [{\citenamefont {Miano}\ \emph {et~al.}(2026)\citenamefont {Miano},
  \citenamefont {Cangemi},\ and\ \citenamefont
  {Forestiere}}]{miano_ECM_PRA_2026}%
  \BibitemOpen
  \bibfield  {author} {\bibinfo {author} {\bibfnamefont {G.}~\bibnamefont
  {Miano}}, \bibinfo {author} {\bibfnamefont {L.~M.}\ \bibnamefont {Cangemi}},\
  and\ \bibinfo {author} {\bibfnamefont {C.}~\bibnamefont {Forestiere}},\
  }\bibfield  {title} {\bibinfo {title} {Modified langevin noise formalism for
  multiple quantum emitters in dispersive electromagnetic environments out of
  equilibrium},\ }\href {https://doi.org/10.1103/ps9g-1b4d} {\bibfield
  {journal} {\bibinfo  {journal} {Phys. Rev. A}\ }\textbf {\bibinfo {volume}
  {113}},\ \bibinfo {pages} {023720} (\bibinfo {year} {2026})}\BibitemShut
  {NoStop}%
\bibitem [{\citenamefont {Zheng}\ \emph {et~al.}(2025)\citenamefont {Zheng},
  \citenamefont {Zheng}, \citenamefont {Hei}, \citenamefont {Qiao},
  \citenamefont {Yao}, \citenamefont {Pan}, \citenamefont {Ren}, \citenamefont
  {Huo},\ and\ \citenamefont {Li}}]{lambda_advanced_2025}%
  \BibitemOpen
  \bibfield  {author} {\bibinfo {author} {\bibfnamefont {J.-C.}\ \bibnamefont
  {Zheng}}, \bibinfo {author} {\bibfnamefont {X.-W.}\ \bibnamefont {Zheng}},
  \bibinfo {author} {\bibfnamefont {X.-L.}\ \bibnamefont {Hei}}, \bibinfo
  {author} {\bibfnamefont {Y.-F.}\ \bibnamefont {Qiao}}, \bibinfo {author}
  {\bibfnamefont {X.-Y.}\ \bibnamefont {Yao}}, \bibinfo {author} {\bibfnamefont
  {X.-F.}\ \bibnamefont {Pan}}, \bibinfo {author} {\bibfnamefont {Y.-M.}\
  \bibnamefont {Ren}}, \bibinfo {author} {\bibfnamefont {X.-W.}\ \bibnamefont
  {Huo}},\ and\ \bibinfo {author} {\bibfnamefont {P.-B.}\ \bibnamefont {Li}},\
  }\bibfield  {title} {\bibinfo {title} {Non-markovian dynamics with
  $\lambda$-type atomic systems in a single-end photonic waveguide},\ }\href
  {https://doi.org/https://doi.org/10.1002/qute.202500133} {\bibfield
  {journal} {\bibinfo  {journal} {Advanced Quantum Technologies}\ }\textbf
  {\bibinfo {volume} {8}},\ \bibinfo {pages} {2500133} (\bibinfo {year}
  {2025})}\BibitemShut {NoStop}%
\bibitem [{\citenamefont {Mazzola}\ \emph {et~al.}(2009)\citenamefont
  {Mazzola}, \citenamefont {Maniscalco}, \citenamefont {Piilo}, \citenamefont
  {Suominen},\ and\ \citenamefont {Garraway}}]{PRA_ESBESD_2009}%
  \BibitemOpen
  \bibfield  {author} {\bibinfo {author} {\bibfnamefont {L.}~\bibnamefont
  {Mazzola}}, \bibinfo {author} {\bibfnamefont {S.}~\bibnamefont {Maniscalco}},
  \bibinfo {author} {\bibfnamefont {J.}~\bibnamefont {Piilo}}, \bibinfo
  {author} {\bibfnamefont {K.-A.}\ \bibnamefont {Suominen}},\ and\ \bibinfo
  {author} {\bibfnamefont {B.~M.}\ \bibnamefont {Garraway}},\ }\bibfield
  {title} {\bibinfo {title} {Sudden death and sudden birth of entanglement in
  common structured reservoirs},\ }\href
  {https://doi.org/10.1103/PhysRevA.79.042302} {\bibfield  {journal} {\bibinfo
  {journal} {Phys. Rev. A}\ }\textbf {\bibinfo {volume} {79}},\ \bibinfo
  {pages} {042302} (\bibinfo {year} {2009})}\BibitemShut {NoStop}%
\bibitem [{\citenamefont {Dung}\ \emph {et~al.}(1998)\citenamefont {Dung},
  \citenamefont {Kn\"oll},\ and\ \citenamefont
  {Welsch}}]{Huttner1992Quantization}%
  \BibitemOpen
  \bibfield  {author} {\bibinfo {author} {\bibfnamefont {H.~T.}\ \bibnamefont
  {Dung}}, \bibinfo {author} {\bibfnamefont {L.}~\bibnamefont {Kn\"oll}},\ and\
  \bibinfo {author} {\bibfnamefont {D.-G.}\ \bibnamefont {Welsch}},\ }\bibfield
   {title} {\bibinfo {title} {Three-dimensional quantization of the
  electromagnetic field in dispersive and absorbing inhomogeneous
  dielectrics},\ }\href {https://doi.org/10.1103/PhysRevA.57.3931} {\bibfield
  {journal} {\bibinfo  {journal} {Phys. Rev. A}\ }\textbf {\bibinfo {volume}
  {57}},\ \bibinfo {pages} {3931} (\bibinfo {year} {1998})}\BibitemShut
  {NoStop}%
\bibitem [{\citenamefont {Miguel-Torcal}\ \emph {et~al.}(2025)\citenamefont
  {Miguel-Torcal}, \citenamefont {Gonz\'alez-Tudela}, \citenamefont
  {Garc\'{\i}a-Vidal},\ and\ \citenamefont
  {Fern\'andez-Dom\'{\i}nguez}}]{Vidal_PRB_fewmode_2025}%
  \BibitemOpen
  \bibfield  {author} {\bibinfo {author} {\bibfnamefont {A.}~\bibnamefont
  {Miguel-Torcal}}, \bibinfo {author} {\bibfnamefont {A.}~\bibnamefont
  {Gonz\'alez-Tudela}}, \bibinfo {author} {\bibfnamefont {F.~J.}\ \bibnamefont
  {Garc\'{\i}a-Vidal}},\ and\ \bibinfo {author} {\bibfnamefont {A.~I.}\
  \bibnamefont {Fern\'andez-Dom\'{\i}nguez}},\ }\bibfield  {title} {\bibinfo
  {title} {Photon-mediated interactions and dynamics of coherently driven
  quantum emitters in complex photonic environments},\ }\href
  {https://doi.org/10.1103/syk9-6dny} {\bibfield  {journal} {\bibinfo
  {journal} {Phys. Rev. B}\ }\textbf {\bibinfo {volume} {112}},\ \bibinfo
  {pages} {235419} (\bibinfo {year} {2025})}\BibitemShut {NoStop}%
\bibitem [{\citenamefont {Dicke}(1954)}]{Dicke_1954}%
  \BibitemOpen
  \bibfield  {author} {\bibinfo {author} {\bibfnamefont {R.~H.}\ \bibnamefont
  {Dicke}},\ }\bibfield  {title} {\bibinfo {title} {Coherence in spontaneous
  radiation processes},\ }\href {https://doi.org/10.1103/PhysRev.93.99}
  {\bibfield  {journal} {\bibinfo  {journal} {Phys. Rev.}\ }\textbf {\bibinfo
  {volume} {93}},\ \bibinfo {pages} {99} (\bibinfo {year} {1954})}\BibitemShut
  {NoStop}%
\bibitem [{\citenamefont {D\"ur}\ \emph {et~al.}(2000)\citenamefont {D\"ur},
  \citenamefont {Vidal},\ and\ \citenamefont {Cirac}}]{PRA_dicke_2000}%
  \BibitemOpen
  \bibfield  {author} {\bibinfo {author} {\bibfnamefont {W.}~\bibnamefont
  {D\"ur}}, \bibinfo {author} {\bibfnamefont {G.}~\bibnamefont {Vidal}},\ and\
  \bibinfo {author} {\bibfnamefont {J.~I.}\ \bibnamefont {Cirac}},\ }\bibfield
  {title} {\bibinfo {title} {Three qubits can be entangled in two inequivalent
  ways},\ }\href {https://doi.org/10.1103/PhysRevA.62.062314} {\bibfield
  {journal} {\bibinfo  {journal} {Phys. Rev. A}\ }\textbf {\bibinfo {volume}
  {62}},\ \bibinfo {pages} {062314} (\bibinfo {year} {2000})}\BibitemShut
  {NoStop}%
\end{thebibliography}%
\end{document}